\documentclass[11pt,a4paper]{article}

\pdfoutput=1
\usepackage{xcolor}
\definecolor{darkblue}{rgb}{0,0.1,0.5}
\usepackage{booktabs}
\usepackage{multirow,rotating}
\usepackage{amsmath,amssymb,bm,cite,graphicx,wasysym,mathrsfs,dsfont,amsbsy}
\usepackage{slashed,relsize}
\usepackage[colorlinks=true,linkcolor=darkblue,anchorcolor=black,citecolor=darkblue,filecolor=cyan,menucolor=red,runcolor=filecolor,urlcolor=darkblue,bookmarks=true,bookmarksnumbered=true]{hyperref}
\usepackage{subfigure}

\newcommand{\dl}{\delta\kern-0.6pt\kappa_3}
\newcommand{\dlq}{\delta\kern-0.6pt\kappa_4}
\def\lra#1{\overset{\text{\scriptsize$\leftrightarrow$}}{#1}}

\newcommand{\llll}{\ell^+\ell^-}
\newcommand{\baslum}{\widehat{{\mathcal{L}}}}
\def\lra#1{\overset{\text{\scriptsize$\leftrightarrow$}}{#1}}
\newcommand{\lyxdot}{.}

\newcommand{\ww}{W^+W^-}
\newcommand{\h}{h}
\newcommand{\OH}{\mathcal{O}_H}
\newcommand{\Osix}{\mathcal{O}_6}
\renewcommand{\O}{\mathcal{O}}

\title{\bf Two Paths Towards Precision at a \\ Very High Energy Lepton Collider}

\date{}

\author{
Dario Buttazzo$^{a}$, Roberto Franceschini$^{b}$,  Andrea Wulzer$^{c,d,e}$ \\ \\
{\small\emph{$^a$ INFN, Sezione di Pisa, Largo Bruno Pontecorvo 3, I-56127 Pisa, Italy}}\\
{\small\emph{$^b$ Universit\`a degli Studi and INFN Roma Tre, Via della Vasca Navale 84, I-00146, Rome}}\\
{\small\emph{$^c$ CERN, 1211 Geneva 23, Switzerland}}\\
{\small\emph{$^d$ Theoretical Particle Physics Laboratory (LPTP), Institute of Physics,}}\\
{\small\emph{EPFL, Lausanne, Switzerland}}\\
{\small\emph{$^e$ Dipartimento di Fisica e Astronomia, Universit\`a di Padova, Italy}}\\
}
\begin{document}
\baselineskip=13pt
\begin{flushright}
CERN-TH-2020-216
\end{flushright}
\vspace{2em}
{\let\newpage\relax\maketitle}
\begin{abstract}
We illustrate the potential of a very high energy lepton collider (from $10$ to $30$~TeV center of mass energy) to explore new physics indirectly in the vector boson fusion double Higgs production process and in direct diboson production at high energy. Double Higgs production is found to be sensitive to the anomalous Higgs trilinear coupling at the percent level, and to the Higgs compositeness $\xi$ parameter at the per mille or sub-per mille level thanks to the measurement of the cross-section in the di-Higgs high invariant mass tail. High energy diboson (and tri-boson) production is sensitive to Higgs-lepton contact interaction operators at a scale of several tens or hundred TeV, corresponding to a reach on the Higgs compositeness scale well above the one of any other future collider project currently under discussion. This result follows from the unique capability of the very high energy lepton collider to measure Electroweak cross-sections at $10$~TeV energy or more, where the effect of new physics at even higher energy is amplified. The general lesson is that the standard path towards precision physics, based on measurements of high-statistics processes such as single and double Higgs production, is accompanied at the very high energy lepton collider by a second strategy based on measurements at the highest available energy. 
\end{abstract}

\newpage

\begingroup
\tableofcontents
\endgroup 

\setcounter{equation}{0}
\setcounter{footnote}{0}
\setcounter{page}{1}

\newpage
\section{Introduction}

A lepton collider operating at a center of mass energy of $10$~TeV or more is currently not technologically feasible. However, such machine might exist in the future in the form of a circular $\mu^+\mu^-$~\cite{Delahaye:2019omf} or of a linear $e^+e^-$ collider based on plasma wake field acceleration~\cite{ALEGRO:2019alc}. It is not our purpose to discuss the technological virtues and the limitations of the different proposals. It suffices here to say that an intense R{\&}D activity is foreseen in the next few years to assess their viability. For the muon collider, this will be performed in the context of the newly-formed international muon collider collaboration~\cite{muon}. A first look at the physics potential of this hypothetical machine is required already at this preliminary stage in order to motivate the R{\&}D effort, and to orient it according to the physics needs.  

Basic considerations~\cite{Delahaye:2019omf,ALEGRO:2019alc} lead to a preliminary target for the total integrated luminosity 
\begin{equation}\label{lumieq}
\baslum=10\,{\rm{ab}}^{-1}\left(\frac{E_{\rm{cm}}}{10\,{\rm{TeV}}}\right)^2\,,
\end{equation}
which we consider as the baseline for our studies. Notice that the luminosity scales like the square of the energy, in order to compensate for the geometric $1/E_{\rm{cm}}^2$ scaling of the $2\rightarrow2$ cross-section. Three energy benchmarks are considered: $E_{\rm{cm}}=10$, $14$ and $30$~TeV. The energies and luminosities above define the Very High Energy Lepton collider (VHEL) which is the subject of the present paper. 

Our results will be independent of the nature of the colliding leptons, because they are based on simple leading-order predictions without Initial State QED Radiation (ISR) and with monochromatic beams. However it should be emphasized that the effect of ISR and the departure from beam monochromaticity (due to Beamstrahlung) are expected to be very significant in the case of electrons and to be small or negligible for muons. The reduction of the luminosity that is effectively available for collisions at the nominal collider energy, which is impossible to quantify at the current stage, should be taken into account at an $e^+e^-$ VHEL. Beam-Induced Background (BIB) is the other important aspect in which $e^+e^-$ and $\mu^+\mu^-$ colliders differ significantly. Based on the studies performed for CLIC (see~\cite{Robson:2018enq}, and references therein, for a recent summary), at an $e^+e^-$ VHEL it should be possible to cope with the BIB and to obtain good detector performances on high-level objects, superior or comparable with those of current LHC experiments. The situation is less clear for a muon collider, where the BIB emerging from the decay of the muons is copious and requires new mitigation strategies. Being able to deal with the BIB is a pre-requisite for the feasibility of the project, and preliminary results are encouraging~\cite{Bartosik:2020xwr,Bartosik:2019dzq}. Therefore, in what follows we assume CLIC-like detector performances in our estimates. However some elements that are specific of the muon collider detector and that are different from CLIC, such as a reduction of the acceptance along the beam axis and a possible degradation of the reconstruction performances for low-$p_T$ objects, will be taken into account when relevant.

\begin{figure}[t]
\begin{center}
\includegraphics[height=.355\textwidth]{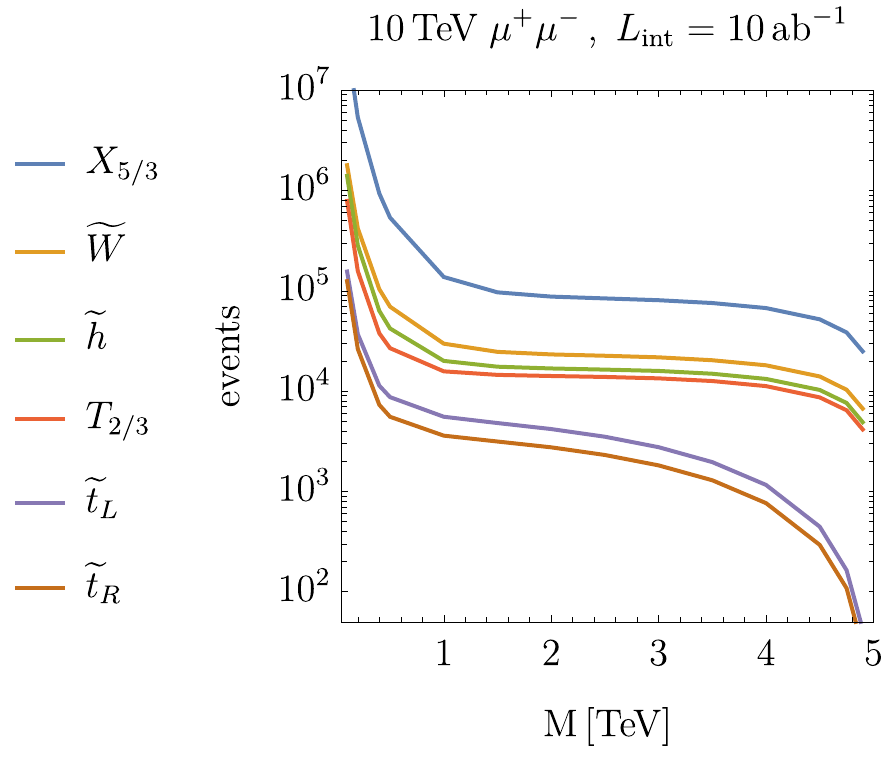}\hfill
\includegraphics[height=.355\textwidth]{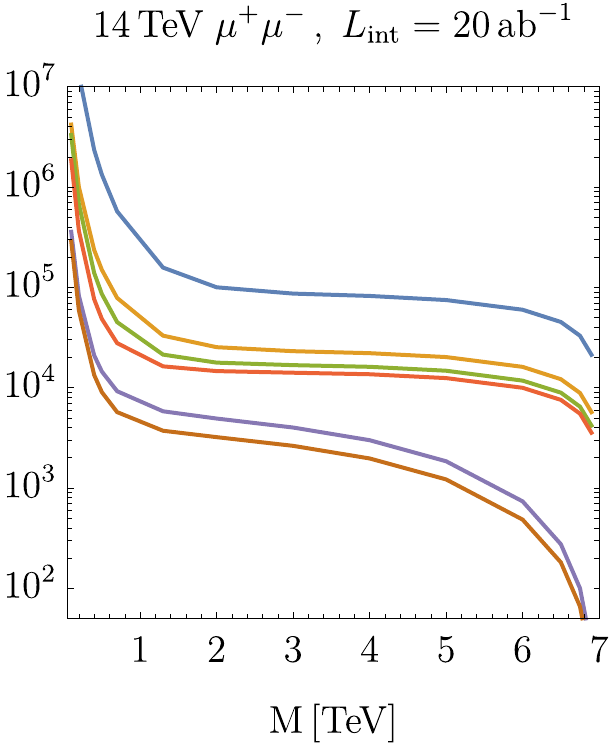}\hfill
\includegraphics[height=.355\textwidth]{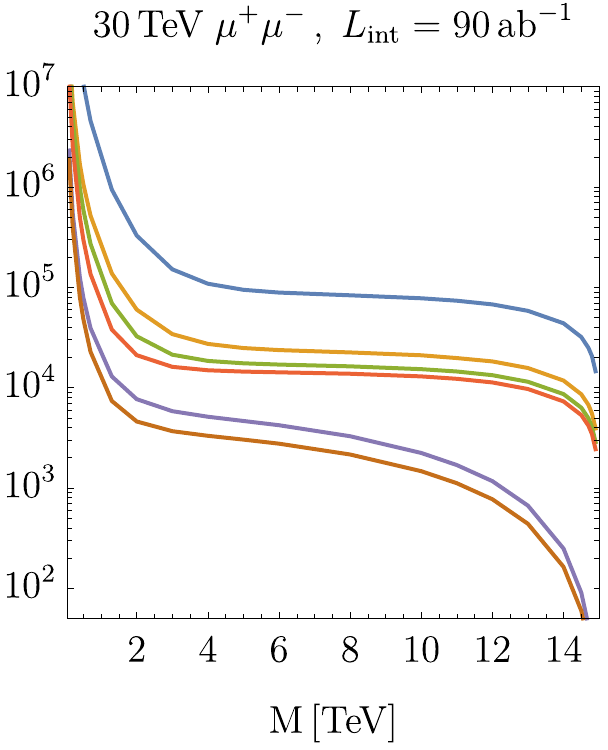}
\caption{\small Number of EW pair-production events, computed with {\sc{MadGraph}}~\cite{Alwall:2014hca}, using the Effective Photon Approximation for the calculation of the neutral VBF production cross-section. Namely, neutral VBF is evaluated as the sum of the $4$ subprocess initiated by $\ell^+\ell^-$, $\ell^+\gamma$, $\gamma \ell^-$, and $\gamma\gamma$, with a $\sqrt{-Q^2}>30$~GeV cut on the virtual photons and the corresponding $Q_{\rm{max}}=30$~GeV cutoff in the photon distribution function. The photon distribution function is the one for muons. The neutral VBF cross-section would thus be larger than what shown in the figure at the $e^+e^-$ VHEL because of the smaller electron mass.\label{NEV}}
\end{center}
\end{figure}

The direct observation of new heavy particles is one of the main physics drivers of the VHEL. While only few concrete reach projections are available (see Ref.s~\cite{Buttazzo:2018qqp,Ruhdorfer:2019utl,Han:2020uak,Capdevilla:2020qel,Chakrabarty:2014pja}), the case for direct searches is straightforward and it can be illustrated by simple plots like the ones in Figure~\ref{NEV}. The figure shows the number of pair-produced hypothetical new particles ``$P$'' as a function of the mass, for the baseline VHEL energies and luminosities. All the pair-production processes induced by Electroweak (EW) interactions are considered in the figure. Namely, the plots include $s$-channel $\ell^+\ell^-\to P\overline{P}$ production plus charged ($W^+W^-\to P\overline{P}$) and neutral ($Z/\gamma\,Z/\gamma\to P\overline{P}$) Vector-Boson Fusion (VBF) production. The particles are labeled with a standard Beyond-the-SM (BSM) terminology, however only gauge interactions are taken into account in the cross-section calculation. Specifically, the ``left-handed stop'' ${\tilde{t}}_L$ is modeled as a scalar degenerate doublet of SU$(2)_L$ with $1/6$ Hypercharge (and in the ${\bf{3}}$ of SU$(3)_c$), and similarly for the other particles. The VBF cross-section is relevant only at very low masses, but it can be enhanced by other (non-gauge) interactions that might be present in specific BSM models~\cite{Costantini:2020stv}. We see that the statistics is sufficient to discover all particles up around the collider mass-threshold $E_{\rm{cm}}/2$, provided they decay to energetic and easily detectable SM particles. By comparing with the reach projections of other future collider projects (see~\cite{Strategy:2019vxc}), this simple plot is sufficient to qualify as striking the direct discovery potential of the VHEL, especially for $E_{\rm{cm}}\geq14$~TeV. On the other hand, detailed detector-level studies including BIB mitigation strategies are compulsory to assess the observability of BSM particles decaying to soft objects (because of, e.g., a compressed spectrum), or displaying disappearing tracks signatures like the Higgsino/Wino ($\widetilde{h}$/$\widetilde{W}$) Minimal Dark Matter candidates. Ref.~\cite{DiLuzio:2018jwd} studied the possibility of observing these candidates indirectly through their radiative effects, bypassing these complications and in some case extending the reach above the mass-threshold. The reach of mono-photon searches has been also studied~\cite{Han:2020uak}.

The VHEL potential for indirect new physics discoveries is equally or perhaps even more striking that the direct one, but it is slightly less trivial to assess and to illustrate. The present paper aims at outlining the elements for this assessment, based on selected sensitivity estimates. 

The indirect physics potential emerges from the combination of two items. The first one is that indirect effects of heavy new physics effects are generically more pronounced on processes that take place at higher energy, i.e.\ closer to the new physics scale. In the Effective Field Theory (EFT) description this is merely the observation that the corrections from operators of dimension larger than $4$ can grow polynomially with the energy. The luminosity benchmark in eq.~(\ref{lumieq}) generically allows for measurements of $2\rightarrow2$ short-distance electroweak scattering processes with percent or few-percent (i.e., moderate) precision.  Still, a dimension-$6$ EFT operator displaying quadratic energy growth, inducing relative corrections to the SM of order $E_{\rm{cm}}^2/\Lambda^2$, could be probed at the VHEL with $E_{\rm{cm}}\geq10$~TeV for an effective interaction scale $\Lambda$ in the ballpark of $100$~TeV (see also~\cite{Buttazzo:2020eyl}). On a process occurring at the EW scale, of $100$~GeV, $\Lambda\sim100$~TeV would instead contribute as an unobservable $O(10^{-6})$ relative correction. The power of precision probes based on high-energy cross-section measurements has been outlined extensively in the context of CLIC studies~\cite{Ellis:2017kfi,deBlas:2018mhx}. They make, for instance, the highest energy stage of CLIC superior or comparable to the other future colliders project on physics targets such as Higgs and Top compositeness~\cite{Strategy:2019vxc}. By rescaling the highest CLIC available energy, of $3$~TeV, to the lowest VHEL energy of $10$~TeV, we immediately conclude that the VHEL  performances are expected to be vastly superior to those of any other project currently under discussion. 

\begin{figure}[t]
\begin{center}
\includegraphics[width=.5\textwidth]{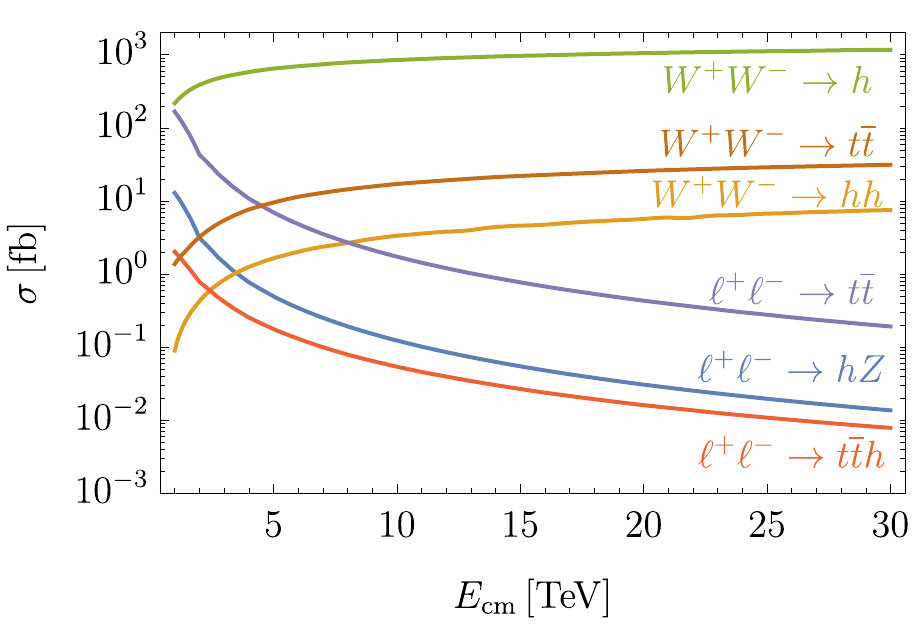}\hfill%
\includegraphics[width=.49\textwidth]{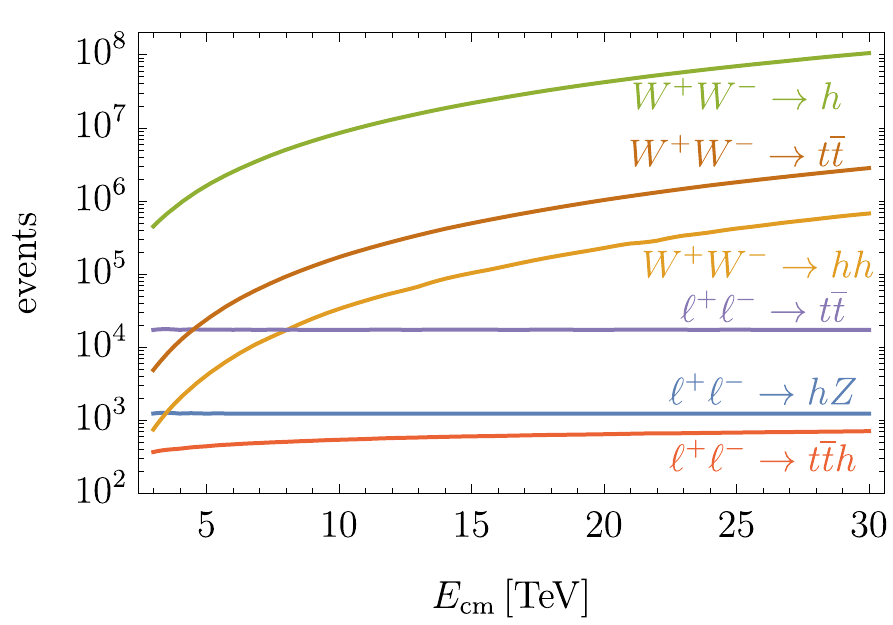}
\caption{\small Cross-sections and total number of expected events, using eq.~(\ref{lumieq}), for selected SM processes. }\label{SM}
\end{center}
\end{figure}

High-energy probes are the first of the two paths towards precision to be explored for the assessment of the VHEL physics potential. It is unique of the VHEL, because of the high collider energy and because the nominal collider energy is entirely available to produce short-distance reactions, unlike for hadron colliders due to the shape of the parton distribution functions. The second path is the more standard approach to precision, based on very accurate measurements of processes with high statistics. There are several high-rate processes at the VHEL, eminently those that proceed through $\ell\rightarrow \ell' V$ collinear splittings, with $V=W,Z,\gamma$ a SM vector boson, followed by a scattering or production process $VV\rightarrow{X}$ occurring at the EW scale. Relevant examples are the VBF production of a single Higgs or of a pair of Higgs bosons. Since all the reactions involved take place at a fixed scale (the EW one) which is much smaller than the collider energy $E_{\rm{cm}}$, the cross-section for the VBF processes is very large and nearly constant with energy up to a mild logarithmic growth. The total number of collected events thus grows quadratically with $E_{\rm{cm}}$ following the luminosity, as Figure~\ref{SM} shows. 

It should be emphasized that there is no direct competition between the high-energy and the high-rate paths towards new physics. Namely, any new physics effect (or, EFT operator) that grows with the energy in a measurable $2\to2$ process is unmistakably probed way more effectively at high energy than in any high-rate VBF process. This is because, as previously mentioned, the reach of the high-energy probes corresponds to $O(10^{-6})$ effect at the EW scale, which is where the high-rate VBF processes take place. On the other hand, not all the EFT operators induce measurable growing-with-energy effects. High-rate probes are thus sensitive to other operators and complementary to the high-energy ones. Furthermore the sensitivity of the high-energy probes is quantified under the assumption that the new physics scale (i.e., the EFT cutoff) is above the collider energy, while the high-rate probes only rely on the EFT validity at or slightly above the EW scale. If the new physics scale is in between, high-rate probes will play a crucial role in the characterization of new physics, together with the high-energy ones and with the direct production of the new states.

The general considerations above are illustrated in the rest of the paper through examples. The pair production of vector or Higgs bosons is studied in Section~\ref{highen} as a high-energy probe for two EFT operators ($\O_W$ and $\O_B$) that grow quadratically with the energy in the diboson processes. We also consider the production of three bosons, which is found to contribute significantly owing to the IR enhancement of soft-collinear massive vector boson emissions. The sensitivity of diboson (and tri-boson) measurements to the operators Wilson coefficients is quantified and compared with the one of other future collider projects. The projected sensitivity is also employed for a first assessment of the VHEL reach on Higgs compositeness. 

\newpage

In light of Figure~\ref{SM}, it is tempting to consider VBF single-Higgs production, and the corresponding projections on precision Higgs couplings measurements, as an illustration of the high-rate potential. However the single-Higgs statistics is so high (even after acceptance and selection cuts~\cite{Han:2020pif}) that systematic and theoretical uncertainties definitely play the dominant role in the assessment of the anomalous Higgs couplings sensitivity. No conclusive evaluation of the experimental systematic uncertainties is currently possible, and a careful investigation of the theoretical uncertainties in the SM predictions and of their impact goes beyond the scope of the present paper. At present we can only conclude that the high single-Higgs statistics enables, in principle, VHEL Higgs coupling measurements at or below the per mille level. Such per mille accuracy, which matches the projections of proposed future Higgs factory, will be taken as reference for semi-quantitative comparisons. On the other hand, for the determination of small couplings such as the one to muons, or for the search of exotic Higgs decays, systematic uncertainties play a minor role and the sensitivity could be realistically estimated on purely statistical bases.

Rather than single Higgs, we consider VBF double Higgs production as an illustration of the high-rate path towards new physics. This process is a good target because the number of events is considerable, but not so large to invalidate statistical sensitivity estimates. Furthermore it is sensitive to new physics effects that do not induce any growth in $2\to2$ processes, hence it does not compete with high-energy probes. One such effect is the anomalous trilinear Higgs coupling $\dl$, which is a standard target for future colliders. The VHEL sensitivity to $\dl$ is estimated in Section~\ref{highrate} and compared with other projects. See~\cite{Costantini:2020stv,Han:2020pif,Chiesa:2020awd} for recent VHEL studies.

While it is useful to distinguish high-energy from high-rate probes, the separation between the two categories is not sharp. Moreover, processes occurring at moderately high energy and with moderately high rate can be also powerful probes of new physics. This is shown in Section~\ref{inbetween} by studying double-Higgs production in the high (TeV-scale) di-Higgs invariant mass tail, which is sensitive to a contact interaction (the $\O_H$ operator) that grows with the energy in the $VV\to hh$ amplitude. The sensitivity to $\O_H$ is compared with the one of single Higgs couplings measurements at Higgs factories, and its impact on Higgs compositeness quantified. 

Finally, a summary of our results, a first assessment of the VHEL potential on precision physics, and future directions of investigation, are discussed in Section~\ref{conc}.

\section{High-energy diboson production}\label{highen}

We consider the direct $2\to2$ production of a pair of SM (vector or Higgs) bosons, and we restrict our attention to BSM effects that grow quadratically with the energy in the zero-helicity (longitudinal polarization) scattering amplitudes.\footnote{Quadratic energy growth in the transverse polarizations could be also studied. However the effects on the longitudinal vectors (and Higgs) amplitudes are directly connected with the Higgs sector, and thus more relevant to probe BSM scenarios such as Composite Higgs.} Following~\cite{Franceschini:2017xkh}, these effects are fully characterized by three ``high-energy primary'' parameters, which are in one-to-one correspondence with the Warsaw-basis \cite{Grzadkowski:2010es} operator coefficients $G_{3L}$, $G_{1L}$ and $G_{lR}$. The growing-with-energy BSM contributions to the different amplitudes are reported in Table~\ref{primaries}, for operators defined as 
\begin{eqnarray}\label{ci}
& {\mathcal{O}}_{3L}=
\left(\bar{{\rm{L}}}_{L}\gamma^{\mu}\sigma^{a}{\rm{L}}_{L}\right)(i H^\dagger\sigma^a \lra D_\mu   H)\,,\;\;\;\;\;
 {\mathcal{O}}_{1L}=
\left(\bar{{\rm{L}}}_{L}\gamma^{\mu}{\rm{L}}_{L}\right)(i H^\dagger \lra D_\mu   H)\,,&\nonumber\\
& {\mathcal{O}}_{lR}=\left(\bar{l}_{R}\gamma^{\mu}l_{R}\right)(i H^\dagger \lra D_\mu   H)\,.&\label{Wop}
\end{eqnarray}
Strictly speaking, the only processes reported in the table that can be measured at the VHEL at the highest available energy $\sqrt{s}=E_{\rm{cm}}$ are the ones initiated by charged leptons $\ell=\mu,e$. However, neutrino-initiated processes can also be effectively probed, at a comparable energy, through the IR-enhanced emission of soft $W$ bosons from the charged initial leptons. The charged-current $\ell\nu\to Wh$ process is discussed in Section~\ref{subleading-processes} as an illustration of this mechanism.

A particularly interesting two-dimensional slice of the high-energy primaries parameter space is the one populated by Universal~\cite{Wells:2015uba} BSM models, in which the heavy particles couple only to the SM Higgs and vector bosons. The lepton currents appearing in the operators of eq.~(\ref{Wop}) are thus generated ``indirectly'', through the SM gauge couplings (i.e., by using the equations of motion of the $W$ and $B$ gauge fields), out of operators that do not contain lepton fields. Since the $B$ field coupling to right-handed leptons is twice the one to left-handed leptons, the ${\mathcal{O}}_{lR}$ operator coefficient is related to the one of ${\mathcal{O}}_{1L}$ by $G_{lR}=2\,G_{1L}$. 

There are four Universal SILH-basis~\cite{Giudice:2007fh} operators, namely ${\mathcal{O}}_{W}$,  ${\mathcal{O}}_{B}$, ${\mathcal{O}}_{HW}$ and ${\mathcal{O}}_{HB}$, that generate the operators in eq.~(\ref{Wop}) by the equations of motion. The Warsaw-basis coefficients read 
\begin{equation}\label{Gs}
G_{3L}=\frac{g^{2}}{4}\left(C_{W}+C_{HW}\right)\,,\;\;\;\;\;G_{1L}=\frac{g^{\prime2}}{4}\left(C_{B}+C_{HB}\right)=\frac12 G_{lR} \,,
\end{equation}
where $C_{(H)W,B}$ are the (dimensionful) coefficients of the ${\mathcal{O}}_{(H)W,B}$ operators defined as in~Table~\ref{primaries}. Our analysis of growing-with-energy effects in dibosons will thus be sensitive only to two linear combinations of the four SILH operators. However since $C_{HW,HB}$ are small in Composite Higgs models, in what follows we set them to zero and illustrate the sensitivity in terms of the $C_W$ and $C_B$ parameters.

In Universal theories, the two parameters combinations $C_{W}+C_{HW}$ and $C_{B}+C_{HB}$ also control other interactions, generated by equations of motion, analog to eq.~(\ref{Wop}) but involving quarks rather than leptons. The latter interactions induce growing-with-energy effects in diboson production at hadron colliders, that can be probed at the HL-LHC and at the FCC-hh~\cite{Franceschini:2017xkh}. This enables a comprehensive comparison of the VHEL sensitivity with the reach (see~\cite{deBlas:2019rxi}) of all the other (hadronic or leptonic) future collider projects. Let us consider for definiteness the single-operator reach on $C_W$. The $1\sigma$ sensitivity is $C_{W,\,1\sigma}^{\text{HL-LHC}}=1/(6.7\,{\rm{TeV}})^2$ at the HL-LHC, $C_{W,\,1\sigma}^{\text{FCC}}=1/(19\,{\rm{TeV}})^2$ after the full FCC program, and $C_{W,\,1\sigma}^{\text{CLIC}}=1/(26\,{\rm{TeV}})^2$ at CLIC. The CLIC sensitivity is driven by high-energy diboson measurements performed at the highest available CLIC center of mass energy of $3~{\rm{TeV}}$~\cite{Ellis:2017kfi}. The FCC reach benefits from high-energy probes in the diboson final state at the FCC-hh, but it is dominated by the FCC-ee accurate measurements of  $Z$ pole and other EW-scale observables. The reach of FCC-ee alone is $C_{W,\,1\sigma}^{\text{FCCee}}=1/(17\,{\rm{TeV}})^2$.

\begin{table}[t]
\begin{center}
\begin{tabular}{c|c}
Process & BSM Amplitude  \\\hline
\rule[-.6em]{0pt}{1.7em}$ \ell_L^+ \ell_L^-\to Z_0 h$& \multirow{ 2}{*}{$\displaystyle{{s}\left(G_{3L}+G_{1L}\right)\sin \theta_\star}$}\\
\rule[-.55em]{0pt}{1.45em}$\bar \nu_L \nu_L\to W_0^+W_0^-$&\\
\hline
\rule[-.6em]{0pt}{1.7em}$ \ell^+_L \ell^-_L\to W_0^+W_0^-$& \multirow{ 2}{*}{$\displaystyle{{s}\left(G_{3L}-G_{1L}\right)\sin \theta_\star}$}\\
\rule[-.55em]{0pt}{1.45em}$\bar \nu_L \nu_L\to Z_0h$&  \\
\hline
\rule[-.7em]{0pt}{1.9em}$\ell^+_R \ell^-_R\to W_0^+W_0^-,Z_0h$& $\displaystyle{{s}\,G_{lR}\sin \theta_\star}$\\
\hline
\rule[-.6em]{0pt}{1.7em}$\bar \nu_L \ell^-_L\to W_0^- Z_0\,/\,W_0^- h$  & \multirow{ 2}{*}{ $\displaystyle{\sqrt{2}\,s\, G_{3L}\sin \theta_\star}$ }\\
\rule[-.55em]{0pt}{1.45em}$\nu_L \ell^+_L\to W_0^+ Z_0\,/\,W_0^+ h$  \\
 \end{tabular}\hspace{40pt}
 \begin{tabular}{|c|}
\hline
SILH Operators \\
\hline
\rule[-1.2em]{0pt}{3em}$\displaystyle{\cal O}_W=\frac{ig}{2}\left( H^\dagger  \sigma^a \lra {D^\mu} H \right )D^\nu  W_{\mu \nu}^a$\\
\rule[-1.2em]{0pt}{3em}$\displaystyle{\cal O}_B=\frac{ig'}{2}\left( H^\dagger  \lra {D^\mu} H \right )\partial^\nu  B_{\mu \nu}$\\
\rule[-.6em]{0pt}{2em}$\displaystyle{\cal O}_{HW}=i g(D^\mu H)^\dagger\sigma^a(D^\nu H)W^a_{\mu\nu}$\\
\rule[-.6em]{0pt}{2em}$\displaystyle{\cal O}_{HB}=i g'(D^\mu H)^\dagger(D^\nu H)B_{\mu\nu}$\\
\hline
 \end{tabular}
  \caption{Left: BSM contributions to diboson production amplitudes that grow with energy. The center of mass energy and scattering angle are denoted as $\sqrt{s}$ and $ \theta_\star$. Right: the relevant SILH basis operators.}
\label{primaries}
\end{center}
\end{table}

It should be emphasized that FCC-ee can be sensitive to such small values of $C_W$ only because of the extreme accuracy of its measurements and of the SM theoretical predictions that are needed to identify the tiny BSM effects due to $C_W$. For EW-scale observables, the relative magnitude of these effects is quantified by the ``hatted'' $S$ parameter~\cite{Barbieri:2004qk} \footnote{The tree-level expression for ${\widehat{S}}$ given below receives large radiative RG-running correction if the EFT scale is as high as tents of TeVs. Nevertheless, it is a valid semi-quantitative estimate of the size of the correction to SM observables at the EW scale.}
\begin{equation}\label{shat}
{\widehat{S}}=m_W^2(C_W+C_B)\,.
\end{equation}
The FCC reach on $C_W$ corresponds (for $C_B=0$) to ${\widehat{S}}={\widehat{S}}_{1\sigma}^{\text{FCC}}=2.2\times10^{-5}$, i.e.\ to measurements and theoretical predictions at the level of $10^{-5}$. This level of accuracy can be considered as the ultimate accuracy for EW-scale measurements. Correspondingly, the FCC reach on $C_W$ can be regarded as the ultimate sensitivity to this operator that can be obtained by high-rate probes at low energy with high precision. High energy probes performed at the VHEL will easily pass this threshold.

\subsection[High-energy ${Zh}$]{High-energy $\mathbf{Zh}$}\label{zh}

We consider the direct $2\to2$ ``Higgs-strahlung'' $Zh$ production. At the high VHEL energies, the process is conveniently described, using the Goldstone Boson Equivalence Theorem, as the production of one Higgs and of one neutral Goldstone boson that corresponds to the longitudinally polarized $Z$ boson. The process is mediated in the SM by a virtual $Z$ or photon, that couples to the Higgs doublet via the regular gauge interaction vertex. The SM amplitude, as well as the BSM contribution in Table~\ref{primaries}, is proportional to the sine of the scattering angle $ \theta_\star$. The amplitude for producing a $Z$ boson with transverse polarization is suppressed by $m_Z/E_{\rm{cm}}$, and thus completely negligible at the VHEL energies. Similarly, the EFT operators we are studying do not produce growing-with-energy effects in the transverse $Z$ boson production amplitudes. 

The considerations above make the phenomenological analysis of growing-with energy effects in $Zh$ a rather trivial task. The dependence on $ \theta_\star$ is the same for the SM and for the BSM contributions, therefore measuring $ \theta_\star$ does not bring any additional discriminating power between the SM and the BSM hypotheses. Since the only non-vanishing amplitude is for longitudinal $Z$ bosons, the angular distributions of the $Z$ (and $h$) decay products are also identical and not worth measuring. All the information about the presence of the BSM effects is thus captured by the measurement of the total high-energy cross-section. Performing such measurement is not extremely challenging because the final states are central and the $Zh$ invariant mass is close to the collider center of mass energy. The background in this kinematical regime emerges from EW $2\to2$ (central) production processes, whose typical cross-section is comparable to the one of the signal.\footnote{The largest cross-section in this regime, which constitutes a background for fully-hadronic $Zh$, is the one for $\llll\to {\overline{q}}q$. Summing over all quarks, this is only $60$ times larger than the signal and it can be vastly reduced by $b$-tagging, jet masses and substructure cuts~\cite{Leogrande:2019dzm}.} 

The perspectives for measuring the high-energy $Zh$ cross-section at the $3$~TeV CLIC, based on CLICdp full detector simulation, has been studied in~\cite{Leogrande:2019dzm} for hadronically decaying $Z$ and $h\to{b}\overline{b}$. Based on this study we expect that it should be possible to eliminate the backgrounds with selection cuts that preserve a considerable fraction of the signal. A total signal efficiency $\epsilon_{{Zh}}=26\%$, including decay branching ratios, is considered in what follows for the estimate of the statistical accuracy of the measurement. Provided the VHEL detector performances are as good as the ones of CLIC, ours is most likely a conservative estimate. Other Higgs decay final states could indeed be included, as well as the channels with leptonically-decaying $Z$ (where the background is lower), with the potential of improving $\epsilon_{{Zh}}$ significantly. In the Conclusions we further comment on the additional studies that are required for a conclusive assessment of the VHEL measurement potential. We will see that they also include an assessment of the impact of soft vector bosons emissions that is more relevant at the VHEL than at CLIC.

For a first illustration of the VHEL sensitivity we focus on the $C_W$ operator, setting $C_B=0$. Up to negligible corrections of order $m_Z^2/E_{\rm{cm}}^2$, the cross-section reads
\begin{equation}\label{xszh}
\sigma(\llll\to Zh)=1220\left[1+\left(\frac{E_{\rm{cm}}}{0.78}\right)^2C_W+\left(\frac{E_{\rm{cm}}}{0.96}\right)^4{C_W}^2\right]\times\left(\frac{10\,{\rm{TeV}}}{E_{\rm{cm}}}\right)^2 0.1\,{\rm{ab}}\,,
\end{equation}
and it corresponds to $1220$ $Zh$ SM events with the baseline integrated luminosity in eq.~(\ref{lumieq}). Notice that the number of expected events is independent of the VHEL energy because the $E_{\rm{cm}}^2$ scaling of the baseline luminosity compensates the $1/E_{\rm{cm}}^2$ of the SM cross-section. The relative effect of $C_W$ grows instead with the energy. The growth is quadratic in the interference (linear in $C_W$) term, owing to eq.~(\ref{primaries}), and quartic in the $C_W^2$ contribution. 

In the vicinity of the SM point, the number of observable $Zh$ events is $1220\times \epsilon_{{Zh}}=317$, which corresponds to a statistical relative uncertainty $\delta=5.6\%$ in the $Zh$ cross-section measurement. This is a good estimate of the total error in the reasonable assumption that the experimental systematic uncertainties could be brought at or below the percent. Assuming that the theoretical errors in the SM prediction will be also irrelevant, and retaining only the interference term in the BSM contribution, we estimate the $1\sigma$ sensitivity reach as $C_{W,\,1\sigma}^{\text{VHEL}}=\delta\times0.78^2/E_{\rm{cm}}^2=1/(5.4\,E_{\rm{cm}})^2$. The sensitivity is much better than the one of the other future colliders already at the lowest VHEL energy $E_{\rm{cm}}=10$~TeV. In terms of the low-energy ${\widehat{S}}$ parameter in eq.~(\ref{xszh}), the $10$~TeV VHEL sensitivity is ${\widehat{S}}_{1\sigma}^{\text{VHEL}_{10}}=2.2\times10^{-6}$, way below what the FCC-ee or any other future project could conceivably achieve with accurate measurements of EW-scale observables. In the Composite Higgs scenario~\cite{Giudice:2007fh}, $C_W$ is of order $1/m_*^2$, where $m_*$ is the Higgs compositeness scale (i.e., the inverse of the Higgs boson geometric size). The VHEL can thus reach $m_*=5.4\,E_{\rm{cm}}$ at one $\sigma$, $m_*=3.8\,E_{\rm{cm}}$ for a $2\sigma$ exclusion and $m_*=2.4\,E_{\rm{cm}}$ for a discovery.

\begin{figure}
\centering{}\includegraphics[width=0.6\linewidth]{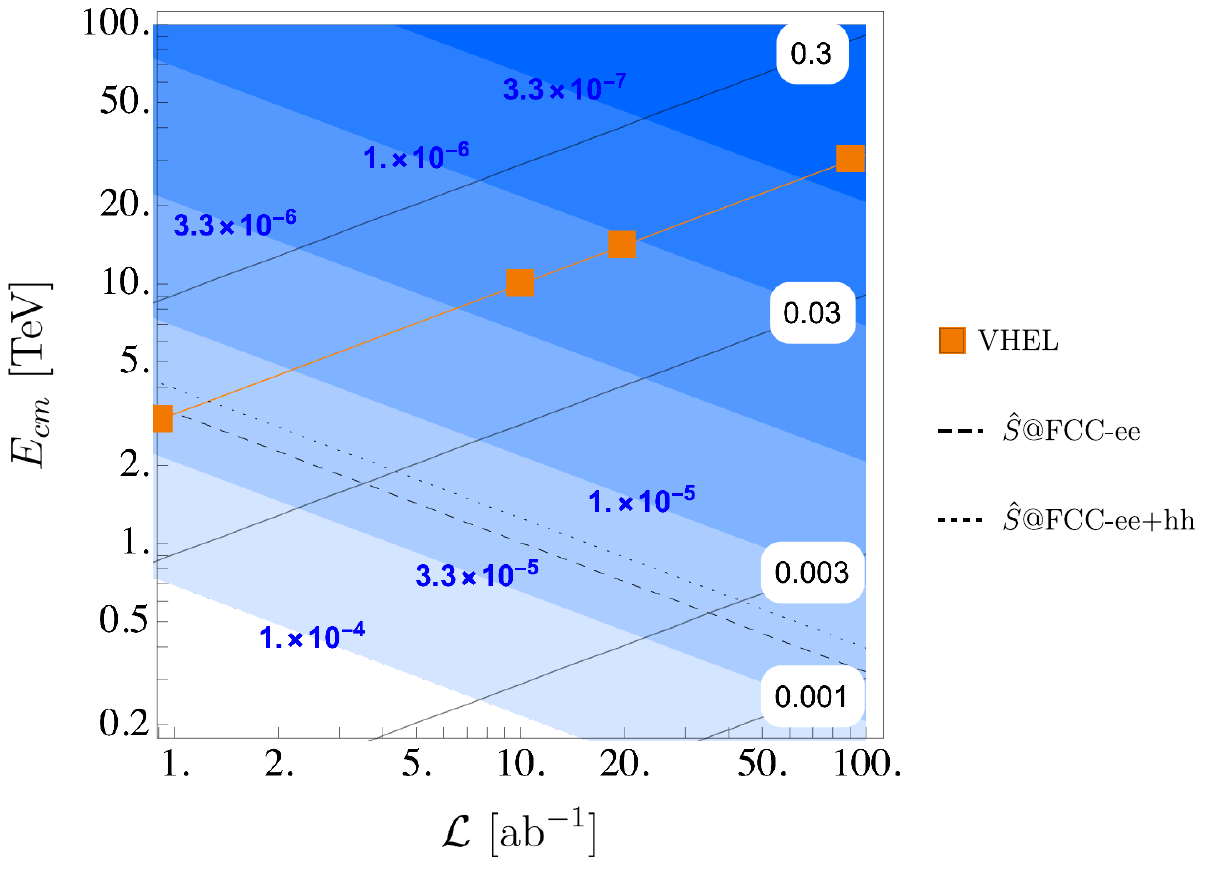}
\caption{\label{fig:Comparison-of-bounds-ZHvsShat} Iso-contours of the $1\sigma$ sensitivity to $C_W$, in terms of the ${\widehat{S}}$ parameter in eq.~(\ref{shat}), in the luminosity-energy plane. The FCC reach is also indicated, as well as the iso-contours of the relative statistical uncertainty of the cross-section measurement.}
\end{figure}

If the integrated luminosity $\mathcal{L}$ is varied around the baseline $\baslum$, the sensitivity scales as
\begin{equation}
C_{W,\,1\sigma}^{\text{VHEL}} = 3.4\times10^{-4}\,{\text{TeV}}^{-2}\left({\baslum}/{\mathcal{L}}\right)^{1/2}\left(\frac{10\,\rm{TeV}}{E_{\rm{cm}}}\right)^2\,.
\end{equation}
The reach iso-contours are shown in Fig.~\ref{fig:Comparison-of-bounds-ZHvsShat} in terms of ${\widehat{S}}$ in the luminosity-energy plane. The contour associated with the FCCee reach is outlined with a black dashed line, showing the superiority of the VHEL sensitivity. The orange line represents the baseline energy-luminosity relation in eq.~(\ref{lumieq}). The black lines are instead the contours of the statistical uncertainties of the measurement, which is equal to $\delta\times({\baslum}/{\mathcal{L}})^{1/2}$ with $\delta=5.6\%$ being the uncertainty on the baseline luminosity line. Increasing the luminosity would be beneficial for the reach until the statistical uncertainty reaches the threshold of the systematic and theoretical uncertainties. If we tentatively set this floor to $1\%$, the sensitivity could improve by a factor up to $5$ with a $25$ times larger luminosity. 

The sensitivity would still be way better than the one of the FCCee if the luminosity was significantly smaller than the baseline. However our results should not be applied blindly if the luminosity is small. At the purely technical level, this is because they neglect the quadratic $C_W^2$ term in the cross-section prediction~(\ref{xszh}). This is a good approximation only if the uncertainty of the cross-section measurement is low such that the reach on $C_W$ is significantly below $1/E_{\rm{cm}}^2$. Otherwise, as eq.~(\ref{xszh}) shows, the quadratic term becomes competitive with the linear one and affects the reach significantly. 

Physically, the problem with extremely inaccurate measurements (as they would result from a very low VHEL luminosity) is that they cannot be interpreted consistently in the EFT language in the case of underlying BSM theories, like Composite Higgs, where $C_W=1/m_*^2$ is directly related to the mass $m_*$ of new BSM particles. An order-one measurement uncertainty would indeed result in a sensitivity $m_*\sim E_{\rm{cm}}$, while the validity of the EFT for the VHEL cross-section prediction relies on a scale separation between the EFT cutoff $m_*$ and the center of mass energy $E_{\rm{cm}}$. In this situation, the prediction should be performed in the underlying BSM model and not in the EFT, duly taking into account the detailed short-distance dynamics of the BSM particles and including potentially prominent effects associated with the direct production of the new particles. In theories unlike Composite Higgs, where $C_W$ is potentially enhanced by a strong coupling relative to $1/m_*$, the EFT analysis would instead be justified even at low luminosity, provided of course the quadratic term is duly included in the cross-section prediction.

It is worth stressing again that the VHEL sensitivity to such high $m_*$ scales emerges from the measurement of the high-energy (Higgs-strahlung) $Zh$ cross-section, not from the measurement of the total $Zh$ production cross-section. The latter is dominated by the VBF process $VV\to hZ$.  Considering $E_{\rm{cm}}=30$~TeV for definiteness, the cross-section of the latter process is
\begin{equation}\label{vvzh}
\sigma(VV \to Zh)\overset{E_{\rm{cm}}=30\,{\rm{TeV}}}{=}5.8\times10^6\left[1-(0.5\,{\rm{TeV}})^2C_W+(1.44\,{\rm{TeV}})^4{C_W}^2\right]
\times\frac1{90}\,{\rm{ab}}\,,
\end{equation}
where the sensitivity to the $\O_W$ operator does not come from the contact interactions in eq.~(\ref{ci}), but rather from the interactions between pairs of Higgs field currents that emerge from $\O_W$ using the equations of motion. The SM $VV \to Zh$ is more than $3$ orders of magnitude larger than the one of the $\llll\to Zh$, but its dependence on $C_W$ is much weaker. The linear $C_W$ term is few times the EW scale (squared) in eq.~(\ref{vvzh}) and of order $E_{\rm{cm}}^2$ in eq.~(\ref{xszh}). Correspondingly, the sensitivity to $C_W$ is of order $(1/25\,{\rm{TeV}})^2$ even in the unrealistic assumption that the experimental and theoretical uncertainties were so low that the statistical potential (corresponding to a relative uncertainty of $4\times10^{-4}$) of the measurement could  be entirely exploited. This should be compared with the $(1/162\,{\rm{TeV}})^2$ reach of the Higgs-strahlung measurement at the $30$~TeV VHEL. The comparison is even less favorable for VBF at lower VHEL energies. 

We discussed in the Introduction that this behavior was expected on general grounds. The $VV \to Zh$ takes place at EW-scale energies. On one hand, this is what makes its cross-section large. On the other hand, it makes its dependence on new physics weak because it probes interactions at the EW scale rather than at the VHEL energy $E_{\rm{cm}}$. The VBF Higgs production process is thus irrelevant as a probe of the $C_W$ (and $C_B$) operator. It does not even constitute a relevant background because it produces the $Zh$ system with low invariant mass and is efficiently eliminated with a lower cut. Of course the conclusion only holds for the specific operators we are considering (notice that we are setting $C_{HW,B}=0$). The $VV \to Zh$ potential to probe other EFT operators  should be investigated.

\begin{table}
\centering%
\renewcommand{\arraystretch}{1.2}
\begin{tabular}{|c|c|c|}
\hline 
$(P_{\ell^-},P_{\ell^+})$ & $\sigma(\llll\to Zh)$/ab & 95\% C.L. $C_W$\\
\hline 
\hline & & \\ [-13pt]
$(0\%,0\%)$ & $\begin{aligned}[t]
&122\times\!\Big(\frac{10\,{\rm TeV}}{E_{\rm cm}}\Big)^2\!\!\times\!\left[ 1 + \Big( \frac{E_{\rm cm}}{0.78} \Big)^2 C_W + \Big( \frac{E_{\rm cm}}{1.64} \Big)^2 C_B\right. \\
&\left. + \Big( \frac{E_{\rm cm}}{0.96} \Big)^4 C_W^2 + \Big( \frac{E_{\rm cm}}{1.17} \Big)^4 C_B^2 - \Big( \frac{E_{\rm cm}}{1.09} \Big)^4 C_W C_B \right]\end{aligned}$
& $(3.8\,E_{\rm{cm}})^{-2}$\\[37pt]
\hline & & \\[-13pt]
$(-30\%,+30\%)$ & $\begin{aligned}[t]
&142\times\!\Big(\frac{10\,{\rm TeV}}{E_{\rm cm}}\Big)^2\!\!\times\!\left[ 1 + \Big( \frac{E_{\rm cm}}{0.65} \Big)^2 C_W - \Big( \frac{E_{\rm cm}}{1.69} \Big)^2 C_B\right. \\
&\left. + \Big( \frac{E_{\rm cm}}{0.87} \Big)^4 C_W^2 + \Big( \frac{E_{\rm cm}}{1.31} \Big)^4 C_B^2 - \Big( \frac{E_{\rm cm}}{0.99} \Big)^4 C_W C_B \right]\end{aligned}$
& $(4.0\,E_{\rm{cm}})^{-2}$
\\[37pt]
\hline & & \\[-13pt]
$(30\%,-30\%)$ & $\begin{aligned}[t]
&123\times\!\Big(\frac{10\,{\rm TeV}}{E_{\rm cm}}\Big)^2\!\!\times\!\left[ 1 + \Big( \frac{E_{\rm cm}}{1.12} \Big)^2 C_W + \Big( \frac{E_{\rm cm}}{0.91} \Big)^2 C_B\right. \\
&\left. + \Big( \frac{E_{\rm cm}}{1.15} \Big)^4 C_W^2 + \Big( \frac{E_{\rm cm}}{1.07} \Big)^4 C_B^2 - \Big( \frac{E_{\rm cm}}{1.3} \Big)^4 C_W C_B \right]\end{aligned}$
& $(2.2\,E_{\rm{cm}})^{-2}$
\\[37pt]
\hline 
\end{tabular}
\caption{\label{tab:ZH} Total $\ell^+\ell^- \to Zh$ cross-section in ab, as a function of the collider energy $E_{\rm cm}$ and the Wilson coefficients $C_W$, $C_B$, for different beam polarizations (polarization $-100\%$ means fully left-handed (right-handed) particles (antiparticles)). The 95\% C.L. symmetrized individual constraint on $C_W$ is also given, as obtained from inclusive $\ell^+\ell^-\!\to Zh$ with $C_B = 0$.}
\end{table}

We now discuss the impact of the Higgs-strahlung cross-section measurement in the $2$-parameters fit of $C_B$ and $C_W$. The cross-section is reported in Table~\ref{tab:ZH} and the measurement produces (assuming that the SM is observed) the red elliptical strip displayed in Figure~\ref{fig:chi2-profiles-cBcW} (panel (a)). The degeneracy along the strip cannot be eliminated by more exclusive or differential cross-section measurements because the sensitivity to new physics is entirely captured by the total cross-section as explained at the beginning of this section. The degeneracy could however be easily lifted (up to a four-fold ambiguity) if some degree of polarization of the VHEL lepton beams could be engineered. We cannot comment on the actual technical feasibility of polarized beams. However we notice (see Table~\ref{tab:CWCB} and Figure~\ref{fig:chi2-profiles-cBcW} , panel (b)) that a reasonably modest degree of polarization of $\pm30\,\%$, with non-optimized splitting (half-and-half) of the total luminosity, would be sufficient to obtain a rather satisfactory simultaneous determination of $C_W$ and $C_B$. If beam polarization is not available (or to get rid the four-fold ambiguity), the degeneracy will be removed by combining with the $W^+W^-$ measurements, to be discussed below.

\subsection[High-energy $W^+W^-$]{High-energy $\mathbf{W^+W^-}$}

The phenomenology of the high-energy $W^+W^-$ production process, $\llll\to W^+W^-$, is slightly more complex than the one of Higgs-strahlung. Like for Higgs-strahlung, the amplitude for longitudinally-polarized $W$ bosons (i.e., charged Goldstone bosons) is central (proportional to $\sin \theta_\star$) both for the SM and in the EFT as in Table~\ref{primaries}. However unlike Higgs-strahlung the amplitude for producing transversely-polarized $W$ bosons is sizable in the SM and furthermore it is enhanced in the forward region by the singularity (cut off by the $W$ boson mass) associated with the neutrino exchange in the t-channel. The transverse amplitudes are sensitive to $C_W$ and $C_B$ only through negligible $m_W/E_{\rm{cm}}$-suppressed effects. Therefore, in first approximation, the transverse vector bosons production process can be treated like a background, as we will do in Section~\ref{subsec:Fiducial-cross-section-analysis}. However one can also exploit the SM transverse amplitude to enhance the sensitivity by exploiting the quantum-mechanical interference between the transverse and the longitudinal production amplitudes. We discuss this possibility in Section~\ref{subsec:Differential-analyses}, following Ref.~\cite{Panico:2017frx}.

\subsubsection{Fiducial cross-section \label{subsec:Fiducial-cross-section-analysis}}

With $\theta_{\star}$ defined as the angle between the incoming $\ell^+$ and the outgoing $W^+$, the t-channel enhancement of the transverse amplitude shows up in the forward (small $\theta_{\star}$) region. We eliminate it by an asymmetric cut on $\cos\theta_{\star}$, which after optimization was set to
\begin{equation}
\cos\theta_{\star}\in[-0.98,0.17]\,.\label{eq:cosThetaStarWW}
\end{equation}
This cut defines the fiducial $\llll\to W^+W^-$ cross-section to be measured at the VHEL. Notice that the lower cut  $\cos\theta_{\star}>-0.98$ was not optimized. Rather, it corresponds to a detector coverage limited to $10$ degrees along the beam axis. Our results are very weakly sensitive to this lower cut, which could be raised to $15$ or $20$ degree without significant sensitivity loss. Similar considerations hold for Higgs-strahlung process discussed in the previous section.

The fiducial cross-section prediction is reported in Table~\ref{tab:WW} for unpolarized and for polarized beams. As expected $C_W$ and $C_B$ enter in combination with $E_{\rm{cm}}^2$, owing to the energy-growing nature of their effect on the longitudinal cross-section. The SM cross-section is somewhat higher than the one of Higgs-strahlung. Around $7000$ events are expected with the baseline luminosity, which would correspond to a $1\,\%$ statistical uncertainty in the fiducial cross-section measurement. Taking also into account that only a fraction of these events will be actually useful for the measurement, it is reasonable to ignore systematical and theoretical uncertainties in the projection of the VHEL sensitivity. In order to measure the fiducial cross-section as defined with the asymmetric cut in eq.~(\ref{eq:cosThetaStarWW}), the final state must include at least one faithful tracker of the $W$ boson charges. We consider the semi-leptonic decay channel, with total branching fraction $\epsilon_{WW}=2\times0.33\times0.67=40\,\%$ including decays to electrons, muons and taus.

\begin{table}
\centering%
\renewcommand{\arraystretch}{1.2}
\begin{tabular}{|c|c|c|}
\hline 
$(P_{\ell^-},P_{\ell^+})$ & $\sigma(\llll\to W^+W^-)$/ab & 95\% C.L. $C_W$\\
\hline 
\hline & & \\[-13pt]
$(0\%,0\%)$ &
$\begin{aligned}[t]
&736\times\!\Big(\frac{10\,{\rm TeV}}{E_{\rm cm}}\Big)^2\!\!\times\!\left[ 1 + \Big( \frac{E_{\rm cm}}{1.77} \Big)^2 C_W + \Big( \frac{E_{\rm cm}}{2.32} \Big)^2 C_B\right. \\
&\left. + \Big( \frac{E_{\rm cm}}{1.69} \Big)^4 C_W^2 + \Big( \frac{E_{\rm cm}}{2.06} \Big)^4 C_B^2 + \Big( \frac{E_{\rm cm}}{1.91} \Big)^4 C_W C_B \right]\end{aligned}$
& $(3.0\,E_{\rm cm})^{-2}$
\\[37pt]
\hline & & \\[-13pt]
$(-30\%,+30\%)$ &
$\begin{aligned}[t]
&1204\times\!\Big(\frac{10\,{\rm TeV}}{E_{\rm cm}}\Big)^2\!\!\times\!\left[ 1 + \Big( \frac{E_{\rm cm}}{1.74} \Big)^2 C_W + \Big( \frac{E_{\rm cm}}{2.81} \Big)^2 C_B\right. \\
&\left. + \Big( \frac{E_{\rm cm}}{1.67} \Big)^4 C_W^2 + \Big( \frac{E_{\rm cm}}{2.52} \Big)^4 C_B^2 + \Big( \frac{E_{\rm cm}}{1.90} \Big)^4 C_W C_B \right]\end{aligned}$
&  $(2.9\,E_{\rm cm})^{-2}$
\\[37pt]
\hline & &\\[-13pt]
$(30\%,-30\%)$ &
$\begin{aligned}[t]
&401\times\!\Big(\frac{10\,{\rm TeV}}{E_{\rm cm}}\Big)^2\!\!\times\!\left[ 1 + \Big( \frac{E_{\rm cm}}{1.87} \Big)^2 C_W + \Big( \frac{E_{\rm cm}}{1.65} \Big)^2 C_B\right. \\
&\left. + \Big( \frac{E_{\rm cm}}{1.73} \Big)^4 C_W^2 + \Big( \frac{E_{\rm cm}}{1.61} \Big)^4 C_B^2 + \Big( \frac{E_{\rm cm}}{1.97} \Big)^4 C_W C_B \right]\end{aligned}$
&  $(2.1\,E_{\rm cm})^{-2}$
\\[37pt]
\hline 
\end{tabular}
\caption{\label{tab:WW} Fiducial $\ell^+\ell^- \to W^+W^-$ cross-section in ab, as a function of the collider energy $E_{\rm cm}$ and the Wilson coefficients $C_W$, $C_B$, for different beam polarizations. The 95\% C.L. symmetrized individual constraint on $C_W$ is also given, as obtained from $\ell^+\ell^-\! \to W^+W^-$ alone, with $C_B = 0$.}
\end{table}

In order to measure $ \theta_\star$ (and the $WW$ invariant mass), the leptonic $W$ momentum needs to be reconstructed. This can be achieved by first reconstructing the neutrino momentum imposing the on-shell $W$ boson condition like at hadron colliders, using the measurement of the missing transverse energy. In spite of the fact that the on-shell condition has $2$ solutions for the neutrino, the $W$ momentum reconstructed with this method becomes exact (see e.g.~\cite{Panico:2017frx}) if the $W$ is boosted in the transverse plane. At the high VHEL energy we can thus rely on an essentially perfect reconstruction (up to detector effects) of the leptonic $W$. In principle at lepton colliders the neutrino momentum could also be reconstructed by exploiting the knowledge of the initial energy. However this is hardly an option at the VHEL because of the potentially large undetected energy in the forward and backward directions.

Notice that the process of interest is central in angle, and the invariant mass for the $WW$ pair is of order $E_{\rm{cm}}$. Background processes with one physical lepton or $W$ boson are tiny in this kinematical regime. Therefore for our sensitivity estimate we ignore the backgrounds and set $\epsilon_{WW}=40\,\%$, purely coming from the branching ratios, as total signal efficiency. In spite of this, we believe that our results are most likely conservative estimates of the sensitivity. Indeed, while the asymmetric cuts in eq.~(\ref{eq:cosThetaStarWW}) are found to produce better $S/\sqrt{B}$ after optimization, a symmetric cut on $|\cos \theta_\star|$ would produce a comparable optimized $S/\sqrt{B}$. The fiducial cross-section defined with a symmetric cut could be measured in the fully hadronic channel and benefit from its higher branching fraction. 

The measurement of the unpolarized fiducial cross-section in the $(C_B,C_W)$ plane corresponds to the blue elliptical shape in panel (a) of  Figure~\ref{fig:chi2-profiles-cBcW}. The combination with the $Zh$ measurement, also shown in the figure, does not allow for a satisfactory simultaneous determination of $C_B$ and $C_W$. This is not a problem for the single-operator sensitivity on $C_W$ (see Table~\ref{tab:CWCB}), nor for the reach in the special direction $C_W=C_B$, which is populated by Composite Higgs models with $P_{LR}$ ``custodial'' symmetry~\cite{Agashe:2006at}. It is more of an issue for $C_B$ and for the marginalized bounds. In order to improve, one option is to measure polarized cross-sections. If this is possible, a significant improvement could be achieved as shown in panel  (b) and in Table~\ref{tab:CWCB}. We also see in panel (a) that the unpolarized cross-section measurements do not resolve the ambiguity between the SM-like region and a second solution with large negative values of $C_W$ and $C_B$. This is arguably not a relevant issue because these large values most likely emerge only in BSM scenarios with relatively light new particles, to be probed directly at the VHEL. On the other hand, it is worth asking if the degeneracy can be eliminated by additional diboson measurements. Notice that the four polarized cross-section measurements are unable to resolve this ambiguity. This fact is readily understood as follows. By setting $C_W=C_B$ in eq.~(\ref{Gs}), and comparing with eq.~(\ref{ci}), we see that the total EFT interaction in this special direction is proportional to $g^2 J_{2,\,l}\cdot J_{2,\,H}+g^{\prime\,2} J_{1,\,l}\cdot J_{1,\,H}$ where $J_{2,1}$ denote, respectively, the \mbox{SU$(2)_L$} and \mbox{U$(1)_Y$} currents of the leptons and of the Higgs field. This is the exact same structure we have in the SM contribution to the amplitudes for longitudinally polarized $W$ and $Z$ (and Higgs) bosons, apart from the factor $1/E_{\rm{cm}}^2$ from the gauge field propagators. If we pick  $C_W=C_B=-2/E_{\rm{cm}}^2$ we can thus set all the longitudinal diboson amplitudes to be equal and opposite to the SM ones, and obtain the SM cross-section. The only way to resolve the ambiguity is to measure a quantity that is sensitive to the sign of the longitudinal amplitudes. In turn, this requires observables that are sensitive to the interference between the longitudinal and the transverse helicity amplitudes. 

\subsubsection{Differential analysis\label{subsec:Differential-analyses}}

\begin{figure}
\begin{centering}
\renewcommand{\subfigcapmargin}{22pt}
\subfigure[Inclusive $Zh$ (red) and fiducial $WW$ (blue) rates for unpolarized beams;]{\includegraphics[clip,width=0.49\textwidth]{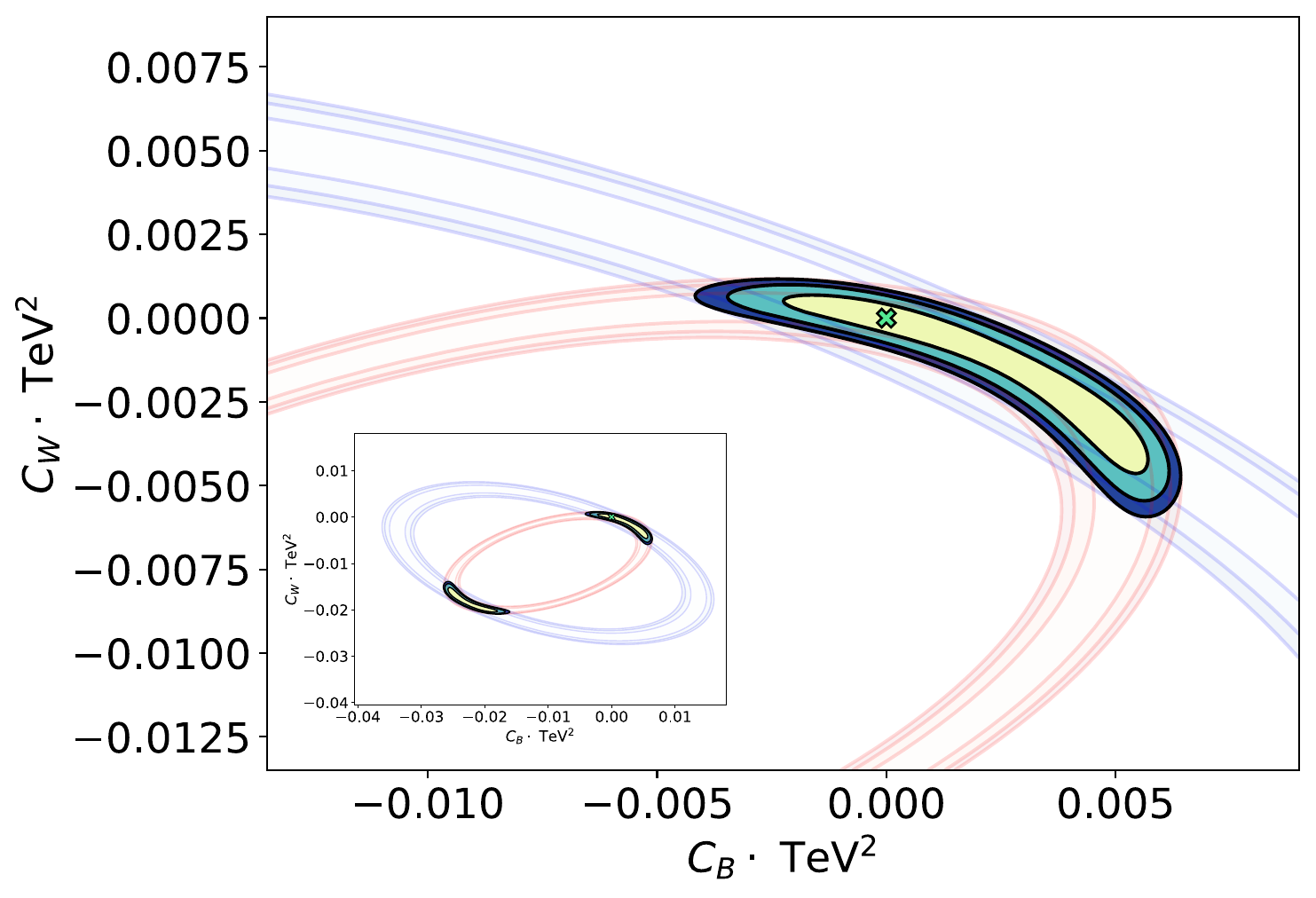}}
\hfill
\subfigure[Polarized inclusive $Zh$ (L: red, R: orange) and fiducial $WW$ (L: blue, R: purple);]{\includegraphics[clip,width=0.49\textwidth]{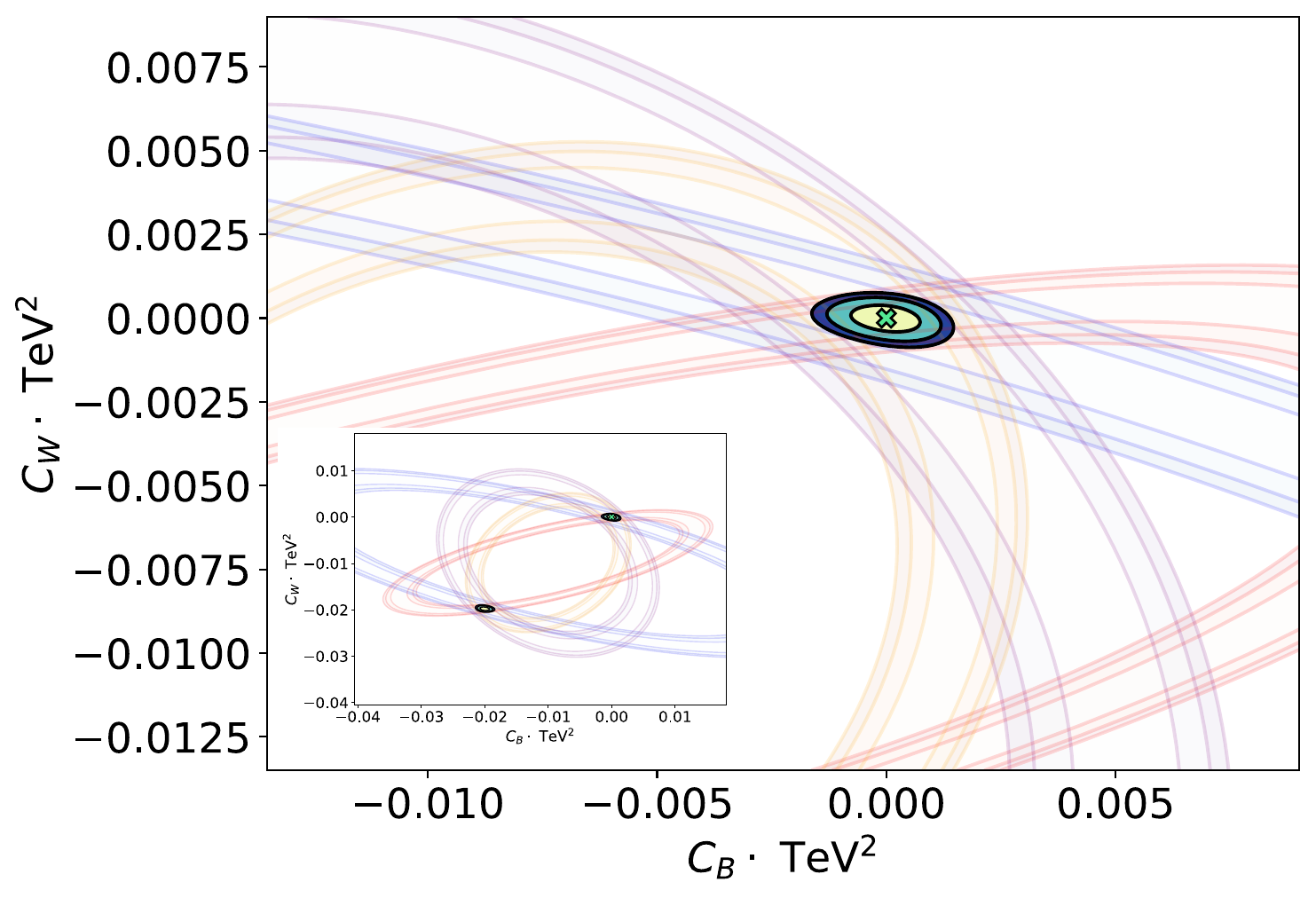}}\\
\subfigure[Same as panel (a), but with differential $WW$ rate (blue) for unpolarized beams.]{\includegraphics[clip,width=0.49\textwidth]{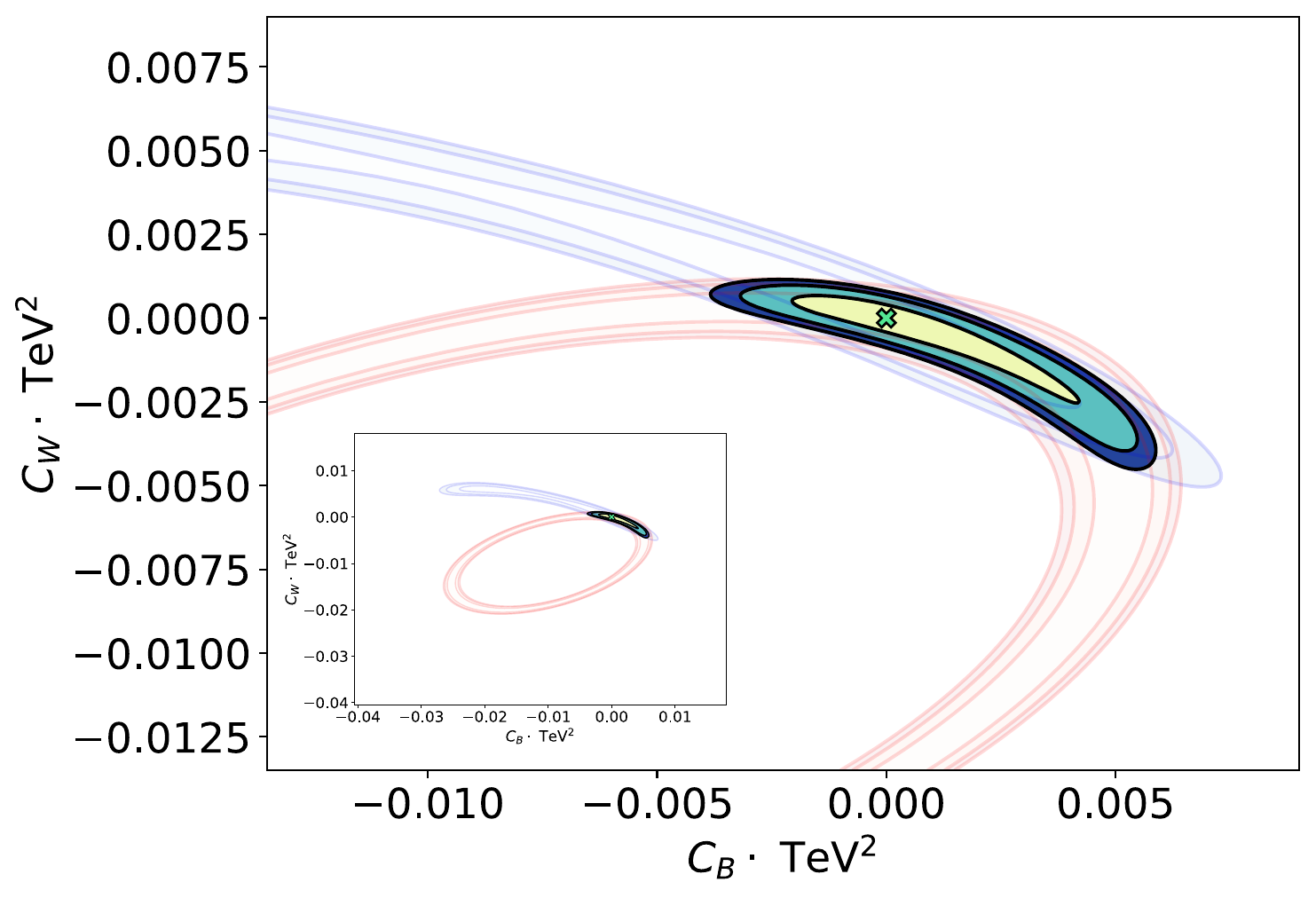}}
\hfill
\subfigure[Same as panel (a), combined with fiducial $WWh$ (green) for unpolarized beams;]{\includegraphics[clip,width=0.49\textwidth]{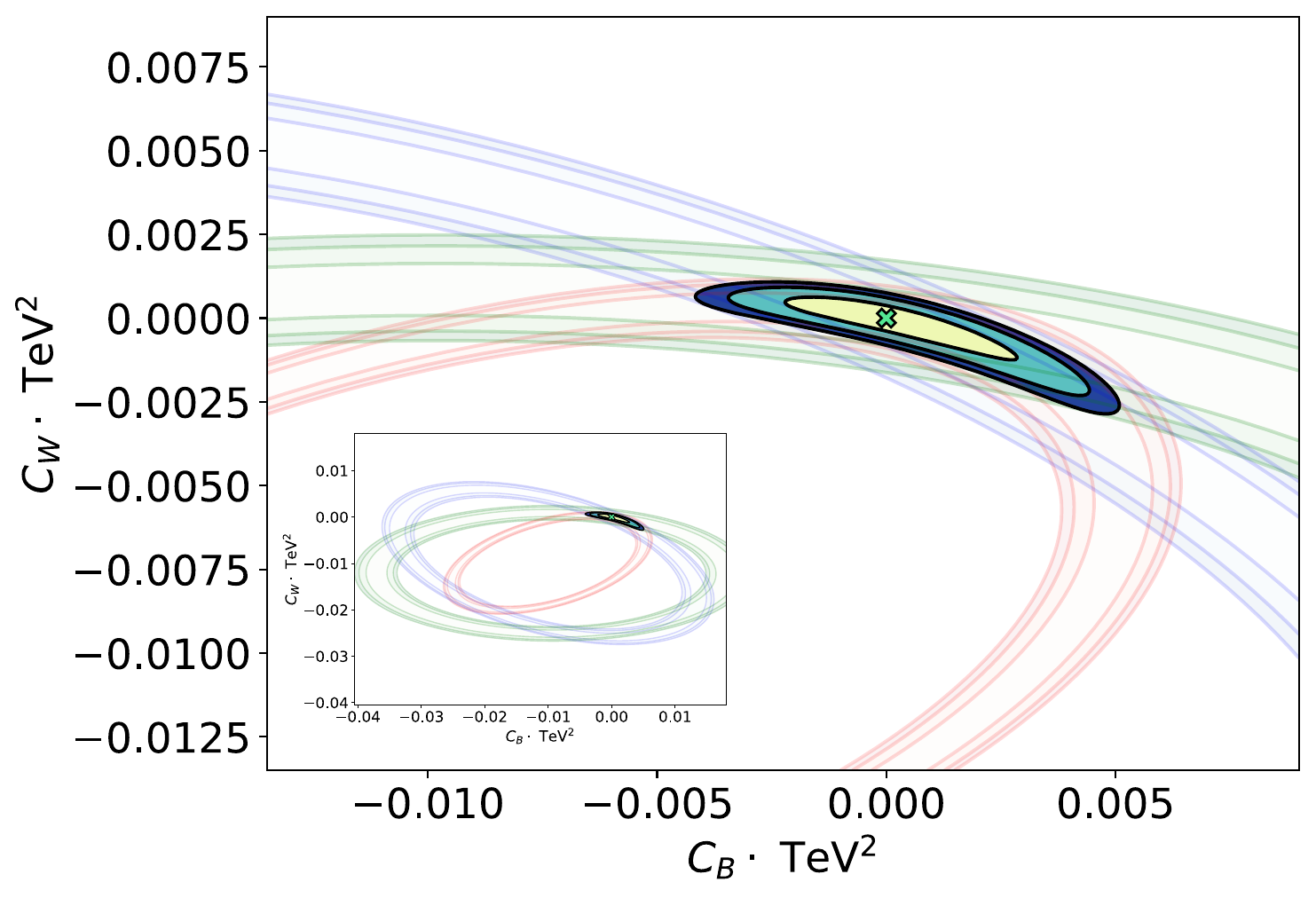}}
\par\end{centering}
\caption{\label{fig:chi2-profiles-cBcW}$\chi^{2}$ profiles in the $(C_{B}, C_{W})$
plane at a $\sqrt{s}=10\text{ TeV}$ muon collider. The four panels combine different inclusive and differential measurements with polarized and unpolarized beams. Solid
filled contours are for the combination of the $\chi^{2}$ relevant for each panel. 
The iso-lines are for $\Delta\chi^{2}$ values equivalent to 67\%, 95\%, and 99\% confidence level. 
For completeness of display, we add the inset in the lower-left corner to show the same quantities on a bigger scale.}
\end{figure}

The design of observables that are sensitive to the interference between different diboson helicity amplitudes has been discussed in Ref.~\cite{Panico:2017frx} (see also~\cite{Hagiwara:1989mx,Duncan:1985ij,Azatov:2017kzw}) in the context of hadron colliders, with the purpose of enhancing the sensitivity to those EFT operators that mostly contribute to helicity amplitudes where the SM is small. The sensitivity improvement associated with measuring such observables is instead expectedly moderate in processes, like the one at hand, where the EFT contributes to an helicity channel that is large also in the SM. This was recently verified for high-energy $WZ$ production at the LHC~\cite{Chen:2020mev}. However these measurements could play an important role in our analysis, because of the stretched shape of the likelihood contours in Figure~\ref{fig:chi2-profiles-cBcW} panel (a) and of the unresolved degeneracy. 

The relevant observables are readily identified as follows (see~\cite{Chen:2020mev} for additional details). In the narrow-width approximation the $2\to4$ differential cross-section including the $W$ bosons decays can be written as
\begin{equation}\label{acs}
d \sigma = \sum d \rho^{\rm hard}_{h_+^{\phantom\prime} h_-^{\phantom\prime} h'_+ h'_-} d \rho^{W^+}_{h_+^{\phantom\prime} h'_+}
d \rho^{W^-}_{h_-^{\phantom\prime} h'_-}\,,
\end{equation}
where the sum runs over two pairs of helicity indices $h_\pm^{\phantom\prime}$ and $h_\pm^{\prime}$ associated with the intermediate $W^\pm$ vector bosons helicities. 

The hard density matrix $d\rho^{\rm hard}$ contains the helicity amplitude of the $\ell^+\ell^-\to W^+W^-$ process with on-shell $W$ bosons. Up to an irrelevant flux factor, it reads
\begin{equation}
d \rho^{\rm hard}_{h_+^{\phantom\prime} h_-^{\phantom\prime} h'_+ h'_-} \propto {\cal M}_{h_+^{\phantom\prime} h_-^{\phantom\prime}} ({\cal M}_{h_+^{\prime} h_-^{\prime}})^*\, d \Phi_{\rm{WW}} \,,
\end{equation}
where $d\Phi_{\rm{WW}}$ is the phase space for the on-shell diboson production. The helicity amplitudes ${\cal M}$ contain both SM and EFT contributions, and they take a very simple form in the high-energy limit. The only relevant (quadratically enhanced with energy) EFT contribution is in the longitudinal amplitude ${\cal M}_{00}$, as in Table~\ref{primaries}, both for Right-handed and for Left-handed initial-state leptons. If the initial leptons are Right-handed, all the helicity amplitudes vanish in the SM apart from the longitudinal one. Consequently, there is no interference contribution. 

If instead the initial leptons are Left-handed, also the SM transverse amplitudes are non-vanishing in the $(\pm,\mp)$ helicity channels. Explicitly
\begin{equation}
\displaystyle
{\cal M}_{+-}=-\frac{g^2}2\sin \theta_\star\,,\;\;\;\;\;{\cal M}_{+-}=g^2\cos^2\frac{ \theta_\star}2\cot^2\frac{ \theta_\star}2\,,
\end{equation}
where $g$ is the SU$(2)_L$ coupling. The longitudinal amplitudes, both in the SM and in the EFT, are proportional to $\sin \theta_\star$. The only relevant interference term in the whole process thus emerges (with Left-handed initial leptons) from the $\pm\mp00$ and $00\pm\mp$ terms in the sum of eq.~(\ref{acs}). 

The density matrices $d \rho^{W^\pm}$ are instead EFT-independent factors that account for the decay of the $W$ bosons. As in~\cite{Panico:2017frx,Chen:2020mev}, we parametrize them in terms of the polar and azimuthal angles ($\theta_{\pm}$ and $\varphi_\pm$) of the helicity-plus fermion or anti-fermion, in the rest frame of the decaying boson. The decay density matrices are readily computed, and the interference due to the $\pm\mp00$ and $00\pm\mp$ terms in eq.~(\ref{acs}) is found to be
\begin{eqnarray}\label{int0}
\displaystyle
&&d\sigma_{\rm{int}}\propto {\cal M}_{00}{\cal M}_{+-}\cos(\varphi_+-\varphi_-) \sin\theta_+(1+\cos\theta_+)\sin\theta_-(1-\cos\theta_-)  \nonumber\\
\displaystyle
&&\;\;\qquad+{\cal M}_{00}{\cal M}_{-+}\cos(\varphi_+-\varphi_-) \sin\theta_+(1-\cos\theta_+)\sin\theta_-(1+\cos\theta_-)\,,
\end{eqnarray}
having exploited the fact that all the hard amplitudes are real.

We can now turn to the definition of the relevant observables. The $\theta_\pm$ and $\varphi_\pm$ angles are not directly observable, for the following reasons. Consider for definiteness the case in which the $W^+$ decays hadronically, to $u\bar{d}$, and $W^-\to \ell^-\bar\nu$. The fermion with helicity $+1/2$ in the $W^+$ decay is the $\bar{d}$ quark, so that $\theta_+$ and $\varphi_+$ are defined as the angles of the $\bar{d}$. However it is very difficult or impossible to tell the $\bar{d}$ from the $u$ quark, therefore the best we can do is to choose at random one of the two jets from the decay, interpret it as the $\bar{d}$ and measure its angles $\theta_{\bar{d}}$ and $\varphi_{\bar{d}}$.\footnote{Equivalently, we might also retain both jets and have two measurements of the angles for each event.} These angles are either equal to $\theta_+$ and $\varphi_+$, or to $\pi-\theta_+$ and $\varphi_++\pi$ with the same probability. The differential cross-section for the $\theta_{\bar{d}}$ and $\varphi_{\bar{d}}$ variables defined in this way is thus the average of eq.~(\ref{int0}) evaluated at $(\theta_+,\varphi_+)=(\theta_{\bar{d}},\varphi_{\bar{d}})$ and at $(\theta_+,\varphi_+)=(\pi-\theta_{\bar{d}},\varphi_{\bar{d}}+\pi)$. The $W^-$ decay angles should instead be defined as those of the $\bar\nu$. However the neutrino momentum is reconstructed imposing the on-shell condition of the $W$ boson, which produces two distinct solutions. The $4$-momenta obtained on two solutions approach each other when the $W$ is boosted in the transverse plane, so that the reconstructed $W$ boson momentum is nearly the same on the two solutions as previously mentioned. The polar angle of the neutrino in the $W$ rest frame also coincides on the two solutions, while the two determinations of the azimuthal angle instead do not coincide, but are related to each other by $\varphi_1=\pi-\varphi_2$~\cite{Panico:2017frx}. If we pick one of the two solutions at random and interpret its angles as $\theta_{\bar\nu}$ and $\varphi_{\bar\nu}$, the distribution for these variables is obtained by further averaging eq.~(\ref{int0}) over $(\theta_-,\varphi_-)=(\theta_{\bar{\nu}},\varphi_{\bar{\nu}})$ and at $(\theta_-,\varphi_-)=(\theta_{\bar{\nu}},\pi-\varphi_{\bar{\nu}})$. After both averages, eq.~(\ref{int0}) becomes
\begin{eqnarray}\label{int}
\displaystyle
&&d\overline\sigma_{\rm{int}}\propto {\cal M}_{00}{\cal M}_{+-}
\sin\varphi_{\bar{d}}\sin\varphi_{\bar{\nu}}\cos\theta_{\bar{d}}\sin\theta_{\bar{d}}
\sin\theta_{\bar{\nu}}(1-\cos\theta_{\bar{\nu}}) \nonumber\\
\displaystyle
&&\;\;\qquad-{\cal M}_{00}{\cal M}_{-+}
\sin\varphi_{\bar{d}}\sin\varphi_{\bar{\nu}}\cos\theta_{\bar{d}}\sin\theta_{\bar{d}}
\sin\theta_{\bar{\nu}}(1+\cos\theta_{\bar{\nu}})\,.
\end{eqnarray}
Since this is non-vanishing, we can access the interference term experimentally by the measurable variables $\theta_{\bar{\nu},\bar{d}}$ and $\varphi_{\bar{\nu},\bar{d}}$.

\begin{table}
\centering%
\renewcommand{\arraystretch}{1.5}
{\begin{tabular}{|c|c|c|cc|c|cc|}
\hline 
& \multirow{2}{*}{$E_{\rm{cm}}$} & \multirow{2}{*}{$\mathcal{L}/{\rm ab}$} & \multicolumn{2}{c|}{Single-operator} & Single-operator & \multicolumn{2}{c|}{Marginalized} \\
& & & $C_W$ & $C_B$ & $C_W = C_B$ & $C_W$ & $C_B$\\
\hline 
\hline 
\multirow{3}{*}[-2pt]{\rotatebox{90}{Inclusive}}
& 10~{\rm TeV} & 10 & {[}-5.9,  5.5{]} & {[}-17,  14{]} & {[}-4.3, 4.2{]} & [-55, 10] & [-35, 62] \\
& 14~{\rm TeV} & 20 & {[}-3.0,  2.8{]} & {[}-8.9,  7.3{]} & {[}-2.2, 2.1{]} & [-28, 5.1] & [-18, 31] \\
& 30~{\rm TeV} & 90 & {[}-0.66 ,  0.61{]} & {[}-1.9,  1.6{]} & {[}-0.48, 0.46{]} &  [-6.1, 1.1] & [-3.8, 6.9] \\
\hline 
\hline 
\multirow{3}{*}[-2pt]{\rotatebox{90}{Polarized}}
& 10~{\rm TeV} & 10 & {[}-5.2,  4.9{]} & {[}-10 ,  9.2{]} & {[}-4.1, 4.0{]} & [-6.9, 6.2] & [-13, 12] \\
& 14~{\rm TeV} & 20 & {[}-2.7,  2.5{]} & {[}-5.1,  4.7{]} & {[}-2.1, 2.0{]} & [-3.5, 3.2] &  [-6.6, 6.1] \\
& 30~{\rm TeV} & 90 & {[}-0.58 ,  0.54{]} & {[}-1.1,  1.0{]} & {[}-0.46, 0.44{]} & [-0.73, 0.66] & [-1.4, 1.3]\\
\hline 
\hline 
\multirow{3}{*}[-2pt]{\rotatebox{90}{Differential}}
& 10~{\rm TeV} & 10 & {[}-5.6,  5.3{]} & {[}-16,  13{]} & {[}-4.1, 3.9{]} & [-40, 9.9] & [-32, 55] \\
& 14~{\rm TeV} & 20 & {[}-2.9,  2.7{]} & {[}-8.0,  6.8{]} & {[}-2.1, 2.0{]} & [-20, 5.0] & [-16, 28] \\
& 30~{\rm TeV} & 90 & {[}-0.62,  0.58{]} & {[}-1.7,  1.5{]} & {[}-0.46, 0.44{]} & [-4.4, 1.1] & [-3.5, 6.1] \\
\hline 
\hline 
\multirow{3}{*}[-2pt]{\rotatebox{90}{Tri-boson}}
& 10~{\rm TeV} & 10 & {[}-5.2,  4.9{]} & {[}-17,  14{]} & {[}-3.9, 3.8{]} & [-23, 9.2] & [-34, 44] \\
& 14~{\rm TeV} & 20 & {[}-2.6,  2.5{]} & {[}-8.5,  7.1{]} & {[}-2.0, 1.9{]} & [-11, 4.6] &  [-18, 22] \\
& 30~{\rm TeV} & 90 & {[}-0.52,  0.51{]} & {[}-1.8,  1.5{]} & {[}-0.41, 0.40{]} & [-1.9, 0.96] & [-3.8, 4.30] \\
\hline 
\hline 
\multirow{3}{*}[-2pt]{\rotatebox{90}{Combined}}
& 10~{\rm TeV} & 10 & {[}-4.9,  4.7{]} & {[}-15,  13{]} & {[}-3.7, 3.6{]} & [-20, 9.1] & [-32, 40]\\
& 14~{\rm TeV} & 20 & {[}-2.5,  2.4{]} & {[}-7.7,  6.6{]} & {[}-1.9, 1.8{]} & [-9.3, 4.6] & [-16, 19]\\
& 30~{\rm TeV} & 90 & {[}-0.51,  0.49{]} & {[}-1.6,  1.4{]} & {[}-0.39, 0.38{]} & [-1.7, 0.95] & [-3.5, 3.9]\\
\hline 
\end{tabular}}
\caption{\label{tab:CWCB} 95\% C.L.\ constraints on $C_W$ and $C_B$, expressed in units of $(100\,{\rm TeV})^{-2}$, for the benchmark VHEL energies and luminosities. The first two columns show the constraints on one coefficient setting the other to zero, the third one is the constraint in the direction $C_W=C_B$. The last two columns show the constraints marginalized in the $(C_W,C_B)$ plane.}
\end{table}

In light of eq.~(\ref{int}), our differential analysis is defined as follows. Both $\varphi_{\bar{\nu}}$ and $\varphi_{\bar{d}}$ should be measured, because the interference vanishes if integrated over any of them. We thus consider a doubly-differential cross-section in $25$ equally-spaced bins in the $(\varphi_{\bar{\nu}},\varphi_{\bar{d}})$  plane. It is also necessary to measure $\theta_{\bar{d}}$ because the interference is odd under $\cos\theta_{\bar{d}}\to-\cos\theta_{\bar{d}}$. We thus bin $\cos\theta_{\bar{d}}$ at $[-1,-0.66,0,0.66,1]$. It is not strictly necessary to measure $\theta_{\bar{\nu}}$, however the peculiar distribution of this variable can improve the sensitivity. We thus bin $|\cos\theta_{\bar{\nu}}|$ at $[0,0.66,1]$, for a total of $8$ regions in the $(\theta_{\bar{\nu}},\theta_{\bar{d}})$ plane. Binning over the $WW$ center of mass scattering angle $ \theta_\star$ could also bring some advantage in terms of sensitivity, because the hard amplitude terms in eq.~(\ref{int}) possess a distinctive angular dependence. We do not consider this possibility for simplicity, and we merely restrict $ \theta_\star$ to the ``fiducial'' region in eq.~(\ref{eq:cosThetaStarWW}). Our analysis, in a total of $200$ bins, is probably close to the statistical optimal analysis that can be achieved with the available statistics, for which a handful of events are found in each bin. Unbinned techniques such as the Matrix Element method could be studied to assess the optimality.

The fully differential cross-section of the process could be obtained analytically by exploiting the narrow-width approximation and the high-energy ($E_{\rm{cm}}\gg m_W$) limit. While these are excellent approximations, we instead employed exact tree-level predictions for the cross-section in the bins as a function of $C_W$ and $C_B$. They have been obtained using {\sc{MadGraph}}~\cite{Alwall:2014hca}, with the EFT operators in eq.~(\ref{ci}) implemented via FeynRules~\cite{Alloul:2013bka}.

The result is shown in panel (c) of Figure~\ref{fig:chi2-profiles-cBcW}, for unpolarized beams. After combining with the $Zh$ cross-section measurement, the differential analysis eliminates the second solution and allows for a better simultaneous determination of $C_W$ and $C_B$ as reported in Table~\ref{tab:CWCB}.

\subsection{High-energy tri-bosons\label{subleading-processes}}

We have seen above that a differential analysis of the $\ww$ process can resolve the degeneracy between $C_W$ and $C_B$ and improve their global determination. However, it is important to highlight that at the high energies that are available at the VHEL a novel approach to the problem becomes viable. Table~\ref{primaries} suggests that we might probe a new direction in the  $(C_B,C_W)$ plane if we could measure charged-current processes such as
\begin{equation}\label{diboson_charged}
 \ell^\pm\nu \to W^\pm h / W^\pm Z\,. 
 \end{equation}
This is possible at the VHEL thanks to the flux of ``effective'' neutrino beams originating from the emission of a charged $W$ of (relatively) low energy from one of the initial leptons. This emission is enhanced by a Sudakov double logarithm of $E_{\rm{cm}}^2/m_W^2$, hence it acquires growing importance at the highest center of mass energies. Notice that the charged $W$ emission is enhanced (though only by a single log) also in the collinear regime where the $W$ is emitted parallel to the beam axis and carries an order-one fraction of the lepton beam energy.  The latter kinematical regime could be described in terms of the neutrino Parton Distribution Function (PDF)~\cite{Han:2020uid}, but it is less relevant for our analysis than the soft (or soft-collinear) regime, which is doubly log-enhanced and maximally benefits from the quadratic energy growth of the EFT contribution because the center of mass energy $\sqrt{s}$ of the $\ell\nu$ scattering essentially coincides with $E_{\rm{cm}}$.

\begin{table}
\begin{centering}
\renewcommand{\arraystretch}{1.2}
\begin{tabular}{|c|c|c|c|c|}
\hline 
$\sqrt{s}$/TeV & $\mathcal{L}/\text{ab}^{-1}$ & $\sigma(\llll\to\ww h)$/ab & $p_T$ cut & 95\% C.L. $C_W$ \tabularnewline
\hline 
\hline 
3 & $0.9$ &
$\begin{aligned}[t]
493\cdot \big[1 &+ (3.09\,{\rm TeV})^2 C_{W} + (2.36\,{\rm TeV})^4 C_{W}^{2} \\
 & + (1.30\,{\rm TeV})^2 C_{B} + (1.68\,{\rm TeV})^4 C_{B}^{2}\big]
\end{aligned}
$ & 0.6 TeV & $(7.0\,{\rm TeV})^{-2}$ \\[18pt]
\hline 
10 & $10$ &
$\begin{aligned}[t]
82.6\cdot \big[1 &+ (11.1\,{\rm TeV})^2 C_{W}+(8.43\,{\rm TeV})^4 C_{W}^{2}\\
 & +(4.68\,{\rm TeV})^2 C_{B} + (5.79\,{\rm TeV})^4 C_{B}^{2}\big] 
 \end{aligned}
$ & 2.2 TeV & $(29.4\,{\rm TeV})^{-2}$ \\[18pt]
\hline
14 & $20$ &
$\begin{aligned}[t]
48.4 \cdot\big[1 & + (15.8\,{\rm TeV})^2 C_{W} + (12.0\,{\rm TeV})^4 C_{W}^{2}\\
 & + (6.64\,{\rm TeV})^2 C_{B} + (8.17\,{\rm TeV})^4 C_{B}^{2}\big]
\end{aligned}
$ & 3.2 TeV & $(43.6\,{\rm TeV})^{-2}$ \\[18pt]
\hline 
30 & $90$ &
$\begin{aligned}[t]
14.7 \cdot \big[ 1 & + (34.4\,{\rm TeV})^2 C_{W} + (26.0\,{\rm TeV})^4 C_{W}^{2}\\
 & + (14.4\,{\rm TeV})^2 C_{B} + (17.6\,{\rm TeV})^4 C_{B}^{2} \big]
\end{aligned}
$ & 6.9 TeV & $(103\,{\rm TeV})^{-2}$ \\[18pt]
\hline 
\end{tabular}
\par\end{centering}
\caption{\label{tab:sigmaWWh}Fiducial $\llll\to\ww h$ cross-section at high energy lepton
colliders. The fiducial phase-space is defined by the cuts on the $p_T$ of the Higgs and hardest $W$ specified in the fourth column.
We also show the 95\% C.L. symmetrized bound on $C_W$, from this measurement alone.}
\end{table}

We focus on the $Wh$ process, rather than on $WZ$, in order to avoid a contamination from SM transverse gauge bosons production, which is negligible for $Wh$ and sizable for $WZ$.
The complete $2\to3$ scattering
$$\ell^+\ell^- \to W^+W^-h\,, $$
is simulated with {\sc{MadGraph}}. The relevant kinematical region is characterized by a hard $W$ and a hard Higgs boson in the central region, with an additional soft or forward $W$. This region is selected by  a lower cut on the $p_T$ of the Higgs and of the hardest $W$ boson. The cut is optimized to maximize the significance of the signal over the SM background, with the results reported in Table~\ref{tab:sigmaWWh} together with the corresponding cross-sections for different VHEL energies. Notice that the soft or forward $W$, which could be difficult to see, does not need to be detected. The high-$p_T$ Higgs and $W$ should instead be relatively easy to detect, with an efficiency in the fully-hadronic channel comparable to the one of the $Zh$ final state discussed in Section~\ref{zh}. For our sensitivity projection estimate we thus set $\epsilon_{{Wh}}=\epsilon_{{Zh}}=26\%$. 

A few comments on the cross-sections in Table~\ref{tab:sigmaWWh} are in order. First of all we remark that in addition to the charged current hard diboson production in \eqref{diboson_charged}, also $\ell^+\ell^-\to Zh$ and $\ell^+\ell^-\to W^+W^-$ with a  $Z\to W W$ or $W\to W \h$  splitting from one of the heavy final states contribute to the $W^+W^-h$ final state. These contributions are responsible for the $C_B$ dependence of the cross-section (notice that the $ \ell^\pm\nu \to W^\pm h$ amplitude depends only on $C_W$), and they tend to align the flat direction of the $WW\h$ cross-section in the ($C_W, C_B$) plane with the one of the inclusive $Zh$ and $WW$ measurements. One could in principle isolate the genuine $\ell\nu\to Wh$ contribution imposing additional cuts on the final state. The contribution from $\ell^+\ell^- \to Z^* h \to W^+W^-h$ can e.g.\ be reduced requiring the two $W$ bosons to be back-to-back, or requiring one of them to have a low $p_T$. This is however obtained at the expense of rate and might require further luminosity to be exploited. 

Figure~\ref{fig:chi2-profiles-cBcW}, panel (d) shows the $\chi^2$ profile for the $WWh$ analysis as a function of $C_B$ and $C_W$ for $E_{\rm{cm}}=10{\rm \, TeV}$, also in combination with the unpolarized and inclusive $WW$ and $Zh$ measurements. Adding the three-body process eliminates the second solution at large negative couplings and improves the determination of the couplings around the SM point. The results at higher VHEL energies are reported in Figures~\ref{APP14} and~\ref{APP30} in Appendix~\ref{appen1}. As expected, the impact of the $WWh$ process becomes more pronounced as the energy increases, due to the soft-collinear logarithm. The $2\sigma$ sensitivity contours in the $(C_W,C_B)$ plane, obtained from the combination of inclusive $Zh$, differential $WW$ and fiducial $WWh$, with unpolarized beams, are summarized for the different VHEL energies in Figure~\ref{occhio} in the Appendix~\ref{appen1}.

The tri-boson $WW\h$ process illustrates an important aspect of the VHEL phenomenology. The real emission of soft or collinear massive vector bosons is IR-enhanced owing to the large scale separation $E_{\rm{cm}}\gg m_W$. Indeed the fiducial tri-boson cross-section in Table~\ref{tab:sigmaWWh} is close to the diboson $Zh$ cross-section~(\ref{xszh}) and only a factor few smaller than the fiducial $WW$ cross-section (see Table~\ref{tab:WW}). It is thus mandatory to take these emissions into account, for an accurate estimate both of the SM background and of the EFT signal, even in total or fiducial cross-section studies. Furthermore, since the emissions mix up the different $2\to2$ diboson subprocesses, a combined study of all ``hard'' ($Zh$, $WW$, $Wh$ and $WZ$) final states will be necessary. The exclusive approach we adopted here can be only regarded as a first estimate of the VHEL sensitivity. Finally, we notice that the Sudakov $ \log^2(E_{\rm{cm}}^2/m_W^2)$ enhancement of the real emissions also controls the enhancement of the virtual corrections to the exclusive diboson cross-sections. Our tree-level predictions are thus expected to receive large NLO EW corrections that will have to be included. We will return to these important methodological considerations in the Conclusions.

\section{Double Higgs production}\label{highrate}

The second way to probe new physics with high precision at the VHEL is to exploit the large number of events produced in vector boson fusion (or scattering) processes. The cross-sections for such processes grow logarithmically with the collider energy, and with the luminosity scaling of eq.~\eqref{lumieq}, the total number of SM VBF events 
thus grows as $E_{\rm cm}^2 \log E_{\rm cm}$ for large energies as in Figure~\ref{SM}. A clear example of this enhancement is the high rate of single Higgs production attainable at a VHEL, which e.g.\ allows to produce $\sim 10^{8}$ Higgs bosons at the $30$~TeV collider. With such a huge number of events, the precision of single Higgs measurements is realistically going to be dominated by systematic uncertainties -- both experimental and theoretical -- which at present can not be quantified. On the other hand, rarer VBF processes like double Higgs production will still be statistically limited, so that a simple estimate of the reach is possible, and might profit fully from the large available rate without hitting the floor of systematic and theoretical uncertainties.

\begin{table}
\centering%
\renewcommand{\arraystretch}{1.3}
\begin{tabular}{|c|c|c|c|c|}
\hline
$E_{\rm{cm}}$/TeV & $\mathcal{L}/{\rm ab}^{-1}$ & $\sigma({\ell^+\ell^-\to hh\nu\bar\nu})$/fb 
& $N_{\rm events}$\\\hline
3 & 5 & $0.82\cdot \big[1 - 0.63\,\dl + 0.48\,\dl^2\big]$ 
& 4k\\\hline
10 & 10 & $3.3\cdot\big[1 - 0.38\,\dl + 0.27\,\dl^2\big]$ 
& 33k\\\hline
14 & 20 & $4.4\cdot\big[1 - 0.34\,\dl + 0.23\,\dl^2\big]$ 
& 88k\\\hline
30 & 90 & $7.4\cdot\big[1 - 0.27\,\dl + 0.18\,\dl^2\big]$ 
& 660k\\\hline
\end{tabular}
\caption{\label{tab:totalHH} Total di-Higgs production cross-section from charged VBF, with its dependence on the trilinear coupling modification $\dl$, and total number of events at different collider energies.}
\end{table}

In this section we study the potential of the VBF double Higgs production process as a probe of two different new physics effects. The first one is an anomalous triple Higgs coupling, which can be tested through the  measurement of the total (or fiducial, in the central angular region) double Higgs production cross-section. The second is the $\OH$ SILH-basis operator, which can be related to the parameter $\xi=v^2/f^2$ of Higgs compositeness~\cite{Giudice:2007fh}. This effect can be probed by the measurement of the differential double Higgs production cross-section in the tail of the di-Higgs invariant mass distribution~\cite{Contino:2013gna} (see also \cite{Han:2020pif} for a recent analysis at the VHEL).

\subsection{Total cross-section and triple Higgs coupling}\label{3h}
Double Higgs production receives a diagrammatic tree-level contribution that depends on the trilinear Higgs coupling. Therefore it provides a so-called ``direct'' measurement of the parameter $\dl$ in the Higgs potential
\begin{equation}
\displaystyle
V(h) = \frac{m_h^2}{2}h^2 + \frac{m_h^2}{2v}\left(1 + \dl \right)h^3 + \frac{m_h^2}{8v^2}\left(1 + \dlq\right)h^4+\ldots\,,
\end{equation}
where $v\simeq246$~GeV is the Higgs VEV and $m_h$ is the physical Higgs boson mass. The trilinear coupling measurement, which is difficult to achieve at the LHC~\cite{Cepeda:2019klc} at a satisfactory level of accuracy, is a standard reference target for future colliders. A $100$~TeV hadron collider is expected to be able to measure modifications in the trilinear coupling $\dl^{\rm FCC} \approx 3.5\%\,\text{--}\,8\%$, depending on the assumptions on detector performance~\cite{Mangano:2020sao}. At lepton colliders, high energies are needed to produce a significant amount of Higgs boson pairs in VBF. The 3 TeV CLIC can reach a precision $\dl^{\rm CLIC} \approx 10\%$~\cite{Roloff:2019crr}. Here we will estimate the sensitivity to $\dl$ at the VHEL.

In the absence of BSM light degrees of freedom, and assuming Custodial Symmetry, double Higgs production is affected by new physics through the following two interactions~\footnote{The third operator that contributes at tree level (called ${\mathcal{O}}_T$ the SILH basis) is neglected because it breaks Custodial Symmetry and is strongly bounded by LEP, at the level $C_Tv^2 \lesssim 10^{-3}$. We estimate that the VHEL sensitivity to this operator could become comparable to the one of LEP only at $E_{\rm{cm}}=30$~TeV.}
\begin{align}\label{SILH}
\displaystyle
\Osix &= -\frac{m_h^2}{2v^2} \left(H^\dag H-\frac{v^2}2\right)^3, & \OH &= \frac{1}{2}\left( \partial_\mu (H^\dag H)\right)^2\,.
\end{align}
The coefficients of these operators, $C_6$ and $C_H$, are related to triple Higgs coupling modifications
\begin{equation}\label{triple_c}
\displaystyle
\dl = v^2\left(C_6 - \frac{3}{2}C_H\right)\,.
\end{equation}
The quartic coupling $\dlq$, studied at VHEL in~\cite{Chiesa:2020awd}, is correlated with $\dl$ in the EFT.
Additionally, the ${\mathcal{O}_H}$ operator also induces a universal rescaling of the single Higgs couplings to vectors and fermions
\begin{equation}\label{single_c}
\displaystyle
\kappa=\kappa_V = \kappa_f = 1 -\frac{C_H\, v^2}2\,.
\end{equation}

The total SM cross-sections for double Higgs production, along with those for several other VBF processes, have been presented in~\cite{Costantini:2020stv} for different multi-TeV collider energies. In Table~\ref{tab:totalHH} we report the cross-sections for the dominant charged VBF $W^+W^- \to hh$ process, which are of the order of a few fb at the various VHEL under consideration, together with their dependence on the anomalous trilinear $\dl$. With the baseline luminosity of eq.~\eqref{lumieq}, around $10^4$ (few $10^5$) events are expected at the $10$~TeV ($30$~TeV) VHEL. Note that the contamination from invisible $Z$ decays to the complete $2\to4$ process, ${\ell^+\ell^-\to hh\nu\bar\nu}$, is very small at these energies, with the $\ell^+\ell^-\to Zhh$ cross-section being of the order of a few ab.

From these numbers one can derive a first rough estimate of the precision attainable on the triple Higgs coupling. For simplicity we consider only the $hh\to 4b$ channel, keeping in mind that a complete analysis could include several other decay channels. Taking into account the branching ratio ${\rm BR}(h\to b\bar b) = 0.58$ for both Higgs bosons, and assuming an overall reconstruction efficiency of $\approx 30\%$ (see the CLIC analysis~\cite{Roloff:2019crr}), one gets $3300$ reconstructed di-Higgs events at the $10$~TeV collider. Neglecting backgrounds, the statistical precision on the cross-section is therefore expected to be around $1.7\%$. With the sensitivity to $\dl$ reported in the table, this corresponds to a $4\%$ precision on the trilinear coupling. At a $30$~TeV collider, instead, one expects around 600'000 events, which correspond to a percent precision on $\dl$. These numbers are in agreement with those of~\cite{Costantini:2020stv,Han:2020pif}, which however do not include reconstruction efficiencies.

These order-of-magnitude estimates could be significantly affected by backgrounds, but even more significantly by detector acceptance. VBF Higgs pair production is a soft process, meaning that the Higgs bosons do not have large transverse momentum. However, they can have considerable longitudinal boost if the collider energy is large. A very large fraction of the Higgs bosons is thus produced in the forward (and backward) region, beyond the detector coverage. The problem could be particularly severe at a $\mu^+\! \mu^-$ VHEL because the radiation-absorbing nozzles might reduce the angular coverage significantly. In addition, soft beam-induced background might affect the ability to reconstruct low-$p_T$ objects from the decay of low-$p_T$ Higgs bosons. 


On the other hand, the contribution of the trilinear coupling to the double Higgs cross-section only comes from the Feynman diagram with a virtual Higgs boson in the s-channel, and therefore it is independent of the scattering angle in the $hh$ rest frame. 
The SM total cross-section instead gets a large contribution from the the t-channel exchange of a virtual $W$, which is enhanced in the kinematical region where the Higgs bosons are produced at small angle in the $hh$ frame. This contribution, however, is insensitive to $\dl$ and it can be regarded as a ``background'' to the $\dl$ determination.
%
The presence of an enhanced signal-free kinematical region explains the relatively low sensitivity of the total cross-section to $\dl$ that we found in Table~\ref{tab:totalHH}.
The sensitivity decreases with $E_{\rm{cm}}$ as the t-channel enhancement of the background gets more significant at higher energy. 
\begin{table}
\begin{centering}
\renewcommand{\arraystretch}{1.2}
\begin{tabular}{|c|c|c|c|c|}
\hline 
$\sqrt{s}$/TeV & $\mathcal{L}/\text{ab}^{-1}$ & $\sigma(\llll\to hh(4b)\nu\bar\nu)$/ab & $N_{\rm SM}$ events & 68\% C.L. $\dl$\tabularnewline
\hline 
\hline 
3 & $5$ &
$\begin{aligned}[t]&132\cdot\big[1 +\left(3.85\, C_{H} - 0.87\, C_{6}\right) v^2\\
 & +\left(26.8\, C_{H}^2 + 0.74\, C_{6}^{2} - 5.52\, C_H C_6 \right) v^4\big]
\end{aligned}$
& 172 & {[}-8.5, 9.9{]} \%\\[18pt]
\hline 
10 & $10$ &
$\begin{aligned}[t]&239\cdot \big[1 + \left(7.25\, C_{H} - 0.80\, C_{6}\right) v^2\\
 & +\left(196\, C_{H}^2 + 0.71\, C_{6}^{2} - 8.40\, C_H C_6\right) v^4\big]
\end{aligned}$
& 621 & {[}-4.9, 5.3{]} \%\\[18pt]
\hline 
14 & 20 &
$\begin{aligned}[t]&257\cdot\big[1 + \left(8.43\, C_{H} - 0.79\, C_{6}\right) v^2\\
 & +\left(300\, C_{H}^2 + 0.68\, C_{6}^{2} - 9.28\, C_H C_6\right)v^4\big]
\end{aligned}$
& 1336 & {[}-3.4, 3.6{]} \%\\[18pt]
\hline 
30 & $90$ &
$\begin{aligned}[t]&271\cdot\big[1 + \left(12.8\, C_{H} - 0.79\,C_{6}\right)v^2\\
 & +\left(1389\, C_{H}^2 + 0.78\, C_{6}^{2} - 13.8\, C_H C_6\right)v^4
\end{aligned}$
& 6341 & {[}-1.6, 1.6{]} \%\\[18pt]
\hline 
\end{tabular}
\par\end{centering}
\caption{\label{tab:hhxsec} Fiducial $\sigma(\llll\to hh(4b)\nu\nu)$ in ab at various high
energy lepton colliders as a function of the new physics couplings
$C_{H}$ and $C_{6}$, normalized with $v = 246$~GeV. The cross-sections are calculated in the fiducial region $10^\circ < \theta_b < 170^\circ$, $p_{T,b} > 10$ GeV. We also report the number of SM events, and the corresponding 68\% C.L.\ bound on the modified trilinear coupling $\dl$, calculated neglecting backgrounds and assuming a selection efficiency $\epsilon_{\rm sig} = 26\%$ on the signal.}
\end{table}
The Higgs bosons produced by the ``background'' SM diagrams are forward (and backward) already in the $hh$ rest frame. After the longitudinal boost, they will thus move almost parallel to the beam axis. The $\dl$ contribution instead is central in the $hh$ frame, producing relatively more central Higgs bosons in the lab frame. Being obliged to restrict the measurement of the cross-section to the central region because of the detector acceptance, eliminating the enhanced background component, could thus even be beneficial for the measurement of the triple coupling, in spite of the radical reduction of the total rate.

We shall now quantify the competing aspects described above by performing a simulation of double Higgs production in the $hh\to 4b$ channel. We compute the total and differential cross-section using {\sc{MadGraph}}~\cite{Alwall:2014hca}. We impose basic detector acceptance cuts, requiring that the $b$ quarks have a transverse momentum larger than $10$~GeV, and exclude a region of $10^\circ$ around the beam axis. The new physics effects are modeled by implementing the $\mathcal{O}_6$ and $\mathcal{O}_H$ operators in eq.~(\ref{SILH}) using FeynRules~\cite{Alloul:2013bka}. The resulting fiducial cross-sections, for various collider energies, are reported in Table~\ref{tab:hhxsec} as functions of the new physics couplings $C_6$ and $C_H$. We first consider new physics that only affects the trilinear Higgs couplings, leaving the single Higgs couplings at their SM values. This corresponds to the configuration $C_H=0$ and $C_6=\dl/v^2$ owing to eq.~(\ref{triple_c}). We will discuss later how to deal with the degeneracy that emerges in the combined determination of $C_6$ and $C_H$ from the measurement of the fiducial cross-section.

\begin{figure}
\includegraphics[width=0.47\textwidth]{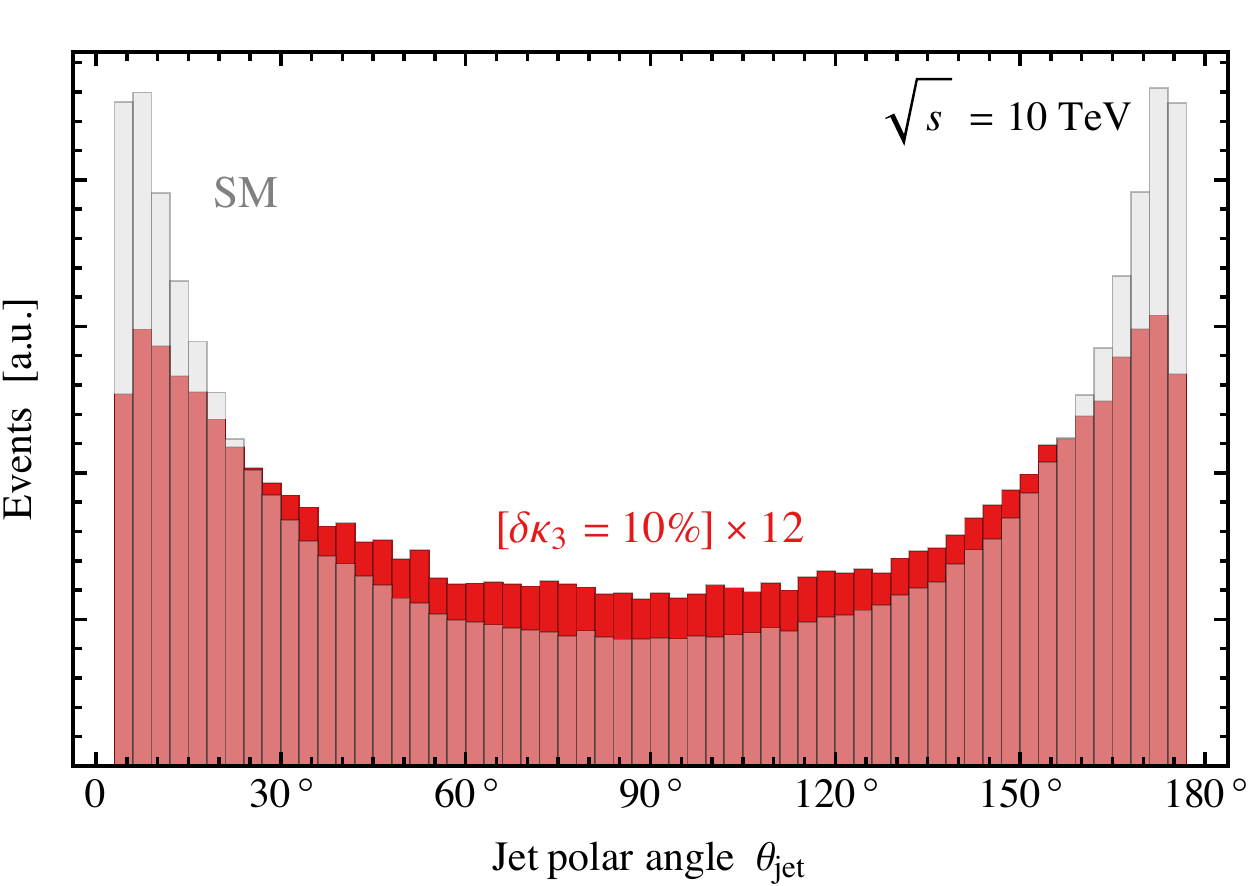}\hfill%
\includegraphics[width=0.48\textwidth]{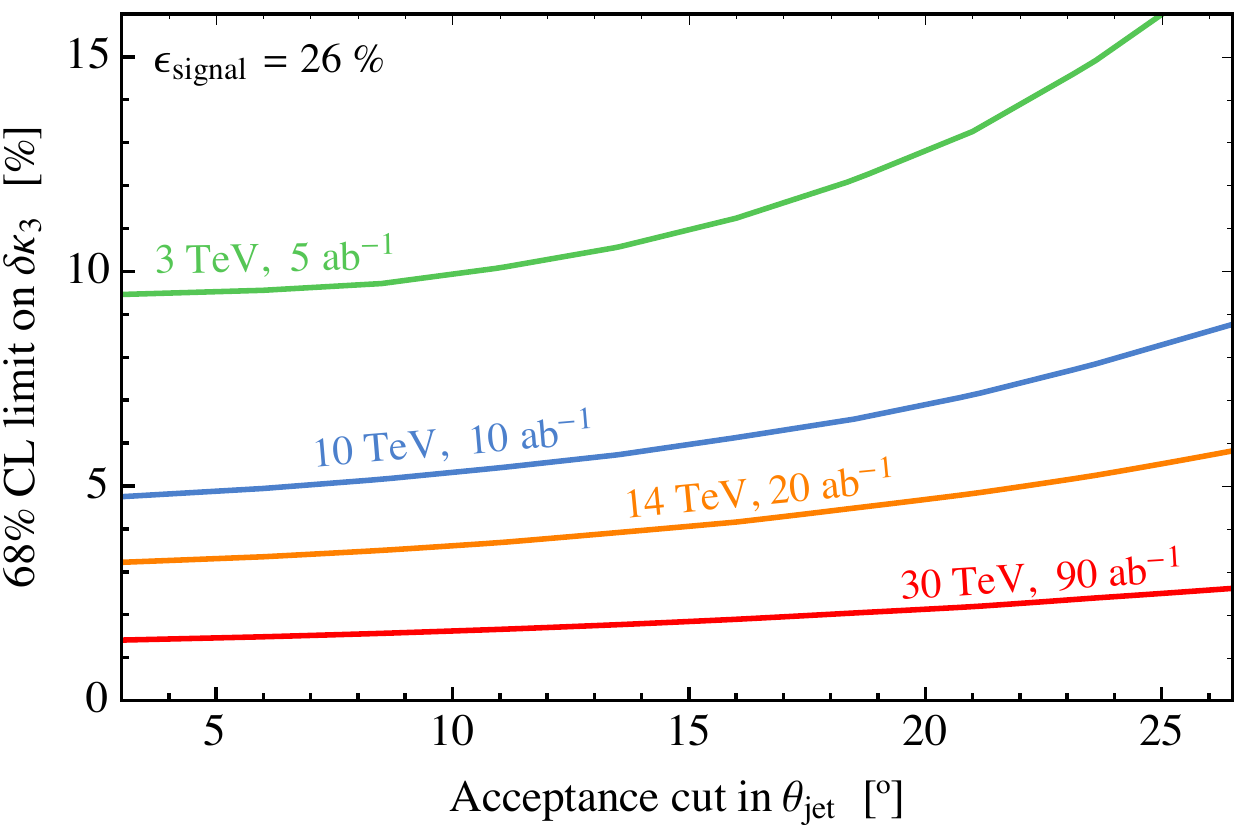}
\caption{\label{fig:angular_lambda}{Left:} Jet polar angle distribution of $\ell^+\ell^- \to hh\nu\nu$ events at $\sqrt{s} = 10$ TeV in the SM (gray) and with $\dl = 10\%$ (red). {Right:} Reach on $\dl$ at 68\% C.L. as a function of the acceptance cut on the jet polar angle at muon colliders with different center of mass energies.}
\end{figure}

One immediately notices that the fiducial SM cross-sections in Table~\ref{tab:hhxsec} do not grow with $E_{\rm{cm}}$ as fast as the total cross-sections of Table~\ref{tab:totalHH}. This is due to the 
enhancements of the cross-section in the forward/backward region discussed before, which become more prominent at higher $E_{\rm{cm}}$, producing a larger total rate but also a larger fraction of Higgs bosons (and, in turn, of $b$-quarks) 
outside the detector acceptance. This enhancement is clearly visible in the distributions of the left panel of Figure~\ref{fig:angular_lambda}. The plot also shows that the contribution from the anomalous trilinear Higgs coupling (defined as the correction to the cross-section in absolute value for $\dl=10\,\%$) is less peaked in the forward detector region, since it lacks the t-channel enhancement of the SM part, and is thus less affected by the acceptance cuts. As a result, the sensitivity of the cross-sections in Table~\ref{tab:hhxsec} to $\dl = C_6 v^2$ is larger and roughly constant for all center of mass energies, as opposed to what we see in Table~\ref{tab:totalHH} for the total cross-section. 

Backgrounds are potentially large and need to be taken into account. The most important source of background is VBF diboson production -- mainly $Zh$, but also $ZZ$, $WW$, or $Wh$, $WZ$ produced in association with a collinear lepton -- where one or two vector bosons are incorrectly reconstructed as a Higgs. Hard $\ell^+\ell^- \to VV$ diboson processes are easily removed as the diboson invariant mass peaks at $E_{\rm{cm}}$ while the di-Higgs invariant mass distribution is much softer. The key factor to isolate the $hh$ signal is thus the ability to distinguish $h\to b\bar b$ from hadronic $Z$ and $W$ decays. We simulate all these VBF diboson processes in the SM with MadGraph, requiring the four jets to be in the same acceptance region in $\theta$ and $p_T$ defined above. In order to assess the background contamination, we apply a gaussian smearing on the jet energy, assuming an energy resolution $\Delta E_{\rm jet}/E_{\rm jet} = 10\%$. Then, we reconstruct the Higgs bosons of the signal pairing the four jets by minimizing $|M_{j_1j_2} - m_h| + |M_{j_3j_4} - m_h|$, where $M_{jj}$ is the invariant mass of two jets and $m_h$ is the Higgs mass. Finally, we select the signal events requiring that for the Higgs candidates the dijet invariant mass $M_{jj} > M_{\rm cut}$, and that at least $N_b$ jets out of four are tagged as $b$-jets (we assume a $b$-tag efficiency of 70\% and a misidentification probability as given in~\cite{Abramowicz:2018rjq}). We optimize the significance of the $hh$ cross-section measurement varying $M_{\rm cut}$ and $N_b$. At $3$~TeV center of mass energy we find the optimal values $M_{\rm cut} = 105$ GeV, $N_b = 3$, with a corresponding signal selection efficiency $\epsilon_{\rm sig} = 25\%$. This result is in perfect agreement with the results of Ref.~\cite{Roloff:2019crr}, based on a full detector simulation and a BDT selection, which quote $\epsilon_{\rm sig} = 26\%$. At $10$~TeV we find a very similar result, with $\epsilon_{\rm sig} = 32\%$. The number of background events that pass the selection cuts is of the same order as the number of signal events. We have also checked that varying the energy resolution on the diboson invariant mass by $\pm 50\%$ has a minor impact on the optimal efficiency, although increasing the background contamination. More details are reported in Appendix~\ref{appen2}.
 
Given these considerations, we simply estimate the error on the cross-section as $\Delta\sigma \sim \sqrt{\mathcal{L}\cdot \epsilon_{\rm sig}\cdot \sigma}$, using the value $\epsilon_{\rm sig} = 26\%$ for all collider energies, but keeping in mind that with a different efficiency the result scales as $\sqrt{\epsilon_{\rm sig}}$. The final precision on the modified trilinear coupling is given in the last column in Table~\ref{tab:hhxsec}.\footnote{For a 3 TeV collider we use a luminosity of 5 ab$^{-1}$, for ease of comparison with the CLIC studies.} A $10$~TeV muon collider could reach a 5\% precision, while a 1.5\% precision can be reached at a $30$~TeV collider. The results are in agreement with the previous rough estimates based on the total number of events, but this is purely accidental. Indeed, the acceptance cuts reduce the cross-section by a very large factor (almost $300$ at $30$~TeV), but the reduced number of events is compensated by a stronger sensitivity to $\dl$.

\begin{figure}[t]
\includegraphics[width=0.48\textwidth]{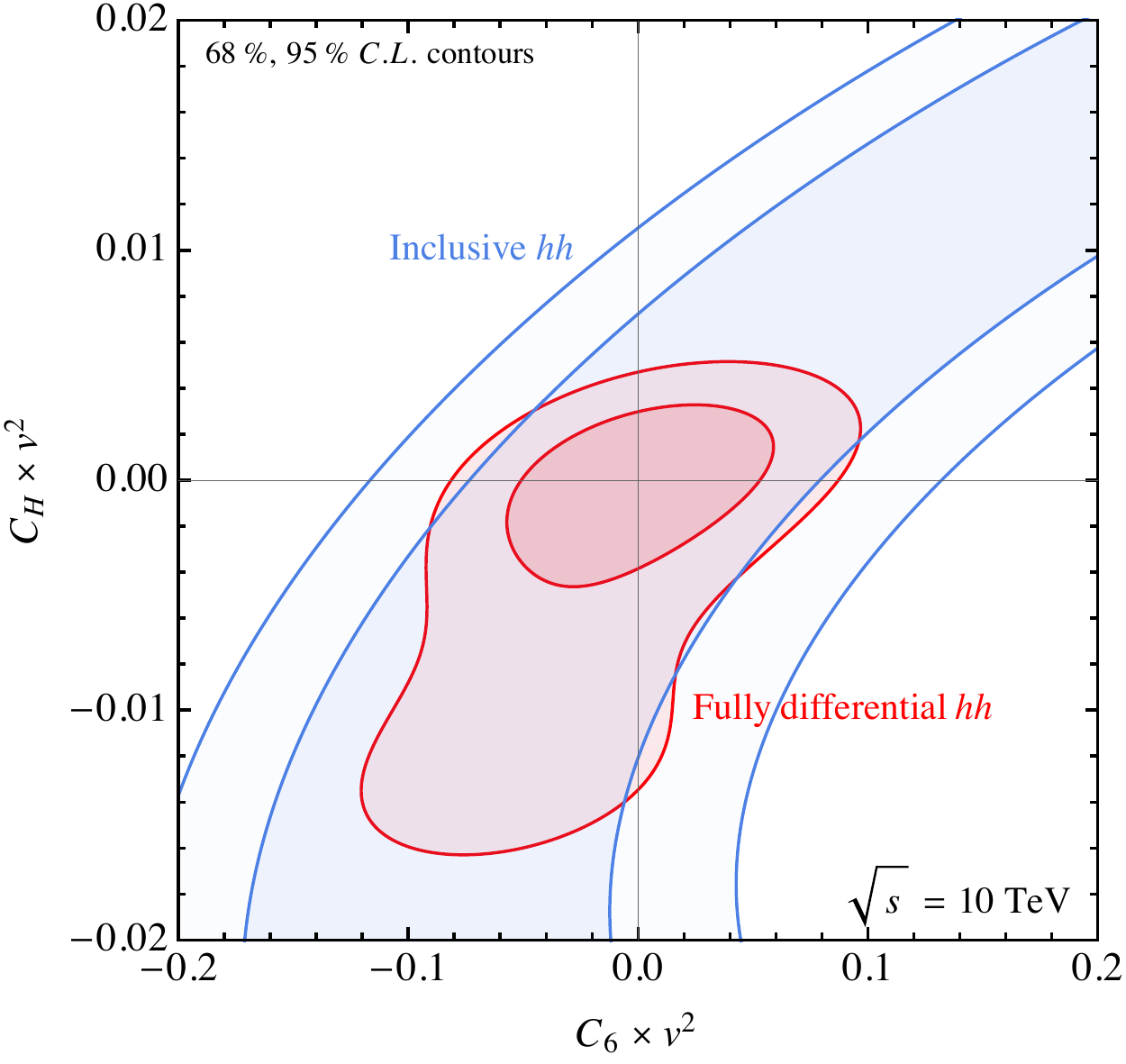}\hfill%
\includegraphics[width=0.48\textwidth]{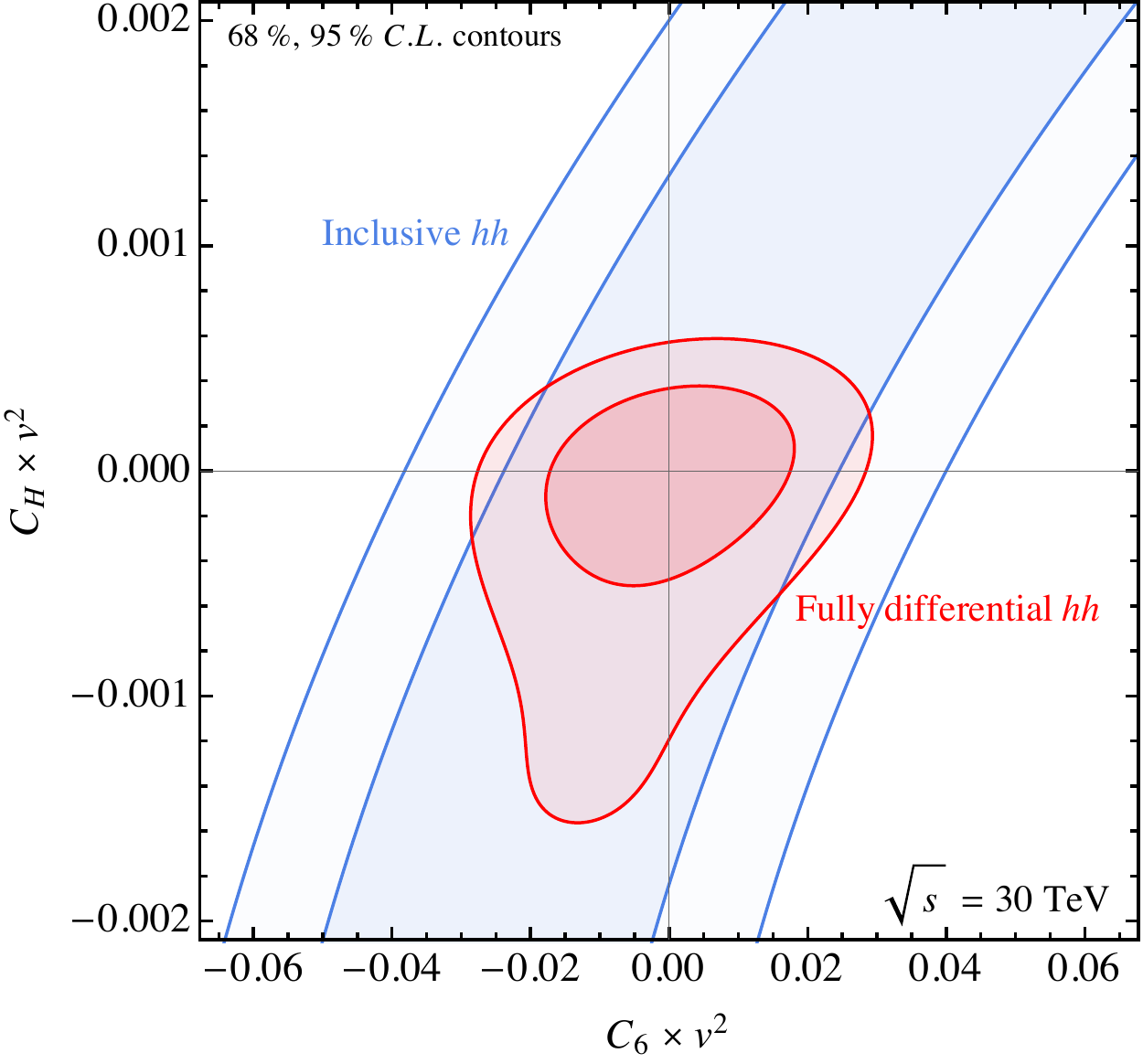}
\caption{\label{fig:constraints_hh} Constraints on the Wilson coefficients $C_6$ and $C_H$ of eq.~\eqref{SILH} from the inclusive $hh\to 4b$ cross-section measurement (blue), and from its differential distribution in di-Higgs invariant mass $M_{hh}$ and Higgs transverse momentum $p_{T,h}$ (red). The contours indicate 68\% and 95\% C.L. constraints (2 d.o.f.). {Left:} $E_{\rm{cm}}= 10$ TeV. {Right:} $E_{\rm{cm}} = 30$ TeV.}
\end{figure}

Since the detector specifications for a VHEL are not known at present, it is interesting to assess the dependence of our results on the assumptions about acceptance and efficiency. In Figure~\ref{fig:angular_lambda} right we show the dependence of the trilinear coupling limit on the acceptance cuts on the polar angle $\theta_{\rm jet}$. We see that the reach is not drastically affected by the angular cut. This is due to the fact that the trilinear coupling contributes to the cross-section mainly in the central region, while the forward events come from SM processes. Restricting to a more central region reduces the total number of events, but increases the sensitivity on $\dl$ as previously explained. Similarly, requiring harder jets does not significantly affect the results, as long as one is able to detect jets with $p_T \gtrsim$ 30--40 GeV (see Appendix~\ref{appen2}).

Finally, we interpret the cross-section measurement as a constraint on the EFT coefficients $C_6$ and $C_H$. The dependence on $C_H$ of the total cross-section is due to both the contribution to $\dl$ in eq.~(\ref{triple_c}), and the modification of the coupling to vector bosons. The resulting bounds are shown as blue contours in the $(C_6, C_H)$ plane in Figure~\ref{fig:constraints_hh}, for collider energies of 10 TeV and 30 TeV. As is well known, the measurement of the total cross-section alone is not able to determine the couplings $C_6$ and $C_H$ simultaneously.

\subsection{High mass tail}\label{inbetween}

The degeneracy between the contributions from $\mathcal{O}_6$ and $\mathcal{O}_H$ to double Higgs production can be resolved considering the differential cross-section ${\rm d}^2\sigma / {\rm d}\hat s\, {\rm d} p_{T,h}$, where $\hat s = M_{hh}^2$ is the di-Higgs invariant mass and $p_{T,h}$ the transverse momentum of the softest Higgs. The invariant mass is the most important variable to discriminate between $C_6$ and $C_H$. The low invariant mass region, which dominates in the total cross-section is mostly affected by the trilinear Higgs coupling, and thus gives a good measurement of only $\dl$ in absence of other new physics effects. However, the $WW\to hh$ amplitude also has a contribution proportional to $C_H$ that grows with energy, $\mathcal{A}_{WW\to hh} \approx C_H \hat s$ for $\hat s \gg m_h^2$. From the viewpoint of the Goldstone Boson Equivalence Theorem, this energy growth emerges from the derivatives in the operator $\mathcal{O}_H$. The highest sensitivity to the interaction $\mathcal{O}_H$ therefore comes from the hard scattering at high di-Higgs invariant masses, while the low invariant mass region is most sensitive to the trilinear coupling $\dl$. A differential measurement of double Higgs production over the full kinematical range can therefore constrain the two new physics couplings $C_H$ and $C_6$ simultaneously.

\begin{table}
{\small
\begin{centering}
\renewcommand{\arraystretch}{1.2}
\begin{tabular}{|c|c|c|c|c|c|c|}
\hline 
\!\!$\sqrt{s}$/TeV\!\! & \!\!$\mathcal{L}/\text{ab}^{-1}$\!\! & $\sigma(\llll\to hh\nu\bar\nu)$/ab & \!\!$\{M_{hh}^{\rm cut}, p_{T,h}^{\rm cut}\}$/GeV\!\! & \!$N_{\rm SM}$\! & \!95\% C.L. $\xi$\!\tabularnewline
\hline 
\hline 
3 & $5$ &
$\begin{aligned}[t]&95.6\cdot\big[1 + \left(10.0\, C_{H} - 0.87\, C_{6}\right) v^2\\
 &+ \left(106\, C_{H}^2 + 0.72\, C_{6}^{2} - 12.7\, C_H C_6 \right) v^4\big]
\end{aligned}$
& \{340, 150\} & 100 & $2.1 \times 10^{-2}$\\[16pt]
\hline 
6 & 3.6 & $\begin{aligned}[t]&45.0\cdot\big[1 + \left(33.1\, C_{H} - 0.98\, C_{6}\right)v^2 \\
 &+\left(986\, C_{H}^2 + 0.63\, C_{6}^{2} - 35.8\, C_H C_6\right) v^4\big]
\end{aligned}$
& \{680, 300\} & 34 & $1.1\times 10^{-2}$\\[16pt]
\hline 
10 & 10 &
$\begin{aligned}[t]& 21.7\cdot\big[1 +\left(77.1\, C_{H} - 0.88\, C_{6}\right) v^2 \\
 &+ \left(5880\, C_{H}^2 + 0.54\, C_{6}^{2} - 81.4\, C_H C_6\right) v^4\big]
\end{aligned}$
& \{1130, 500\} & 46 & $3.9\times 10^{-3}$\\[16pt]
\hline
14 & $20$ &
$\begin{aligned}[t]&\!\! 13.7\cdot\big[1 +\left(131\, C_{H} - 0.81\, C_{6}\right)v^2 +\\
  & \left(1.86\!\times\! 10^4\, C_{H}^2 + 0.51\, C_{6}^{2} - 137.9\, C_H C_6\right) v^4\big]\!\!
\end{aligned}$
& \{1500, 690\} & 58 & $2.0\times 10^{-3}$\\[16pt]
\hline 
30 & $90$ &
$\begin{aligned}[t]&\!\! 5.1\cdot\big[1 +\left(409\, C_{H} - 0.66\, C_{6}\right) v^2 +\\
& \left(2.36\!\times\! 10^5\, C_{H}^2 + 0.41\, C_{6}^{2} - 428\, C_H C_6\right) v^4\big]\!\!
\end{aligned}$
& \{2800, 1360\} & 96 & $4.7\times 10^{-4}$\\[16pt]
\hline 
\end{tabular}
\par\end{centering}}
\caption{\label{tab:hhhighmass} High-mass $\sigma(\llll\to hh\nu\nu)$ in ab at various high
energy lepton colliders as a function of the new physics couplings
$C_{H}$ and $C_{6}$, normalized with $v = 246$~GeV. The cross-sections are calculated in the high invariant mass region defined by the cuts given in the fourth column, and requiring $\eta_h < 2$. We also report the total number of SM $h\to jj$ events, and the corresponding 95\% C.L. bound on the parameter $\xi \equiv C_H v^2$ ($C_H>0$), calculated assuming a di-Higgs reconstruction efficiency $\epsilon_{\rm sig} = 30\%$ for boosted hadronic Higgs bosons.}
\end{table}

The Higgs bosons produced in the hard scattering will be boosted, so a strategy based on reconstructing the Higgs decay products (e.g., the $4$ $b$-jets, like in the previous section) individually might not be effective. We therefore estimate the reach in the high invariant mass kinematical region by simulating $\ell^+\ell^-\to hh\nu\bar\nu$ events without decays, and assigning an overall efficiency for correctly tagging a pair of boosted Higgs bosons. We consider the cross-section as a function of $M_{hh}$ and $p_{T,h}$, and we perform a differential analysis dividing the phase space in 9 bins -- three bins in each variable -- chosen in order to maximize the sensitivity to the new physics coefficients. Furthermore, we require the Higgs bosons to be in the central region with rapidity $\eta_h < 2$ (i.e., around $15^\circ$). The cuts that define the bin of highest invariant mass and $p_T$ are reported in Table~\ref{tab:hhhighmass}, together with the corresponding $\ell^+\ell^- \to hh\nu\bar\nu$ cross-sections as functions of $C_H$ and $C_6$, for the different collider benchmarks. This bin is the one that dominates the single-operator $C_H$ sensitivity for $C_H>0$. The other bins are important for the global sensitivity in the $(C_6,C_H)$ plane.  Notice that the optimal cuts in $(M_{hh}, p_{T,h})$ scale roughly linearly with the collider energy, as one would na\"ively expect in the very high energy regime where all the masses can be neglected. Also note that the SM cross-section decreases with increasing $E_{\rm{cm}}$ since no logarithmic enhancement is present in the high-mass region. We then compute the overall event yield, including the hadronic dijet decay modes $h\to b\bar b, c\bar c, gg$, that add up to ${\rm BR}_{h\to jj} = 70\%$. We do not include hadronic $WW$, $ZZ$ and $\tau\tau$ modes, but they could also be considered in order to increase the number of events. The di-Higgs tagging efficiency is taken to be $\epsilon_{hh} = 30\%$. The resulting number of reconstructed SM events is reported in the fifth column of Table~\ref{tab:hhhighmass}.

\begin{figure}\begin{center}
\includegraphics[width=0.49\textwidth]{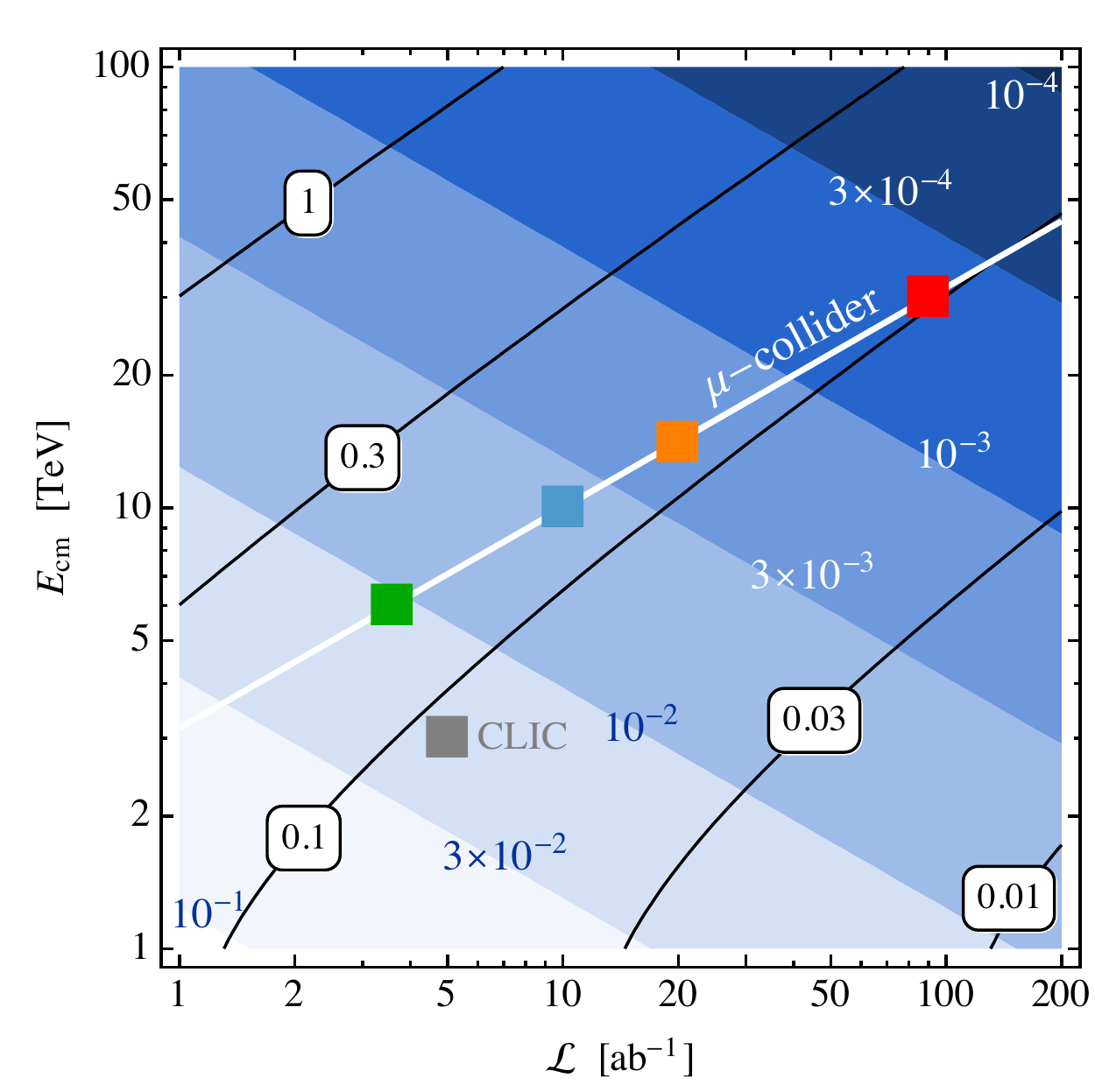}\hfill%
\includegraphics[width=0.5\textwidth]{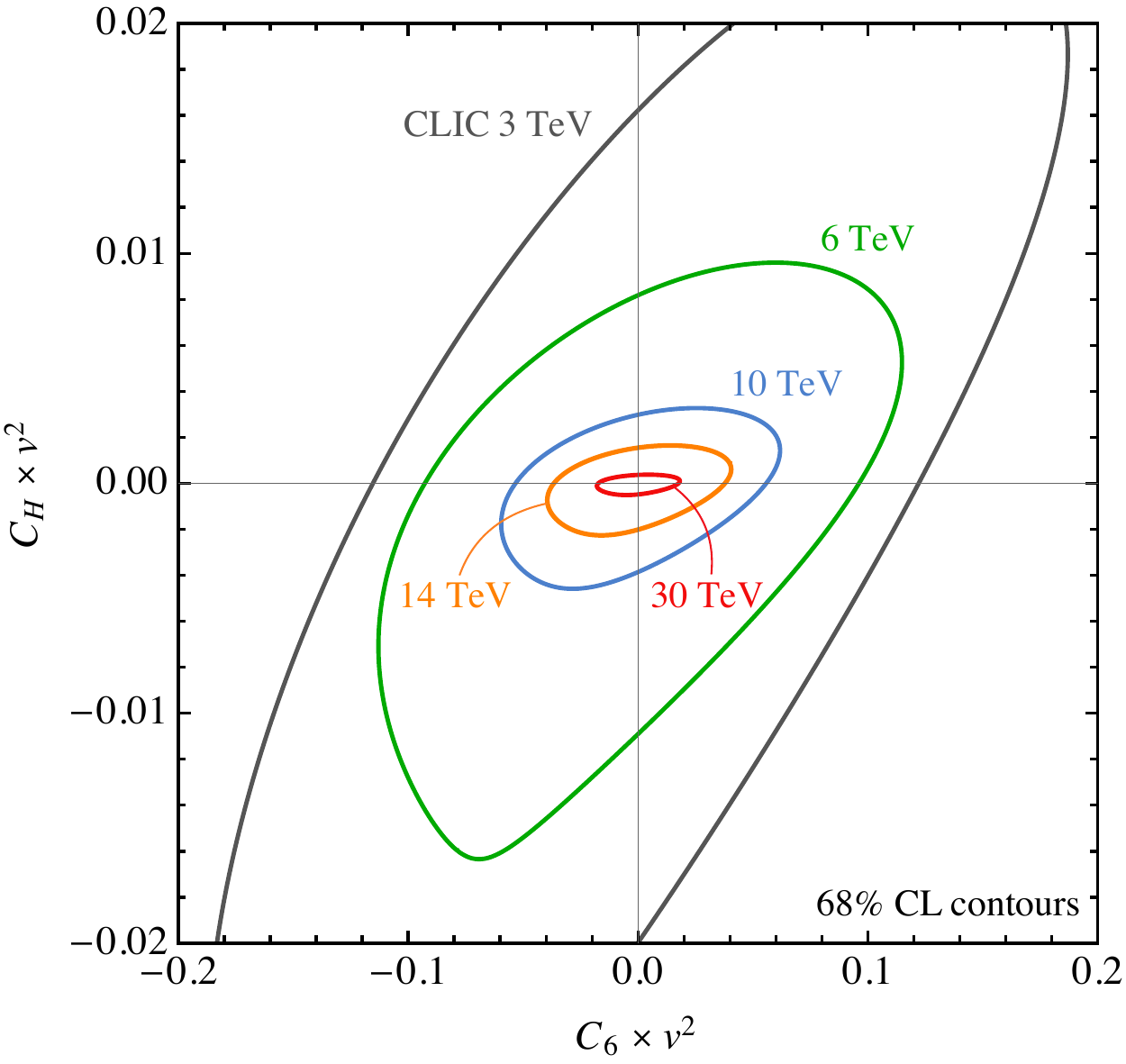}
\caption{{Left:} 95\% C.L.\ reach on $\xi\equiv C_H v^2$ (blue contours), and isolines of $S/B$ (black~contours) as a function of collider luminosity and energy. The baseline luminosity in eq.~\eqref{lumieq} is highlighted as a white line. {Right:} Combined constraints at 68\% C.L.\ in the $(C_H, C_6)$ plane from double Higgs production, for $E_{\rm{cm}}=3$ (gray), $6$~(green), $10$ (blue) and $14$ (orange), and $30$ TeV (red).\label{fig:hhcombined}}
\end{center}\end{figure}

The single-operator sensitivity to $C_H$ (assuming $C_H>0$)  is given in the last column of Table~\ref{tab:hhhighmass}, expressed in terms of the 95\% C.L.\ bound on the parameter $\xi \equiv v^2/f^2 \equiv C_H v^2$, which is related to the sigma-model scale $f$ in theories where the Higgs is a composite pseudo-Goldstone boson~\cite{Giudice:2007fh}. The $\xi$ parameter (i.e., $C_H$) also controls single-Higgs coupling modifications $\delta\kern-0.6pt\kappa = - \xi/2$ (see eq.~(\ref{single_c})), that can be probed at the permille level at future Higgs factories such as CLIC, FCCee and ILC~\cite{deBlas:2019rxi}. A similar sensitivity to $\xi$ can be achieved already at the $10$~TeV VHEL by ``directly'' measuring the effect of ${\mathcal{O}}_H$ in double Higgs production. The $14$ and $30$~VHEL sensitivity on $\xi$ exceeds the one of Higgs factories. For instance a the $30$~VHEL sensitivity corresponds to Higgs coupling modification of $\delta\kern-0.6pt\kappa_{2\sigma}^{\text{VHEL30}}\simeq 2\!\times\!10^{-4}$, which would require exquisite experimental and theoretical precision to be detected. The ``direct'' VHEL sensitivity is instead obtained from measurements with $O(10\%)$ precision in the di-Higgs high mass tail, thanks to the enhancement of the new physics effect. The left panel of Figure~\ref{fig:hhcombined} displays the contours of the 95\% C.L. ($2\sigma$) reach on $\xi$ as a function of collider luminosity and energy. The reach closely follows the na\"ive scaling $E_{\rm c.m.}^{-1}\mathcal{L}^{-1/2}$, which holds in the high energy regime if the analysis cuts are scaled linearly with $E_{\rm c.m.}$. We also plot contours of constant $S/B$, which show that precisions better than 10\% are never required. The contours deviate from straight lines because of logarithmic corrections in the cross-section, which become more important for low invariant masses, hence at low collider energy. We did not perform a detailed study of backgrounds. It is however clear that achieving the precisions quoted here is subject to the ability of distinguishing boosted Higgs bosons from hadronically decaying $Z$ and $W$ bosons. 

The combined constraints on the new physics couplings are shown in Figure~\ref{fig:constraints_hh} as red regions in the $(C_6, C_H)$ plane, for a 10 TeV and a 30 TeV muon collider, together with the limit from the total cross-section only (blue regions). Thanks to the differential analysis, the flat direction of the total cross-section is lifted, and the two couplings can be determined independently. The combined 68\% C.L.\ limits are also given in Figure~\ref{fig:hhcombined} (right) for the various collider benchmarks under consideration. Notice that at low center of mass energies, a correlation between $C_H$ and $C_6$ remains even with the differential analysis. This is due to the fact that the high invariant mass tail, most sensitive to $C_H$, is less separated from the low-mass region, and is therefore more affected by the trilinear coupling, as can also be seen from the cross-sections in Table~\ref{tab:hhhighmass}.

\section{Conclusions}\label{conc}

We have studied the potential of a Very High Energy Lepton collider (VHEL) to probe indirectly non-SM phenomena through precision measurements, outlining the coexistence of two distinct approaches. One is the habitual approach to precision physics, that targets those observables 
which can be measured with the highest possible accuracy thanks to the large available statistics. This is a valid strategy at the VHEL because of the high rate of VBF processes. The double-Higgs production process was studied in Section~\ref{highrate} as an illustration of the potential of such high-rate probes. The second approach is to select those observables that are potentially most sensitive to new physics. The high energy available at the VHEL gives access to a large set of observables with superior sensitivity to very heavy new physics. Total, fiducial and differential cross-section measurements in the production of a pair of bosons at the highest available energy was studied in Section~\ref{highen} as an illustration of the high-energy path towards new physics. Other obvious candidate processes in this class are dilepton, diquark, and ditop high-energy productions.

Our results can be summarized as follows. In Section~\ref{3h} we found that the VHEL could measure the anomalous triple Higgs coupling with an uncertainty $\delta\kern-0.6pt\kappa_3^{1\sigma}=\{5\%,3.5\%,1.6\%\}$ for VHEL energies $E_{\rm{cm}}=\{10,14,30\}$~TeV, respectively. We consider these projections rather robust. They are based on the $h\to4b$ decay channel and take into account realistic detector acceptance and (physics) background mitigation strategies such as $b$-tagging and a cut on the reconstructed Higgs bosons invariant mass. CLIC-like detector performances are assumed for the jet energy resolution and for the $b$-tagging efficiency/rejection performances. After optimizing the number of $b$-tags and the invariant mass cut, this simple strategy has been validated against the corresponding full simulation CLIC study finding perfect agreement. However it would be interesting to confirm these findings by a more complete analysis and by employing the ``target'' muon collider DELPHES card~\cite{delphescard} for the detector simulation rather than smearings. 

In Section~\ref{inbetween} we studied the physics potential of differential cross-section measurements probing the high energy tail of the di-Higgs distribution. The tail is a powerful probe of the operator ${\mathcal{O}_H}$ in eq.~(\ref{SILH}), whose effect grows quadratically with the center of mass energy of the $hh$ pair. In Composite Higgs theories where $C_6$ is predicted to be negligible and $C_H=\xi/v^2>0$, the measurement can be translated into a $2\sigma$ reach $\xi_{2\sigma}=\{4\permil,2\permil,0.5\permil\}$, for energies $E_{\rm{cm}}=\{10,14,30\}$~TeV, on the Composite Higgs $\xi=v^2/f^2$ parameter. A non-vanishing $\xi$ (or, $C_H$) also induces a universal rescaling of the single Higgs coupling eq.~(\ref{single_c}). It can thus be probed at future Higgs factories, or potentially also at the VHEL itself, by very accurate 
measurements of single-Higgs cross-sections. The sensitivity we have found here is not only potentially superior ($\xi_{2\sigma}^{\text{FCC}}=3.6\permil$), it is also easier to attain, not being extracted from extremely accurate measurements combined with accurate SM theoretical prediction, but from the relatively inaccurate few-percent level determination of the high-mass cross-section. Furthermore, a positive hint for $\xi\neq0$ would be more easily turned into a new physics discovery by exploiting the measurement of the differential di-Higgs cross-section and the characteristic energy growth of the signal. For the purpose of the present study, it should also be noted that not relying on extremely accurate measurements and SM predictions adds robustness to the sensitivity estimates we presented.

The enhanced sensitivity to $C_H$ in the high energy di-Higgs measurement can be regarded as toy version of the mechanism we saw at work in the study of diboson processes in Section~\ref{highen}. 
Di-Higgs invariant masses from 1 to 3 TeV can be probed, depending on the VHEL energy. This is much above the EW scale and explains the enhanced sensitivity to $C_H$ relative to single-Higgs cross-section measurements. Energies as large as $E_{\rm{cm}}$, i.e.\ $10$~TeV or more, can instead be probed by $2\to2$ direct production processes. The sensitivity to growing-with-energy new physics effects in these processes is thus incomparably superior to the one of EW-scale high-rate probes. This has been demonstrated in Section~\ref{highen} for the SILH operators ${\mathcal{O}_W}$ and ${\mathcal{O}_B}$. Several analysis strategies of increasing level of sophistication have been presented in Section~\ref{highen}. The simplest approach, in Sections~\ref{zh} and~\ref{subsec:Fiducial-cross-section-analysis}, is to rely on total or fiducial cross-sections. If it was possible to engineer polarized lepton beams, we have shown that this would be sufficient to obtain a satisfactory simultaneous determination of the two operator Wilson coefficients, even with a moderate ($30\%$) degree of polarization. Otherwise, improvements might come from studying highly differential cross-sections as in Section~\ref{subsec:Differential-analyses}, or tri-boson processes like in Section~\ref{subleading-processes}. 

The relevance of the tri-boson processes outlines an important aspect and marks a methodological difference between the phenomenology of the VHEL and the one of lepton colliders of lower energy like CLIC. Even at the highest available CLIC energy of $3$~TeV, there is a considerable gap between the tri-boson and the diboson hard production cross-sections. Indeed, while some tri-boson processes have very large total cross-section at CLIC~\cite{deBlas:2018mhx}, this is due to ``soft'' kinematical regions where all bosons are emitted parallel to the beam axis. This kinematical regime is hardly observable and uninteresting as a probe of short-distance physics. If restricted to the region where some of the bosons are emitted with high energy and in the central region, instead, tri-boson production is subdominant at CLIC. On the contrary, at the VHEL the tri-boson hard cross-section is sizable, due to the double-logarithmic Sudakov enhancement of soft and soft-collinear vector bosons emissions from the initial leptons and from the final bosons. In the resulting kinematical configuration, one boson is soft, potentially close to the beam axis and difficult to detect. The others are instead central and energetic and they bring valuable information on short-distance physics. We discussed in Section~\ref{subleading-processes} how the charged amplitude $\ell\nu\to Wh$ can be probed by exploiting this mechanism. 
This outlines a new handle for new physics exploration at the VHEL, but also the need of revisiting the analysis of Section~\ref{highen} taking the massive vector boson radiation fully into account. Indeed since there is no sizable gap in cross-section, the diboson processes cannot be studied separately from the tri-boson ones. Tri-boson production should be included for the estimate of both the SM background and the new physics signal. Furthermore, it should be noted that the Sudakov enhancement of the real emissions also controls the virtual loop corrections. Therefore EW loop corrections are expected to be sizable and to affect the tree-level cross-section predictions we relied on in our study, possibly at order one. Logarithmically enhanced virtual corrections are known up to two-loops order~\cite{Denner:2006jr}, and it is relatively easy to simulate one (or possibly two) real emission at tree-level. Including these effects,  providing a more refined sensitivity estimate than the one presented here, is thus possible with current technologies. However the situation is different if we consider the accuracy of the theoretical predictions that will be needed in order to concretely exploit the VHEL measurement potential. If NLO corrections are sizable, even if somewhat smaller than order one, higher order effects are probably needed to bring the theoretical uncertainties to percent or sub-percent level. This calls for developing new calculation and simulation tools to model the real and virtual radiation of massive vector bosons accurately. EW PDFs will definitely play a role in this context~\cite{Han:2020uid}, however it should be taken into account that the soft-collinear region that is most relevant for high-energy tri-boson production is not modeled by PDFs.

\begin{figure}\begin{center}
\includegraphics[height=0.35\textwidth]{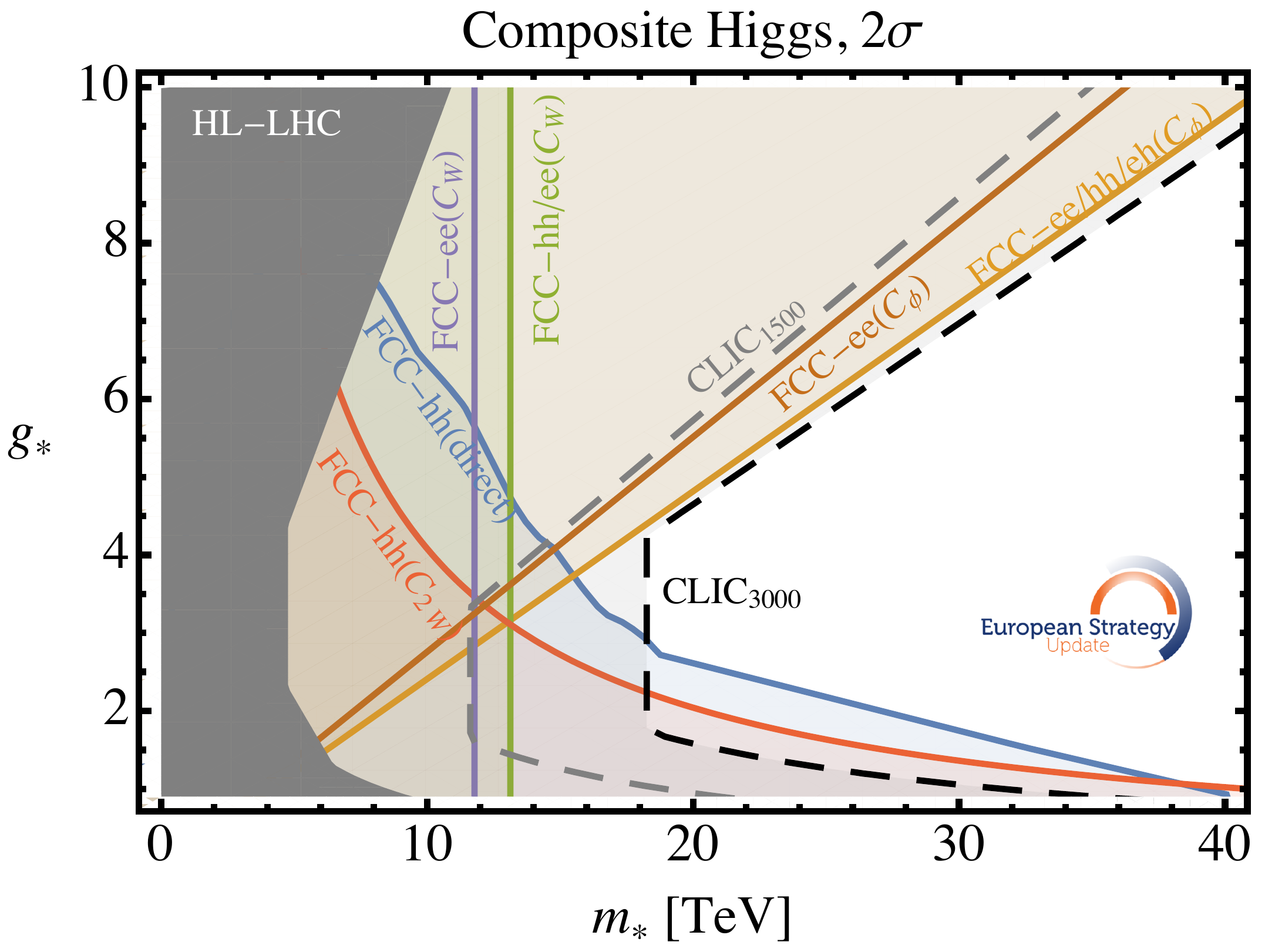}\hfill%
\includegraphics[height=0.35\textwidth]{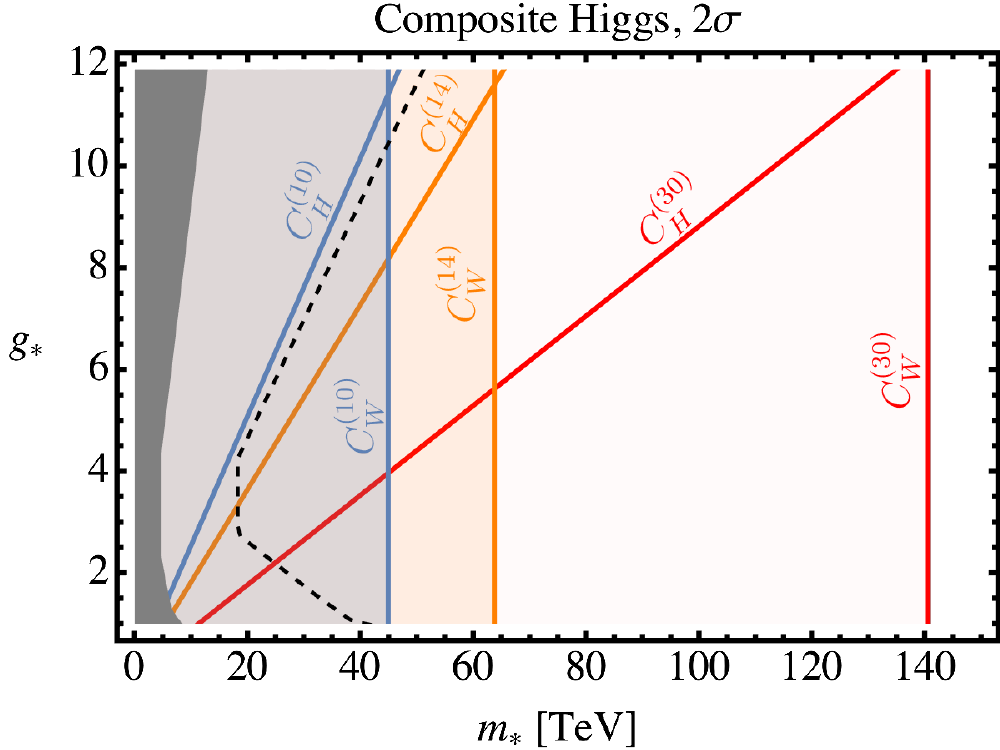}
\caption{{Left:} $2\sigma$ sensitivity of future collider projects to Higgs compositeness, from Ref.~\cite{Strategy:2019vxc}. Right: the VHEL projections based on the sensitivity to $C_H$ in Table~\ref{tab:hhhighmass} and to $C_W$ (single operator reach) from the ``Combined'' fit (which excludes polarized measurements) in Table~\ref{tab:CWCB}.\label{CH}}
\end{center}\end{figure}

We are clearly very far from a complete assessment of the VHEL precision potential to probe new physics. However some of our findings can already be used to quantify the VHEL reach on Higgs compositeness, as in Figure~\ref{CH}. The left panel shows the $2\sigma$ reach on Higgs compositeness at several future collider projects, in the $(m_*,g_*)$ plane~\cite{Strategy:2019vxc}. The parameter $m_*$ is the Higgs compositeness scale, i.e.\ the inverse of the geometric size of the Higgs particle. The parameter $g_*$ is the typical coupling strength of the composite sector the Higgs is part of. It ranges from above the Weak coupling to the maximal coupling $\sim4\pi$. The $\xi$ parameter previously mentioned is $\xi=g_*^2v^2/m_*^2$. The plot is obtained by comparing the future colliders sensitivity to several EFT operators with the estimate of the size of these operators in terms of $m_*$ and $g_*$. For the ${\mathcal{O}}_W$ operator, the estimate reads $C_W\simeq1/m_*$, while the ${\mathcal{O}}_H$ operator scales like $C_H=\xi/v^2=g_*^2/m_*^2$. The impact of other operators, as well as of direct searches at the FCChh for the resonances of the new composite sector, are also reported. The right panel of Figure~\ref{CH} shows the VHEL potential, based on the sensitivity estimate reported in this paper. The envelop of the reach of the other future colliders is displayed as a dashed line, while the one of the HL-LHC is in dark grey. Already at $10$~TeV, the VHEL can probe the scale of Higgs compositeness a factor $2$ better than the other colliders at intermediate $g_*$. The direct sensitivity to $C_H$ (i.e., $\xi$) from $hh$ production at $10$~TeV is somewhat inferior to the indirect one of Higgs factories such as CLIC and FCCee. However this little gap could be bridged by accurate measurements of the Higgs couplings as previously mentioned. At $14$ and at $30$~TeV, the VHEL potential on Higgs compositeness exceeds the one of the other future collider projects even more significantly.

The VHEL can probe Higgs compositeness scales in the ballpark of many tens of TeV. This might confirm the SM point-like nature of the Higgs way beyond current knowledge, or it might reveal Higgs compositeness at scales that are way too high to be probed directly. The evidence for Higgs compositeness in this case will be indirect, but still robust and easy to interpret. 
Indeed, its manifestation 
in high-energy probes will be a 
sizable correction to the SM predictions,
with several peculiar features among which the quadratic energy growth that could also be tested with a scan in energy. New physics characterization would be arguably easier for high-energy probes than for regular precisions tests. Finally, if the Higgs compositeness scale is $10$~TeV or less, its direct signatures will be discovered at the VHEL and the precision probes described in this paper will provide handles for the characterization of the newly discovered Composite~Sector.

\section*{Acknowledgments:}
We acknowledge support from the Swiss National Science Foundation under contract 200021-178999, and by the MIUR under contracts 2017FMJFMW and 2017L5W2PT (PRIN2017). RF acknowledges support by ``Programma per Giovani Ricercatori Rita
Levi Montalcini'' granted by Ministero dell'Istruzione, dell'Universit\`a 
e della Ricerca (MIUR). The research of RF was supported in part by
the National Science Foundation under Grant No. NSF PHY-1748958. The research of DB is supported in part by the PRIN 2017L5W2PT and the INFN grant FLAVOR. RF is grateful to the New York University and UC Santa Barbara KITP for
hospitality and support during the stages of this work. DB is grateful to CERN for hospitality. Computations
in this work have been carried out using free software including~\cite{Tange2011a,Hunter:2007,10.7717/peerj-cs.103,van_der_Walt_2011,2020SciPy-NMeth}.

\appendix

\section[Diboson Likelihood contours at ${\mathbf{14}}$ and ${\mathbf{30}}$~TeV]{Diboson Likelihood contours at ${\mathbf{14}}$ and ${\mathbf{30}}$~TeV}\label{appen1}

\begin{figure}[h]
\begin{centering}
\renewcommand{\subfigcapmargin}{22pt}
\subfigure[Inclusive $Zh$ (red) and fiducial $WW$ (blue) rates for unpolarized beams;]{\includegraphics[clip,width=0.49\textwidth]{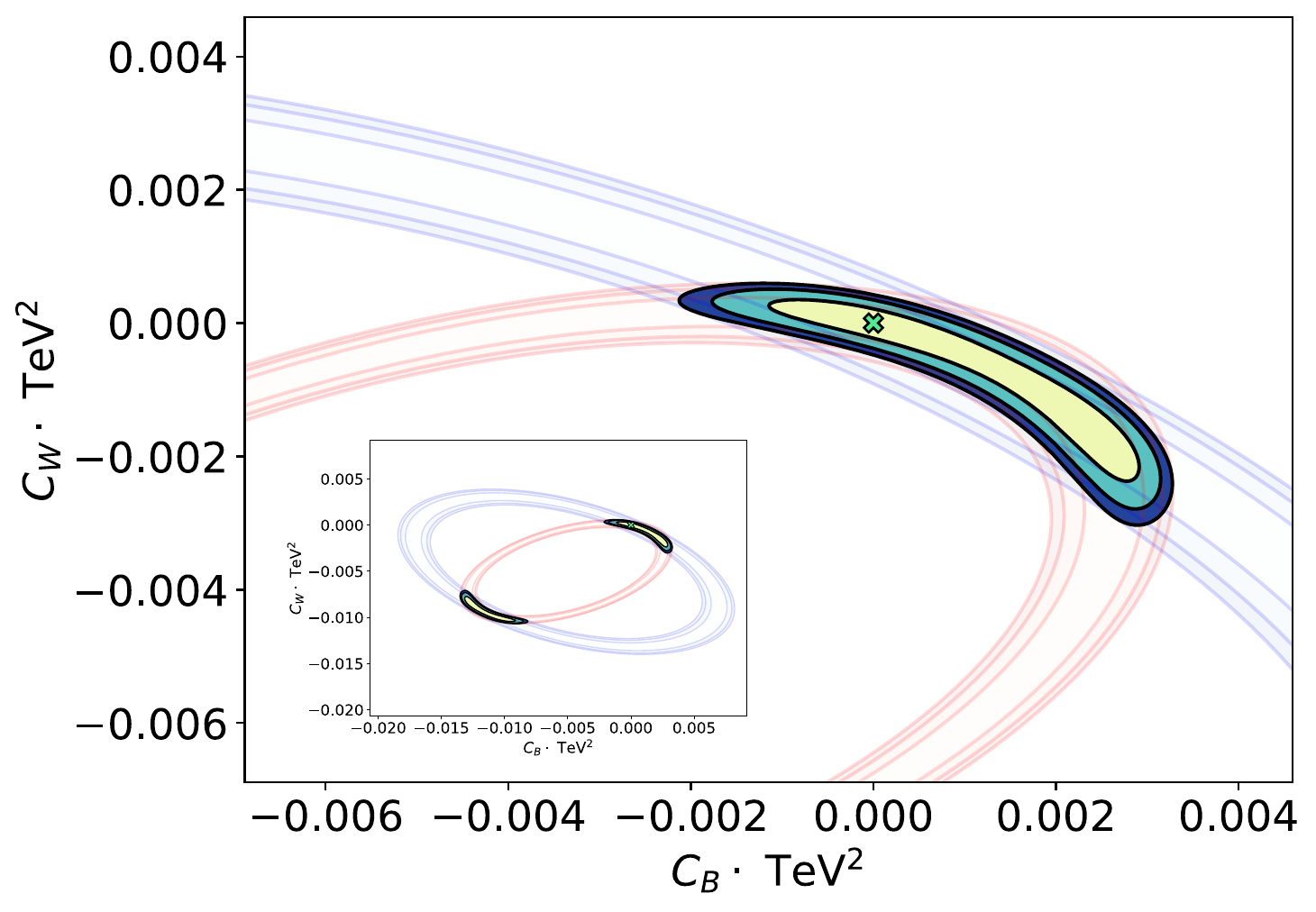}}
\hfill
\subfigure[Polarized inclusive $Zh$ (L: red, R: orange) and fiducial $WW$ (L: blue, R: purple);]{\includegraphics[clip,width=0.49\textwidth]{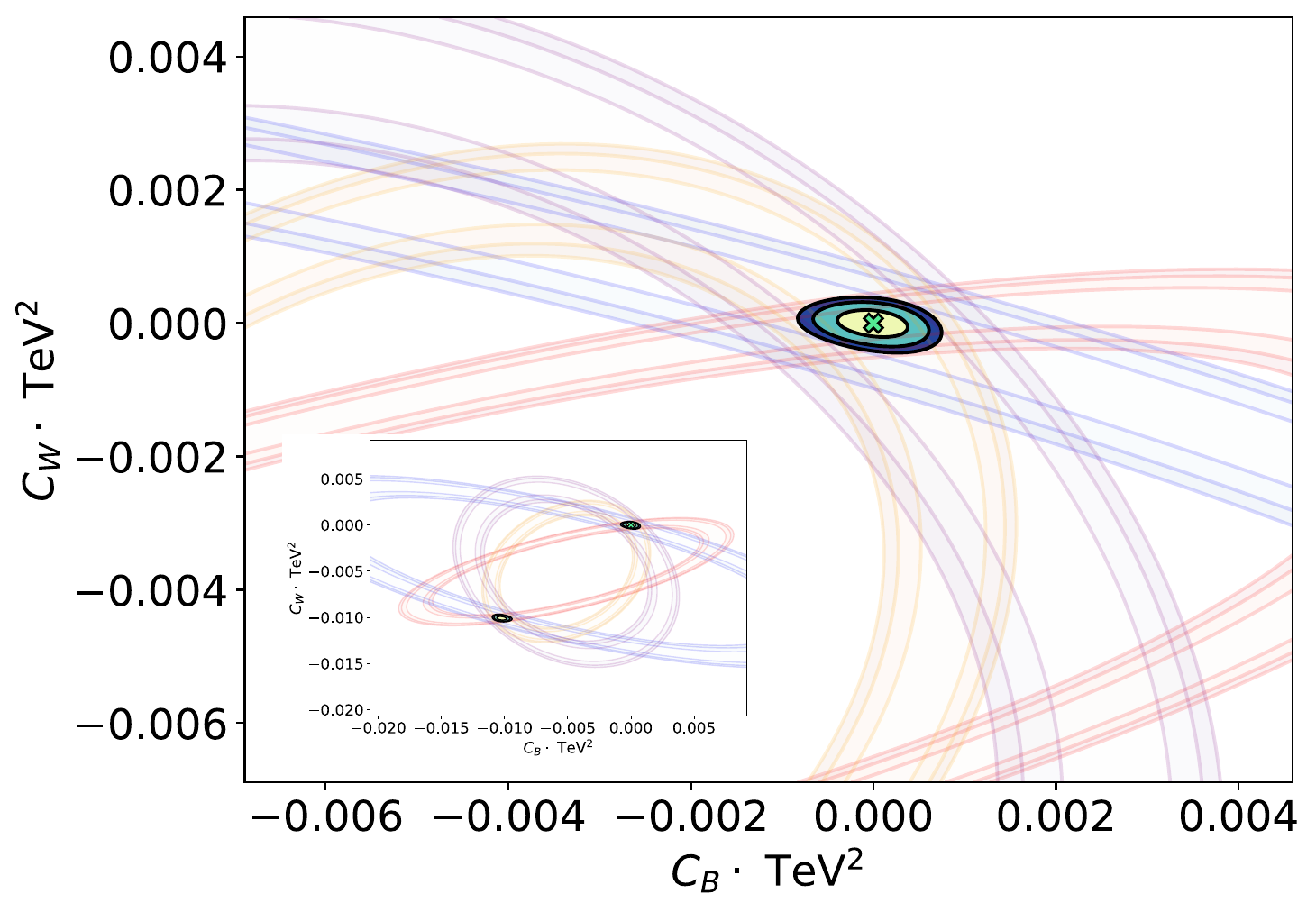}}\\
\subfigure[Same as panel (a), but with differential $WW$ rate (blue) for unpolarized beams.]{\includegraphics[clip,width=0.49\textwidth]{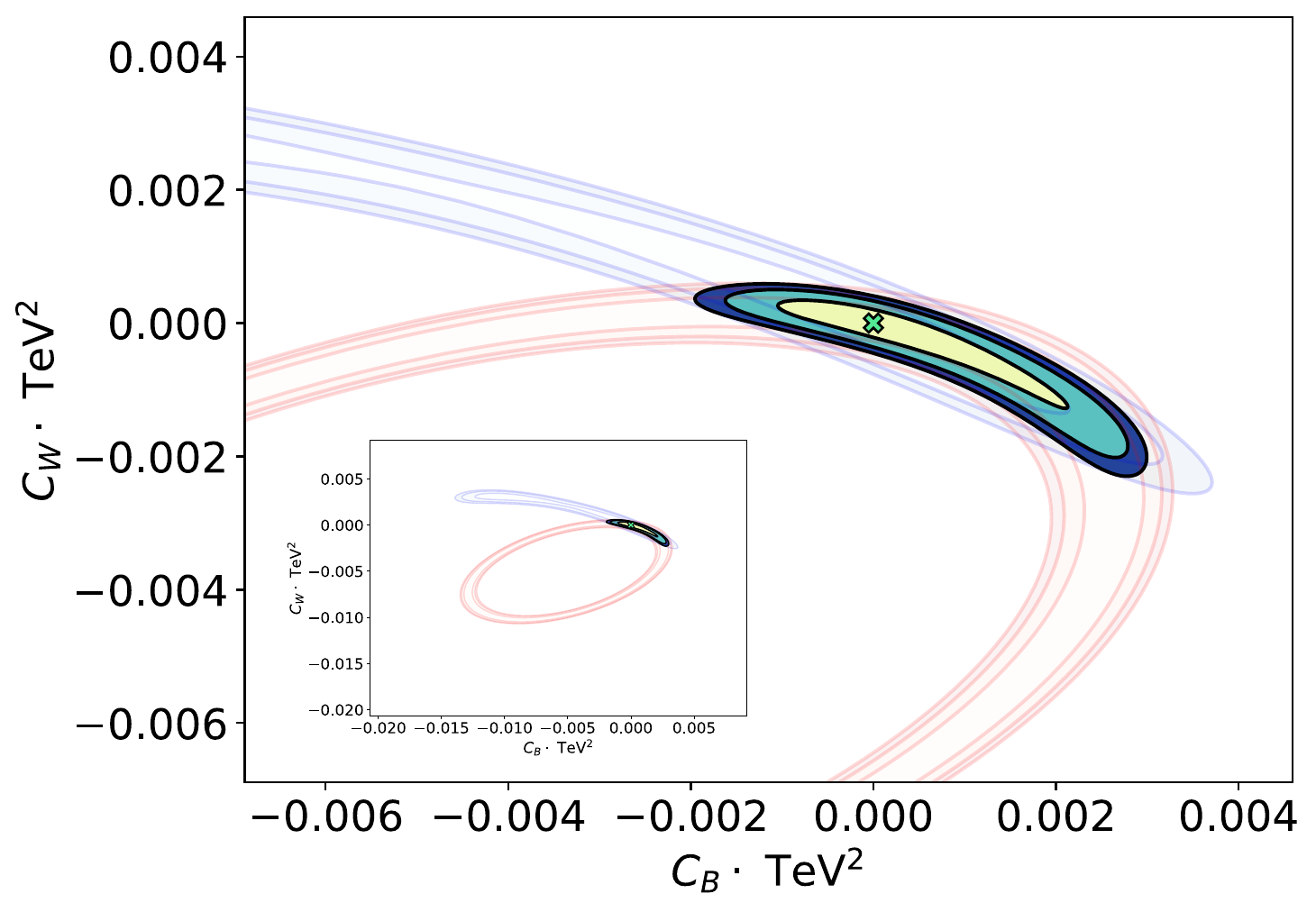}}
\hfill
\subfigure[Same as panel (a), combined with fiducial $WWh$ (green) for unpolarized beams;]{\includegraphics[clip,width=0.49\textwidth]{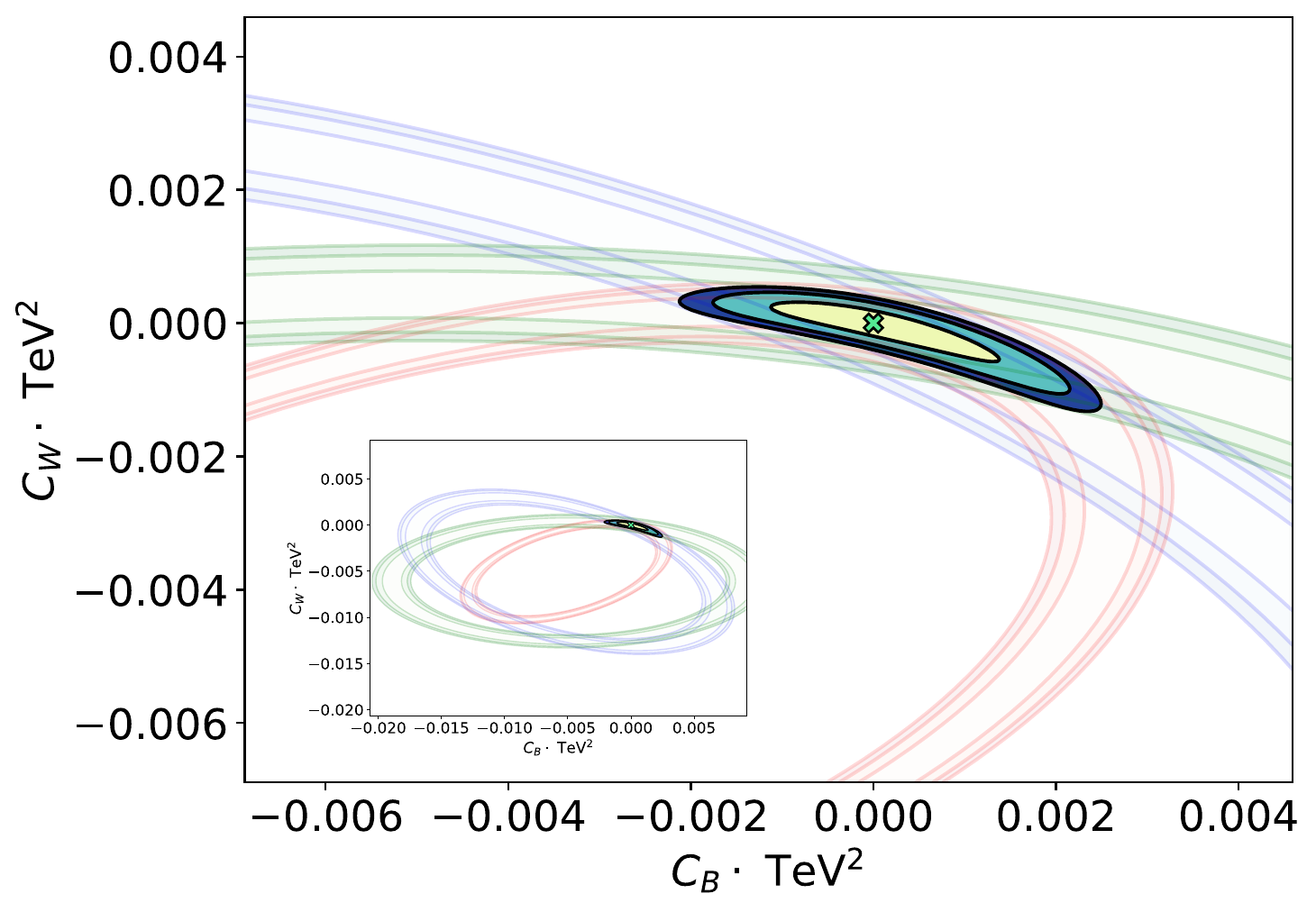}}
\par\end{centering}
\caption{Same as Figure~\ref{fig:chi2-profiles-cBcW}, but for $E_{\rm{cm}}=14$~TeV.\label{APP14}}
\end{figure}

\begin{figure}[t]
\begin{centering}
\renewcommand{\subfigcapmargin}{22pt}
\subfigure[Inclusive $Zh$ (red) and fiducial $WW$ (blue) rates for unpolarized beams;]{\includegraphics[clip,width=0.49\textwidth]{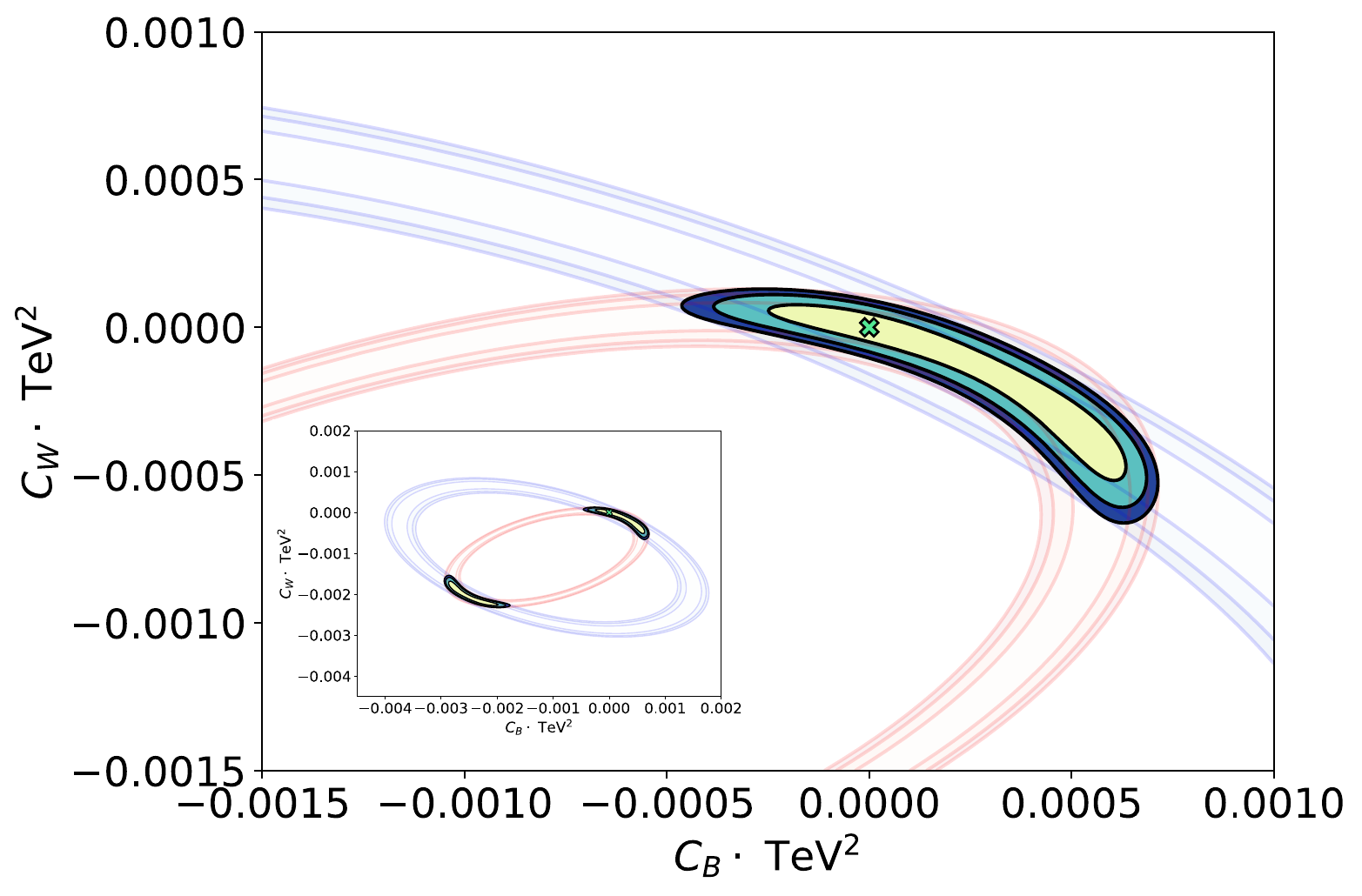}}
\hfill
\subfigure[Polarized inclusive $Zh$ (L: red, R: orange) and fiducial $WW$ (L: blue, R: purple);]{\includegraphics[clip,width=0.49\textwidth]{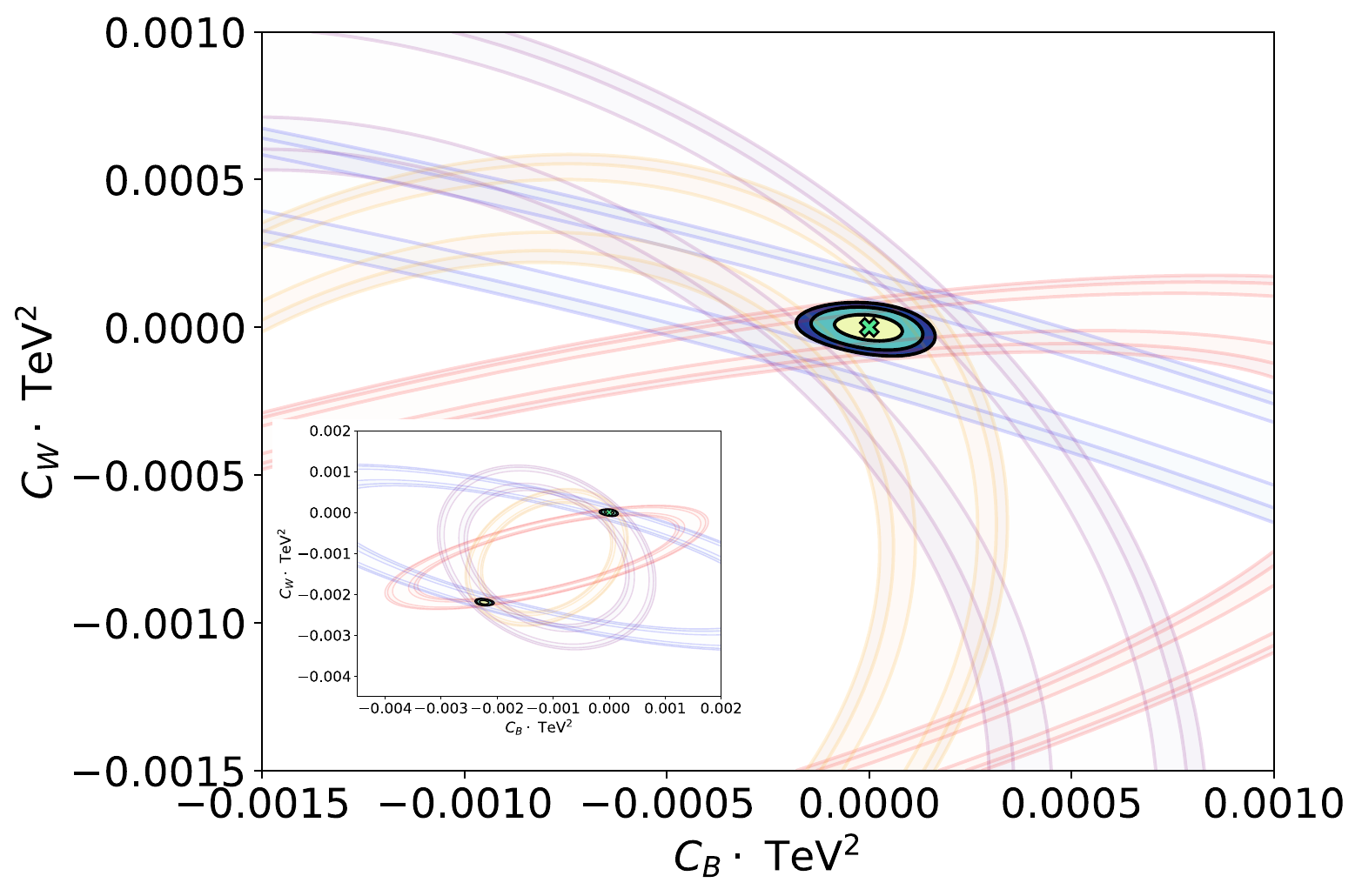}}\\
\subfigure[Same as panel (a), but with differential $WW$ rate (blue) for unpolarized beams.]{\includegraphics[clip,width=0.49\textwidth]{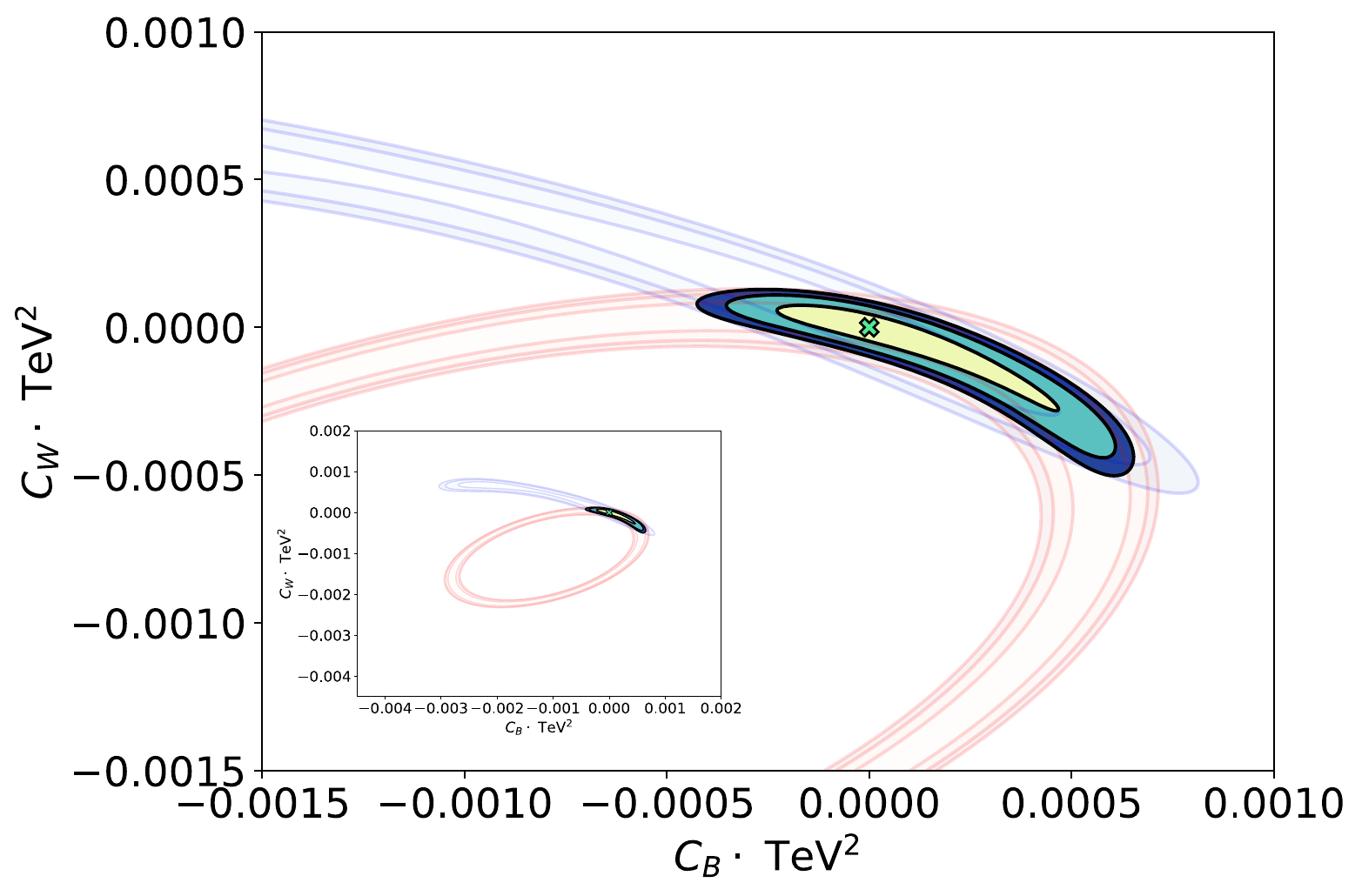}}\hfill
\subfigure[Same as panel (a), combined with fiducial $WWh$ (green) for unpolarized beams;]{\includegraphics[clip,width=0.49\textwidth]{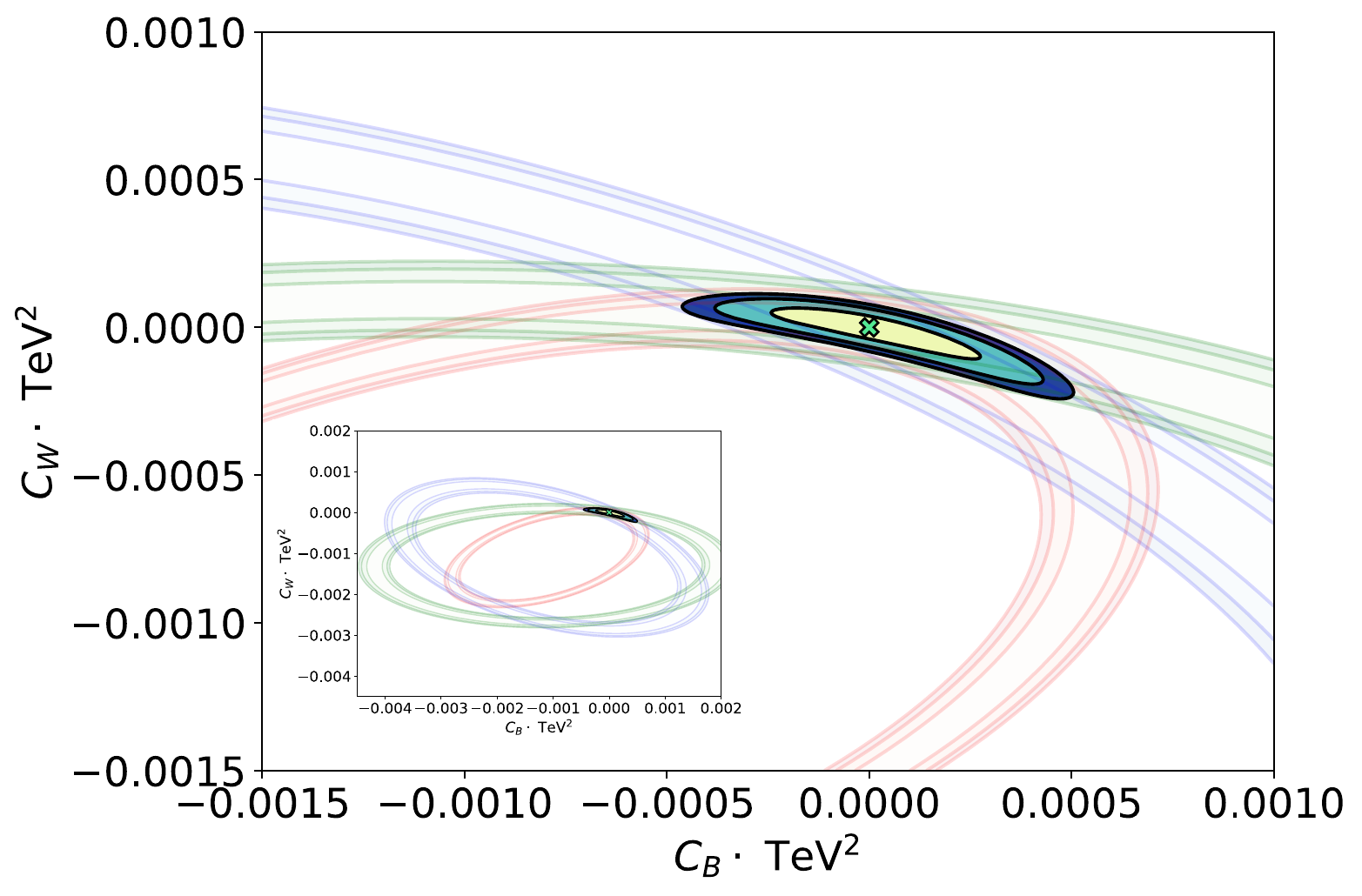}}
\par\end{centering}
\caption{Same as Figure~\ref{fig:chi2-profiles-cBcW}, but for $E_{\rm{cm}}=30$~TeV.\label{APP30}}
\end{figure}

\section{More details on double Higgs production}\label{appen2}

\begin{figure}[b!]
\includegraphics[height=0.3\textwidth]{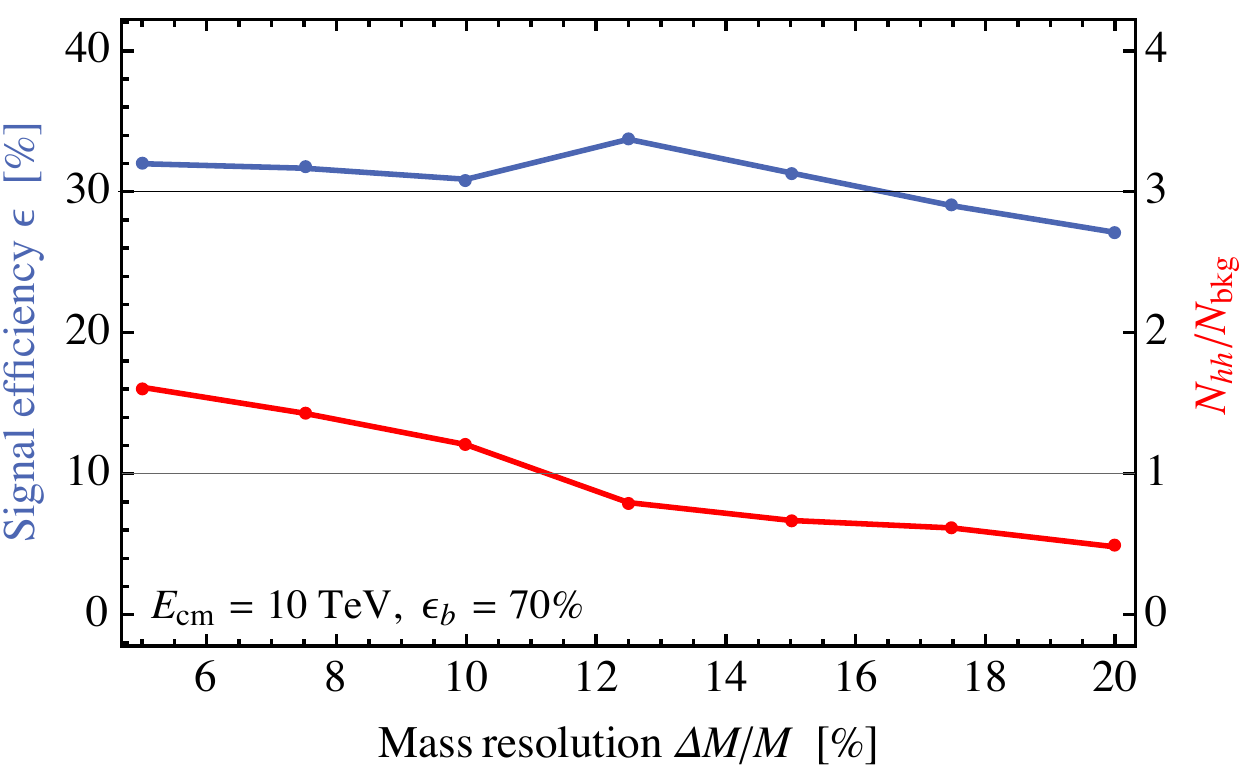}\hfill%
\includegraphics[height=0.3\textwidth]{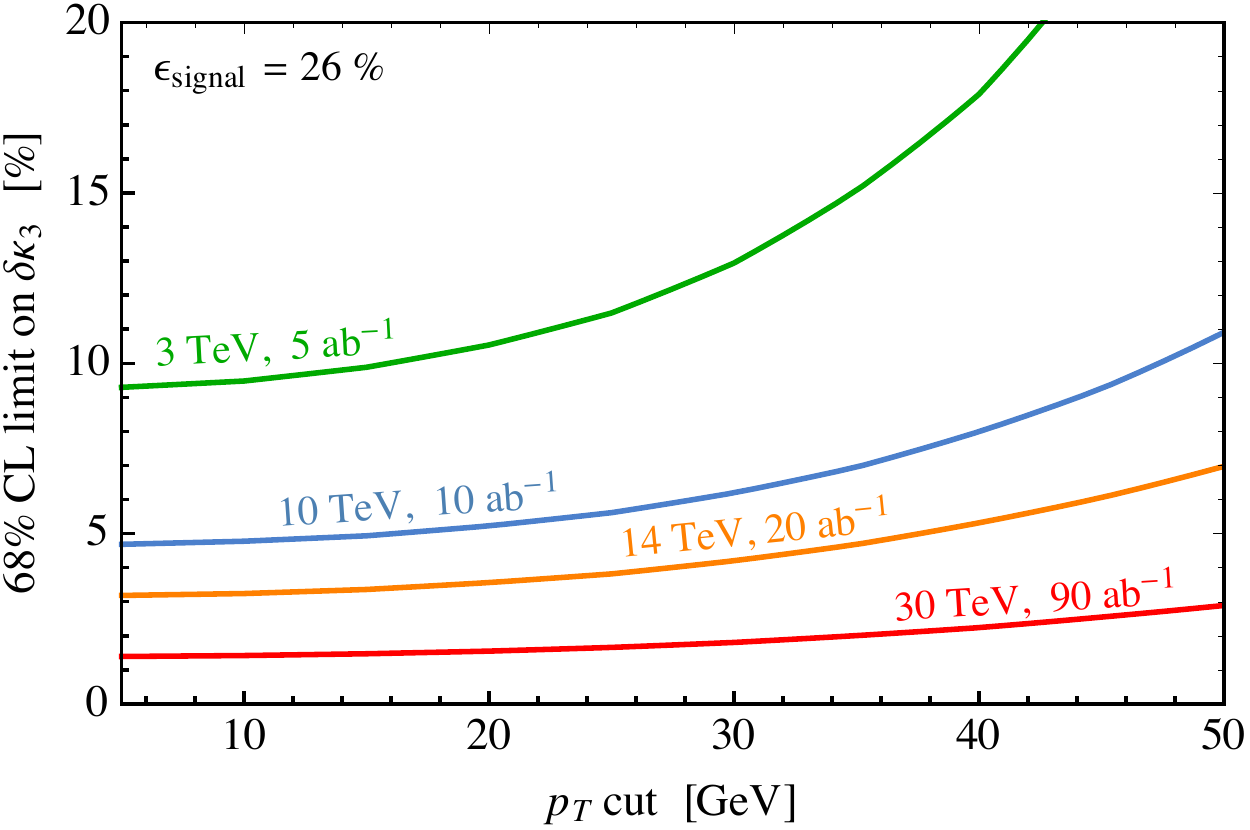}
\caption{\label{fig:hhApp} Left: $hh$ signal efficiency and ratio of signal over background events, as functions of the di-jet mass resolution, for $E_{{\rm cm}} = 10\,{\rm TeV}$, and $\epsilon_b = 70\%$. Right: 68\% C.L.\ limit on $\dl$ as a function of the acceptance cut on the jet transverse momentum, for different VHEL energies.}
\end{figure}

In Table~\ref{tab:hhcuts} we show the values of the di-jet invariant mass cut and the number of $b$-tags that optimize the significance of the $\mu^+\mu^- \to hh\nu\bar\nu$ signal. We assume a $b$-tag efficiency $\epsilon_b = 70\%$, with charm- and light-jet rejection as in~\cite{Abramowicz:2018rjq}. We also report the number of events for the signal and the dominant backgrounds from VBF di-boson productions.
The left panel of Figure~\ref{fig:hhApp} shows the dependence of the signal efficiency $\epsilon_{\rm sig}$ and of the ratio of signal over background events on the di-jet invariant mass resolution $\Delta M_{jj}$. Notice that the efficiency depends weakly on the mass resolution.
We also repeat the analysis by allowing the $b$-tag efficiency to vary,  following the curve of Ref.~\cite{Abramowicz:2018rjq}. We find that a higher significance for the $hh$ signal can be attained by choosing a looser $b$-tag working-point with $\epsilon_b \approx 90\%$. At $E_{\rm cm} = 10\,{\rm TeV}$ the corresponding total signal efficiency is $\epsilon_{\rm sig} \approx 50\%$, as shown in Table~\ref{tab:hhcuts} below.

\begin{table}[h!]
\centering%
\begin{tabular}{|c|c|c||c|c|c|c|c|c|}
\hline
$N_b$ & $M_{jj}^{\rm cut}$ & $\epsilon_{\rm sig}$ & $N_{hh}$ & $N_{Zh}$ & $N_{ZZ}$ & $N_{Wh}$ & $N_{WZ}$ & $N_{WW}$\\
\hline
3 & 106 GeV & 32\% & 1369 & 546 & 451 & 261 & 95 & 1\\
\hline
\hline
$\epsilon_b$ & $M_{jj}^{\rm cut}$ & $\epsilon_{\rm sig}$ & $N_{hh}$ & $N_{Zh}$ & $N_{ZZ}$ & $N_{Wh}$ & $N_{WZ}$ & $N_{WW}$\\
\hline
87\% & 106 GeV & 55\% & 2302 & 1016 & 933 & 334 & 146 & 10\\
\hline
\end{tabular}
\caption{\label{tab:hhcuts} Top: Optimal di-jet invariant mass cut and number of $b$-tags 
for $E_{\rm cm} = 10$~TeV, with $\epsilon_b = 70\%$ and $\Delta M_{jj}/M_{jj} = 10\%$. Bottom: the same, but requiring $N_b = 4$ and varying $\epsilon_b$.}
\end{table}

Finally, the right panel of Figure~\ref{fig:hhApp} shows how the reach on the trilinear coupling $\dl$ depends on the minimum jet $p_T$. The sensitivity to $\dl$ is not significantly affected as long as one is able to reconstruct jets with $p_T \approx 30$\,--\,$40$~GeV. Notice that for energetic jets the $p_T$ dependence is correlated with the $\theta_{\rm jet}$ dependence shown in Figure~\ref{fig:angular_lambda}.


\begin{figure}[t!]
\begin{centering}
\includegraphics[height=0.325\linewidth]{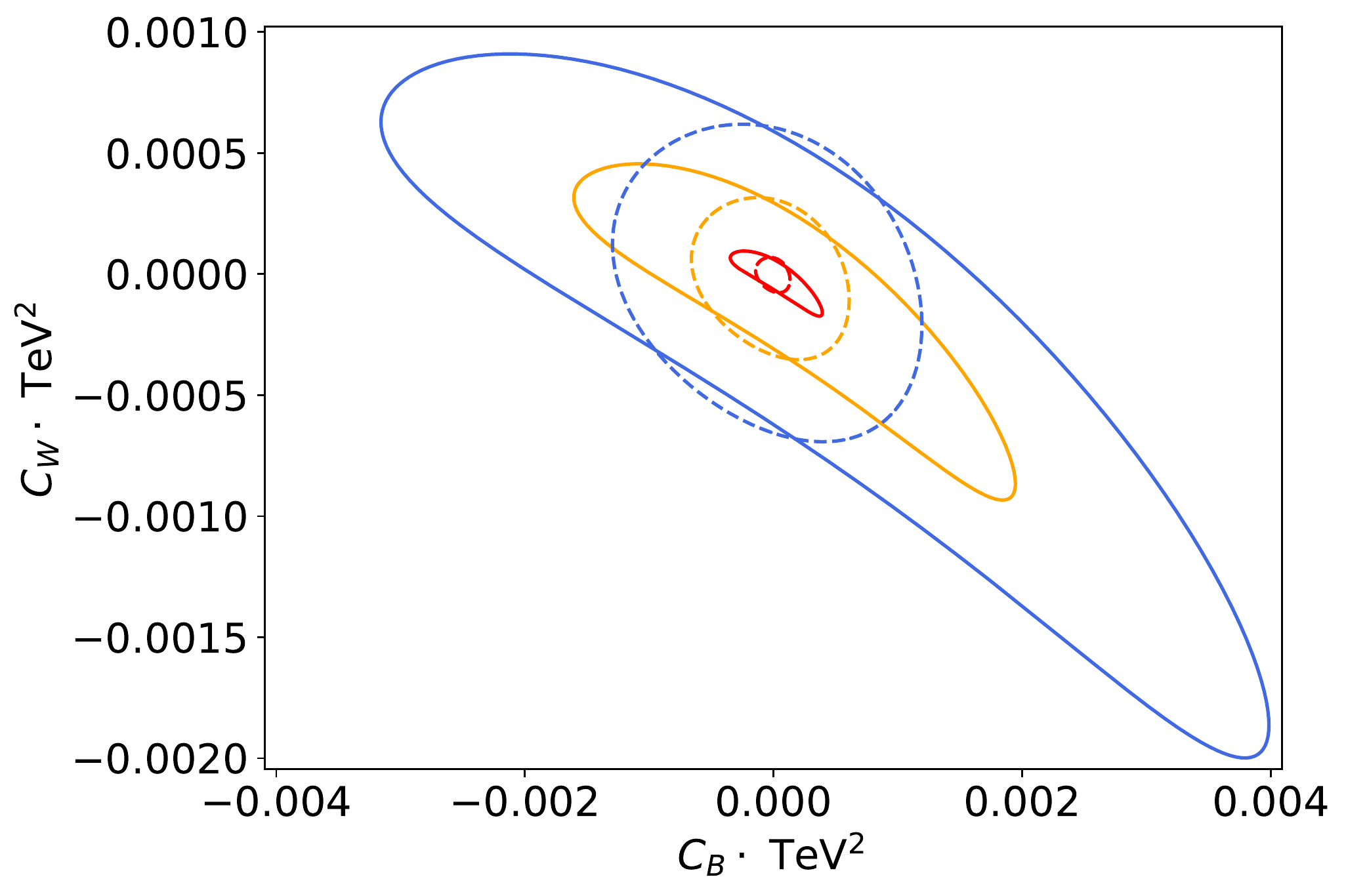}%
\includegraphics[height=0.322\linewidth]{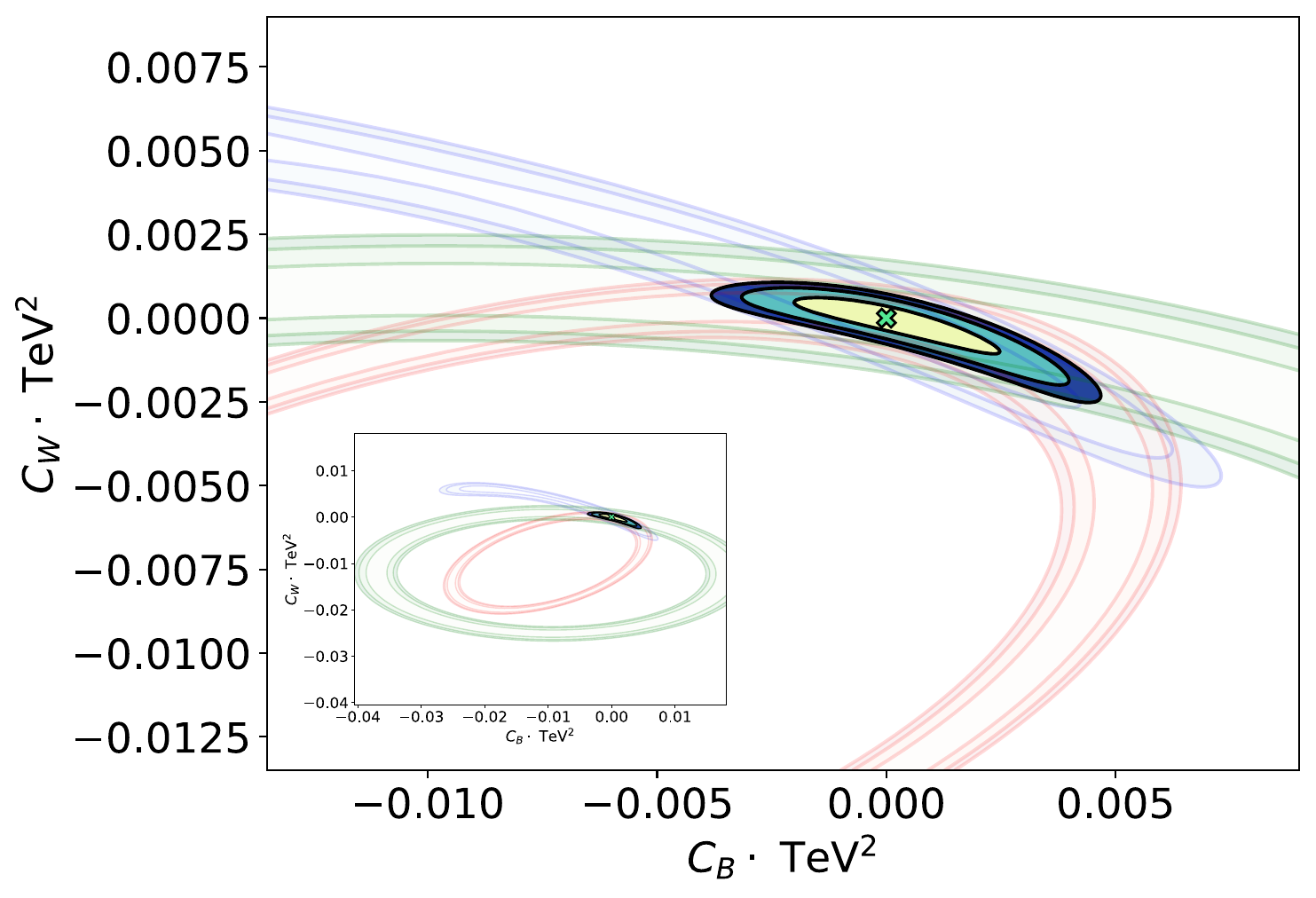}\hfill
\par\end{centering}
\begin{centering}
\hspace{13pt}\includegraphics[height=0.32\linewidth]{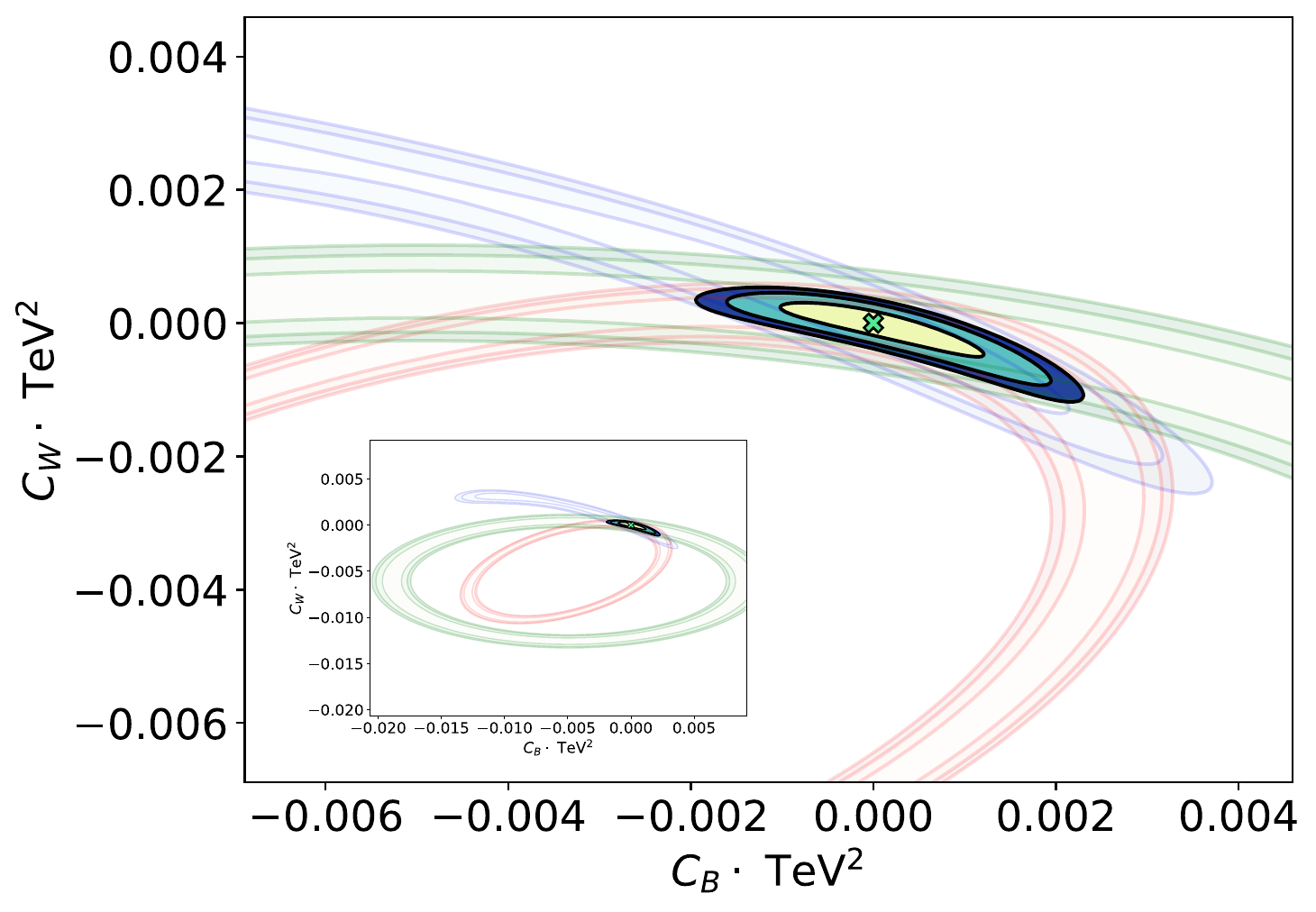}\hfill
\includegraphics[height=0.326\linewidth]{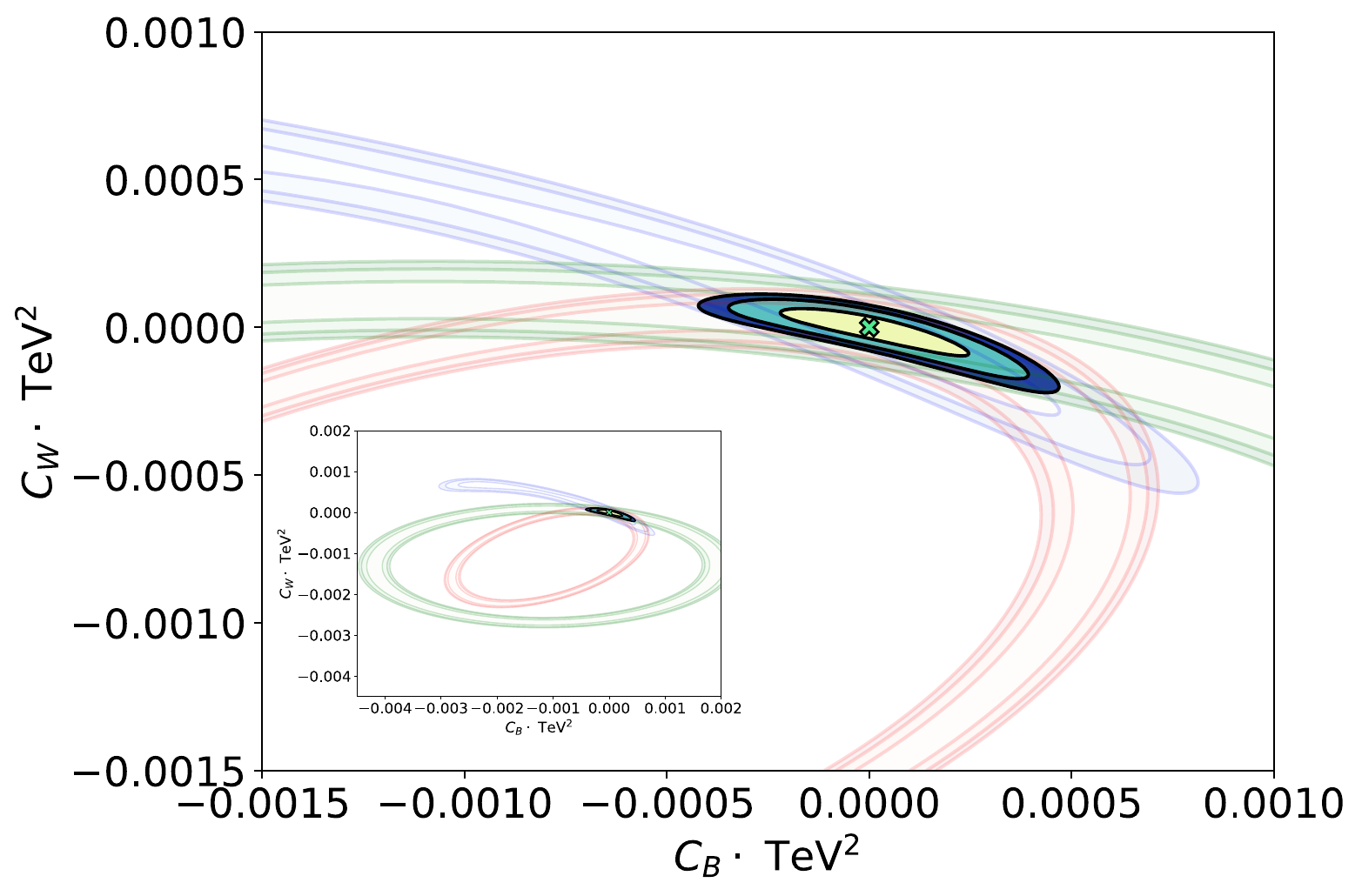}
\par\end{centering}
\caption{\label{occhio} 
{{Top-Left:}}
Constraints at $95\%$~C.L. from diboson and tri-bosons production, for $E_{\rm{cm}}=10$~TeV (blue), $E_{\rm{cm}}=14$~TeV (orange), and $E_{\rm{cm}}=30$~TeV (red), obtained from polarized inclusive $Zh$ and fiducial $WW$ measurements (dashed lines), and from the combination of inclusive $Zh$, differential $WW$ and fiducial $WWh$ rate with unpolarized beams (solid lines). The likelihood contours and the individual contributions to the likelihood from $Zh$ (red), differential $WW$ (blue) and $WWh$ (green) are shown for each collider energy in the other panels.
}
\end{figure}

\bibliographystyle{JHEP}
\newpage

\bibliography{bibliography}

\end{document}